\documentclass[useAMS,usenatbib]{mnras}
\usepackage{natbib}
\usepackage[english]{babel}
\usepackage{graphicx,amsmath,amsfonts,amssymb}
\usepackage{colortbl}
\usepackage{tabu,booktabs}
\usepackage{xcolor}
\usepackage{xspace}

\usepackage{enumerate}
\usepackage{listings}
\usepackage{subfig}
\usepackage[justification=centering]{caption}

\usepackage{gensymb,fontawesome}
\definecolor{linkcolor}{rgb}{0.1216,0.4667,0.7059}
\newcommand{\codeicon}{{\faCloudDownload}}
\newcommand{\codelink}{\href{https://github.com/TobiBu/picassso}{\faGithub}\,\,}
\newcommand{\plotlink}[1]{\href{https://github.com/TobiBu/picassso/blob/master/paper_analysis/#1.ipynb}{\codeicon}\,\,}
\newcommand{\oscaption}[2]{\caption{#2 \plotlink{#1}}}

\newcommand{\plotlinkpy}[1]{\href{https://github.com/TobiBu/picassso/blob/master/paper_analysis/#1.py}{\codeicon}\,\,}
\newcommand{\oscaptionpy}[2]{\caption{#2 \plotlinkpy{#1}}}

\newcommand{\plotlinknet}{\href{https://github.com/Steffen-Wolf/picasso_training}{\codeicon}\,\,}
\newcommand{\oscaptionnet}[1]{\caption{#1 \plotlinknet}}

\def \etal {et~al.~}

\newcommand{\hMpc}{{\ifmmode{h^{-1}{\rm Mpc}}\else{$h^{-1}$Mpc}\fi}}
\newcommand{\hkpc}{{\ifmmode{h^{-1}{\rm kpc}}\else{$h^{-1}$kpc}\fi}}
\newcommand{\kpc}{{\ifmmode{ {\rm kpc} }\else{{\rm kpc}}\fi}}
\newcommand{\kms}{{\ifmmode{ {\rm km\,s^{-1}} }\else{ ${\rm km\,s^{-1}}$ }\fi}}
\newcommand{\hMsun}{{\ifmmode{h^{-1}{\rm {M_{\odot}}}}\else{$h^{-1}{\rm{M_{\odot}}}$}\fi}}
\newcommand{\Msun}{{\ifmmode{{\rm M}_{\odot}}\else{${\rm M}_{\odot}$}\fi}}
\newcommand{\Mhalo}{{\ifmmode{M_{\rm halo}}\else{$M_{\rm halo}$}\fi}}
\newcommand{\Rvir}{{\ifmmode{R_{\rm vir}}\else{$R_{\rm vir}$}\fi}}
\newcommand{\Mstar}{{\ifmmode{M_{\rm star}}\else{$M_{\rm star}$}\fi}}
\newcommand{\Vrot}{{\ifmmode{V_{\rm rot}}\else{$V_{\rm rot}$}\fi}}
\newcommand{\ltsima}{$\; \buildrel < \over \sim \;$}
\newcommand{\gtsima}{$\; \buildrel > \over \sim \;$}
\newcommand{\lsim}{\lower.5ex\hbox{\ltsima}}
\newcommand{\gsim}{\lower.5ex\hbox{\gtsima}}

\def\lesssim{\mathrel{\hbox{\rlap{\hbox{\lower4pt\hbox{$\sim$}}}\hbox{$<$}}}}
\def\gtrsim{\mathrel{\hbox{\rlap{\hbox{\lower4pt\hbox{$\sim$}}}\hbox{$>$}}}}

\newcommand{\beq}{\begin{equation}}
\newcommand{\eeq}{\end{equation}}
\def\beqa{\begin{eqnarray}}
\def\eeqa{\end{eqnarray}}
\def\LCDM{\ensuremath{\Lambda}CDM}

\def\head{ \vbox to 0pt{\vss \hbox to 0pt{\hskip 440pt\rm
      LA-UR-10-07069\hss} \vskip 25pt}}

\def \kms {\ifmmode  \,\rm km\,s^{-1} \else $\,\rm km\,s^{-1}  $ \fi }
\def \kpc {\ifmmode  {\rm kpc}  \else ${\rm  kpc}$ \fi  }  
\def \hkpc {\ifmmode  {h^{-1}\rm kpc}  \else ${h^{-1}\rm kpc}$ \fi  }  
\def \hMpc {\ifmmode  {h^{-1}\rm Mpc}  \else ${h^{-1}\rm Mpc}$ \fi  }  
\def \Mpch {\ifmmode  {h^{-1}\rm Mpc}  \else ${h^{-1}\rm Mpc}$ \fi  }  
\def \Msun {\ifmmode {\rm M}_{\odot} \else ${\rm M}_{\odot}$ \fi} 
\def \hMsun {\ifmmode h^{-1}\,\rm M_{\odot} \else $h^{-1}\,\rm M_{\odot}$ \fi}

\def \LCDM {\ifmmode \Lambda{\rm CDM} \else $\Lambda{\rm CDM}$ \fi}
\def \sig8 {\ifmmode \sigma_8 \else $\sigma_8$ \fi} 
\def \OmegaM {\ifmmode \Omega_{\rm m} \else $\Omega_{\rm m}$ \fi} 
\def \Omegab {\ifmmode \Omega_{\rm b} \else $\Omega_{\rm b}$ \fi} 
\def \OmegaL {\ifmmode \Omega_{\rm \Lambda} \else $\Omega_{\rm \Lambda}$\fi} 
\def \Deltavir {\ifmmode \Delta_{\rm vir} \else $\Delta_{\rm vir}$ \fi}
\def \rhocrit {\ifmmode \rho_{\rm crit} \else $\rho_{\rm crit}$ \fi}
\def \rhou {\ifmmode \rho_{\rm u} \else $\rho_{\rm u}$ \fi}
\def \zc {\ifmmode z_{\rm c} \else $z_{\rm c}$ \fi}

\def\aj{AJ}%
\def\apj{ApJ}%
\def\apjl{ApJ}%
\def\apjs{ApJS}%
\def\aap{A\&A}%
\def\jcap{J. Cosmology Astropart. Phys.}%
\def\mnras{MNRAS}%
\def\pasp{PASP}%
\def\ssr{Space~Sci.~Rev.}%
\def\nat{Nature}%
\def\head{ .ps \vbox to 0pt{\vss \hbox to 0pt{\hskip 440pt\rm
      LA-UR-10-07069\hss} \vskip 25pt}} 

\def \spose#1{\hbox  to 0pt{#1\hss}}  
\def \lta{\mathrel{\spose{\lower 3pt\hbox{$\sim$}}\raise 2.0pt\hbox{$<$}}}
\def \gta{\mathrel{\spose{\lower 3pt\hbox{$\sim$}}\raise 2.0pt\hbox{$>$}}}

\title[{\begin{minipage}{.15\textwidth}
\includegraphics[width=\textwidth]{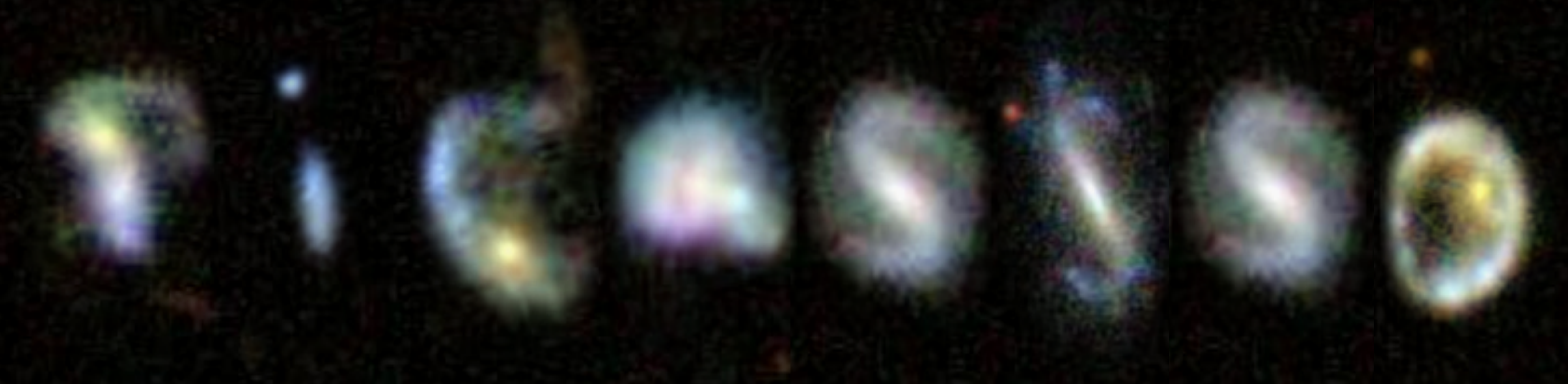}
\end{minipage}}]
{Predicting resolved galaxy properties from photometric images using convolutional neural networks}

\author[Buck \etal] {Tobias Buck$^{1}$\thanks{E-mail:
    tbuck@aip.de},
    Steffen Wolf$^{2}$
    \\
$^1$Leibniz-Institut f\"ur Astrophysik Potsdam (AIP), An der Sternwarte 16, D-14482 Potsdam, Germany\\
$^2$MRC Laboratory of Molecular Biology, Cambridge Biomedical Campus, Cambridge CB2 0QH, United Kingdom 
}

\begin{document}

\date{Accepted XXXX . Received XXXX; in original form XXXX}

\pagerange{\pageref{firstpage}--\pageref{lastpage}} \pubyear{2021}

\maketitle

\label{firstpage}


\begin{abstract}
Multi-band images of galaxies reveal a huge amount of information about their morphology and structure. However, inferring properties of the underlying stellar populations such as age, metallicity or kinematics from those images is notoriously difficult. Traditionally such information is best extracted from expensive spectroscopic observations. Here we present the \emph{Painting IntrinsiC Attributes onto SDSS Objects} (PICASSSO) project and test the information content of photometric multi-band images of galaxies. We train a convolutional neural network on 27,558 galaxy image pairs to establish a connection between broad-band images and the underlying physical stellar and gaseous galaxy property maps. We test our machine learning (ML) algorithm with SDSS \emph{ugriz} mock images for which uncertainties and systematics are exactly known. We show that multi-band galaxy images contain enough information to reconstruct 2d maps of stellar mass, metallicity, age and gas mass, metallicity as well as star formation rate. We recover the true stellar properties on a pixel by pixel basis with only little scatter, $\lesssim20\%$ compared to $\sim50\%$ statistical uncertainty from traditional mass-to-light-ratio based methods. We further test for any systematics of our algorithm with image resolution, training sample size or wavelength coverage. We find that galaxy morphology alone constrains stellar properties to better than $\sim20\%$ thus highlighting the benefits of including morphology into the parameter estimation. The machine learning approach can predict maps of high resolution, only limited by the resolution of the input bands, thus achieve higher resolution than IFU observations. 
The network architecture and all code is publicly available on GitHub \codelink.
\end{abstract}

\noindent
\begin{keywords}
  methods: statistical – 
  techniques: image processing -
  galaxies: abundances -
  galaxies: structure -
  galaxies: fundamental parameters -
  galaxies: photometry
\end{keywords}

\section{Introduction} \label{sec:intro}

The Sloan Digital Sky Survey \citep{SDSS} has revolutionised our understanding of the Universe by creating detailed three-dimensional maps of the Universe. Additionally, this survey provides deep multi-band/wavelength images of about one third of the sky, supplemented by spectra for more than three million astronomical objects. In particular, our understanding of galaxy formation and evolution was strongly impacted by the characterisations of the galaxy population either via the color bi-modality of galaxies \citep[e.g.][]{Strateva2001,Baldry2006,Wilman2010,Peng2010,Peng2012,Arora2021} which divides them into a blue star-forming cloud and a red, passively evolving sequence or via the galaxy mass-metallicity relation \citep{Tremonti2004}.

Large scale surveys such as the Dark Energy Survey \citep{DES2005}, EUCLID or the upcoming Large Synoptic Survey Telescope \citep{LSST2019} and the Nancy Grace Roman Telescope will super-cede the achievements of SDSS by scanning even larger areas of the sky and including even fainter objects. With these tremendous datasets, astronomy is entering the era of big (imaging) data. Spectroscopic follow-up observations for such huge datasets will become increasingly impractical and new, innovative analysis methods need to be explored in order to exploit the datasets in full depth.

One example of such an innovative analysis method has been explored in the crowed-sourced Galaxy Zoo project \citep{galaxyzoo2008} for the classical SDSS imaging dataset. In this citizen science project the public was presented with galaxy images and asked to classify them into different groups (edge-on spiral, face-on spiral, barred spiral, ellipticals, etc.). In case enough people participate, each galaxy will be classified by several individuals and a relatively robust classification can be achieved. One drawback of this approach is that it relies on the pattern recognition abilities of individual humans. In this way, no strong conclusions/quantification of the features that lead to the final decision can be made. And more importantly, even these concepts can not cope with the vastness of future datasets.

Fortunately, the field of machine learning (ML) has matured the tasks of image recognition, classification and segmentation \citep[e.g.][]{Ronneberger2015} and by now, several groups have tested ML techniques on astronomical data with great success. ML techniques have been applied in a large number of astronomical use-cases such as galaxy morphology classification \citep[e.g.][]{Dieleman2015,Huertas-Company2015,Beck2018,Hocking2018}, gravitational lensing analysis \citep[e.g.][]{Hezaveh2017,Lanusse2018,Petrillo2017,Petrillo2019,Jeffrey2020}, galaxy cluster mass estimates \citep[e.g.][]{Ntampaka2015,Ntampaka2018,Ntampaka2019,Ho2019} or cold gas kinematics \citep{Dawson2020}. Furthermore, ML has proven useful for star-galaxy separation \citep[e.g.][]{Kim2017,Bai2019}, galaxy de-blending \citep{Lanusse2019}, super resolution imaging \citep[e.g.][]{Falahkheirkhah2019} and outlier detection \citep{Margalef2020}. Unsupervised learning methods have been used to define and study kinematic structures of galaxies \citep{Domenech-Moral2012,Obreja2018,Obreja2019,Buck2018,Buck2019b} or identify accreted stars from disrupted satellites of the MW \citep[e.g.][]{Buder2021b} in the GALAH survey data \citep{Buder2021}. Hierarchical clustering on the other hand was employed to investigate the connection between stellar birth radii and their abundances \citep[e.g.][]{Ratcliffe2020,Ratcliffe2021}.
ML has further been used to estimate galaxy redshifts \citep{Soo2018,Menou2019,Wilson2020,Campagne2020}, to reconstruct star cluster parameters \citep{Pasquato2016,Pasquato2019} or galaxy merger states \citep{Bottrell2019}, to identify asteroids \citep[e.g.][]{Smirnov2017}, or to transfer knowledge between different galaxy surveys \citep{Dominguez2019,Perez2019}. It has been successfully used to investigate the star formation and quenching processes \citep{Lovell2019,Bluck2020}, estimate cooling rates in cosmological simulations \citep{Galligan2019} or interpret the turbulent inter-stellar medium \citep[ISM,][]{Peek2019,vanOort2019,Buck2019}. Other works have explored dark matter halo formation \citep{Lucie-Smith2019}, created galaxy mock catalogs \citep[e.g.][]{Xu2013,Kamdar2016}, inferred the halo mass distribution function \citep{Charnock2020} or predicted the galaxy-halo connection \citep{Agarwal2018,Jo2019} using ML techniques. Finally, using ML cosmological parameters could be derived from redshift surveys \citep{Ramanah2019,Ntampaka2020}, supernova data \citep{Escamilla2020,Wang2020,Boone2021} or maps of e.g. gas and stars  \citep{Villaescusa-Navarro2021}.
	
From observations it is well established that galaxy color is a good predictor for its stellar mass \citep[e.g.][]{Bell2003,Zibetti2009} and its stellar mass in turn correlates well with the total stellar or gaseous metallicity \citep[e.g.][]{Tremonti2004,Gallazzi2005}. Combining this with the color or morphological bi-modality of spirals vs. ellipticals then implies that there might also exist a correlation between morphology, color and fundamental properties such as stellar mass, gas mass, SFR or metallicity. An immediate question arising from this is: \textit{Can we estimate galaxy properties typically derived from spectroscopy using broad band photometry? And how much more accurate can stellar mass estimates based on color get if we include morphological galaxy properties?}  

Recently, such questions have been studied by \citet{Bonjean2019} using integrated UV, optical and IR luminosities to derive star formation rate (SFR) estimates via a random forrest algorithm and by \citet{Dobbels2019} who investigated a morphology assisted mass-to-light ratio estimate via deep learning methods. \citet{Wu2019} on the other hand used RGB 3-color images from SDSS to infer the spectroscopically derived metallicity solely from imaging data using neural networks. And \citet{Diego2021} combine morphological classification and photometric redshift determination using ML techniques to prevent catastrophic outliers.

In this work we take previous successes of ML techniques in image processing and classification one step further. Whereas previous astronomical studies looked at tasks such as either classifying galaxies \citep[e.g.][]{Dominguez2018} or connecting three-color images and scalar values such as total gas phase metallicity \citep[e.g.][]{Wu2019} or redshift. Here we propose to use ML techniques to establish a connection between astronomical broad band imaging data and the underlying resolved physical properties on a pixel-by-pixel basis. This means, we propose to train a convolutional neural network (CNN) to \textit{predict} resolved maps of physical properties such as surface mass density, metallicity or star formation rate (SFR) solely from photometric images. This enables to obtain resolved maps of galaxy properties from relatively cheap photometric images otherwise only possible to derive from expensive integral field unit (IFU) spectroscopic surveys such as SDSS-MANGA \citep{manga2015}, SAMI \citep{sami2015} or CALIFA \citep{califa2012}.

The paper is organised as follows: In \S2 we present a short summary of the data set used, SDSS mock images created for model galaxies from the Illustris simulation. In \S3 we describe the neural network and the training procedure and proceed to present the main results in \S4. In this section we further discuss the accuracy and possibilities of our method. Section \S5 describes how the methods presented here can be applied to observational datasets and we conclude our work in \S6 with a summary.

\section{Methods} \label{sec:method}

\begin{figure*}
\begin{center}
\includegraphics[width=.49\textwidth]{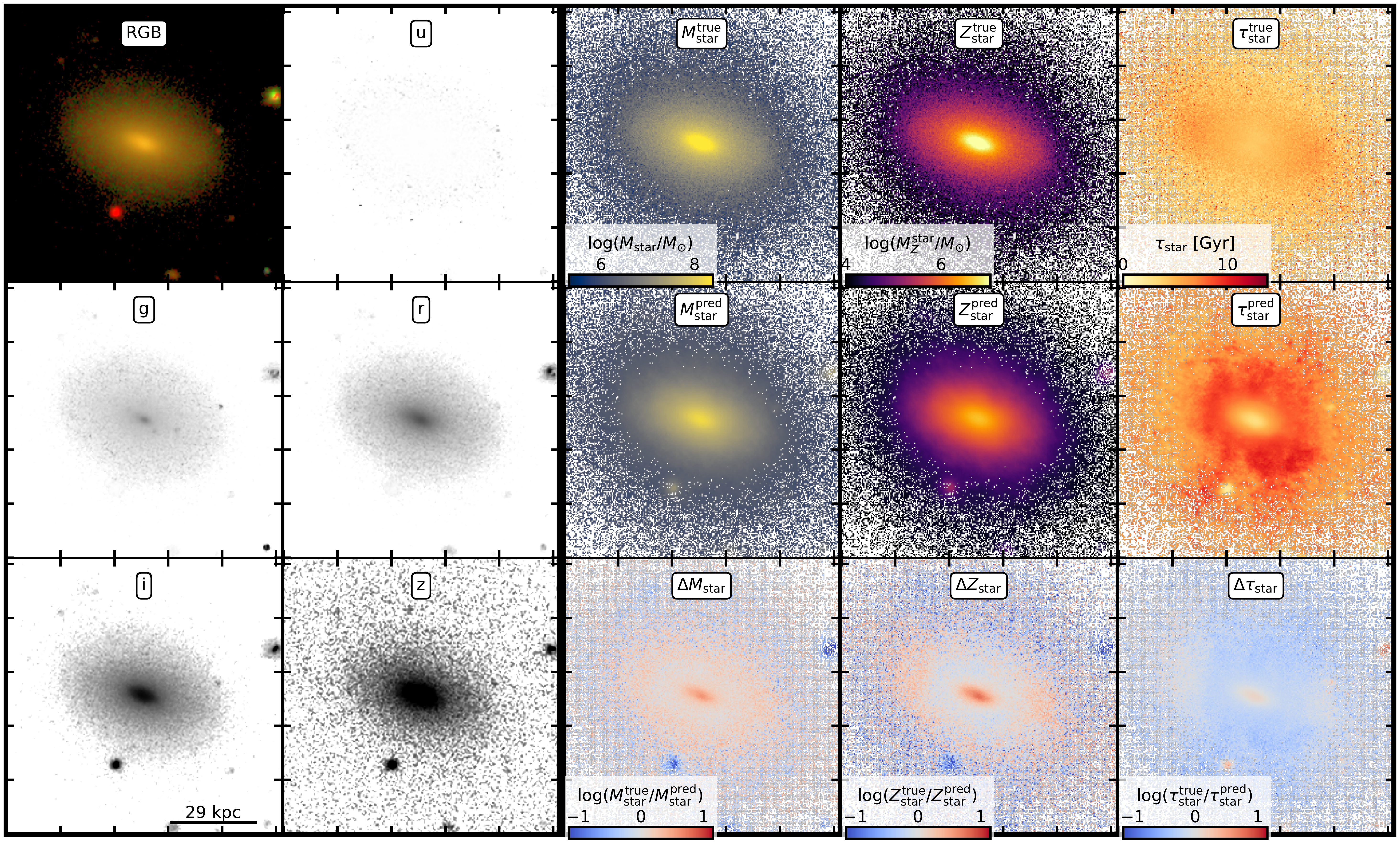}
\includegraphics[width=.49\textwidth]{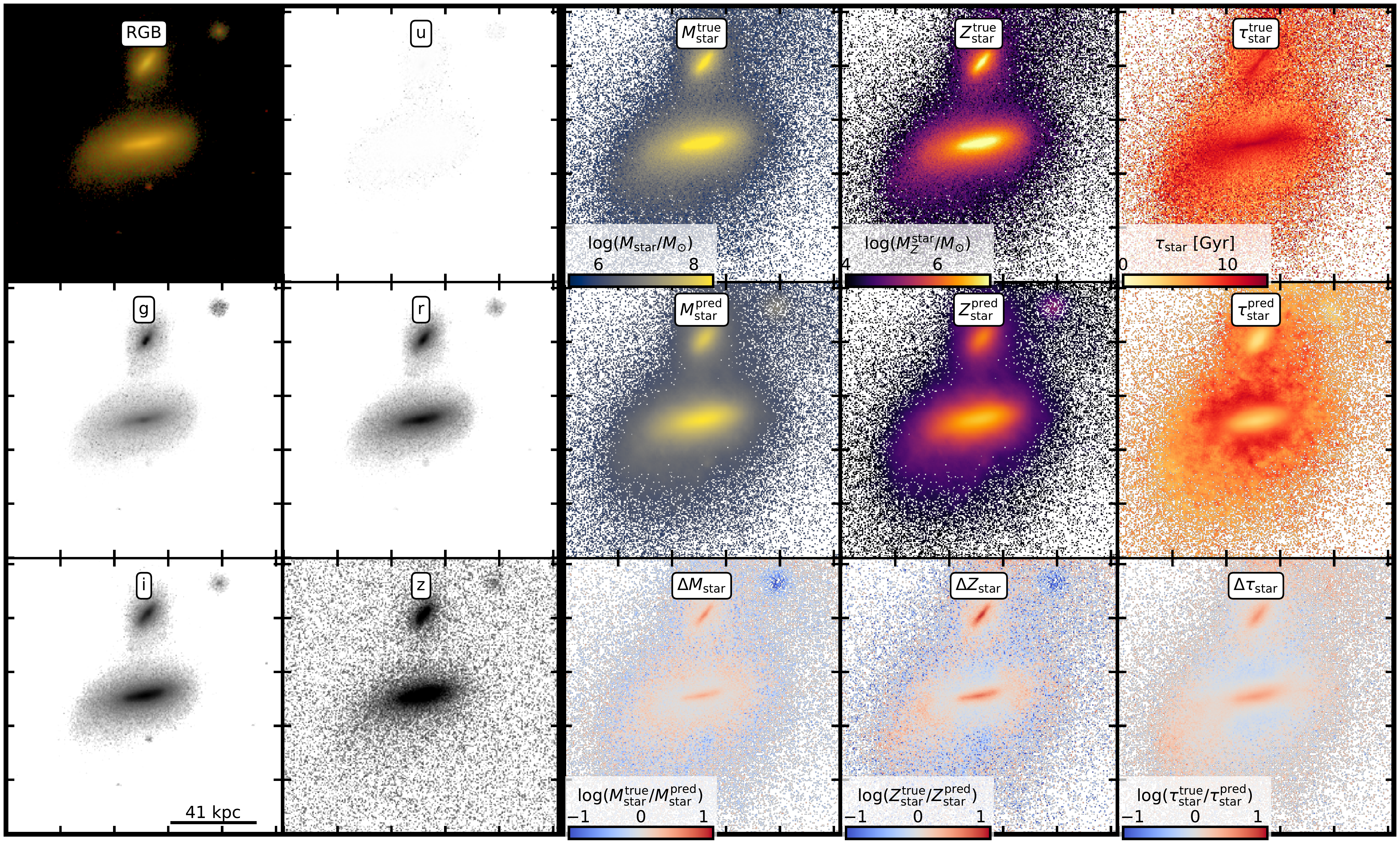}
\includegraphics[width=.49\textwidth]{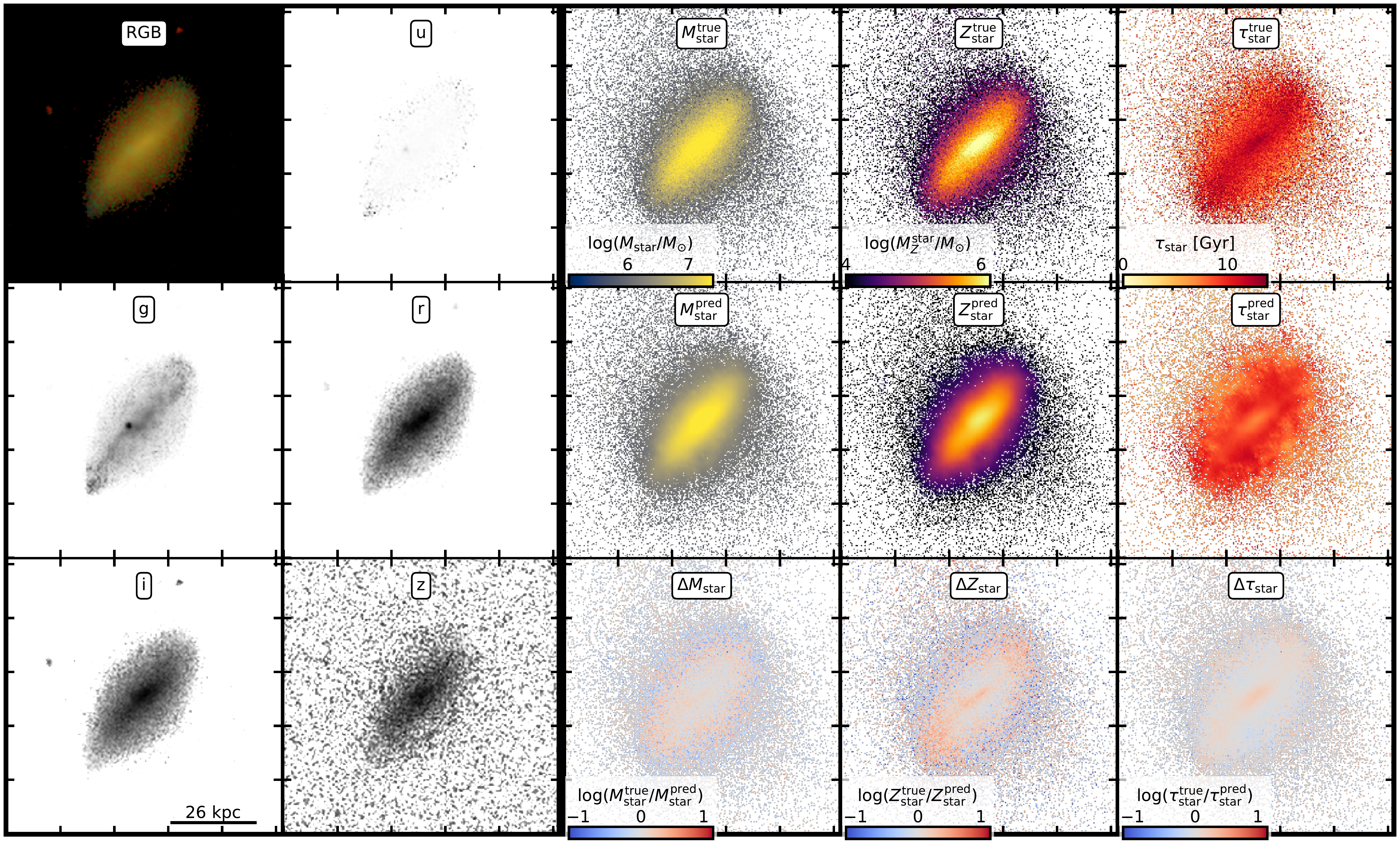}
\includegraphics[width=.49\textwidth]{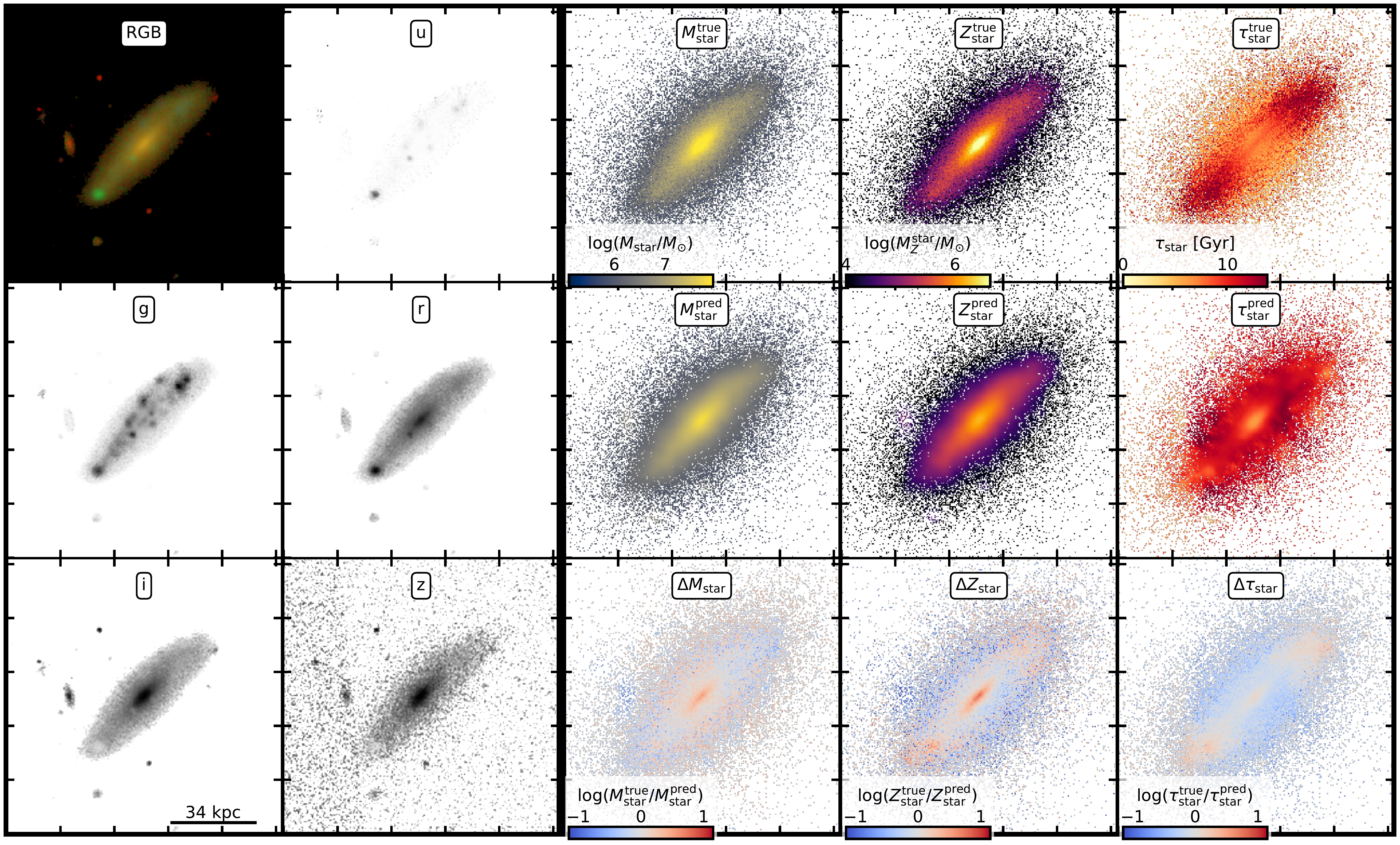}
\includegraphics[width=.49\textwidth]{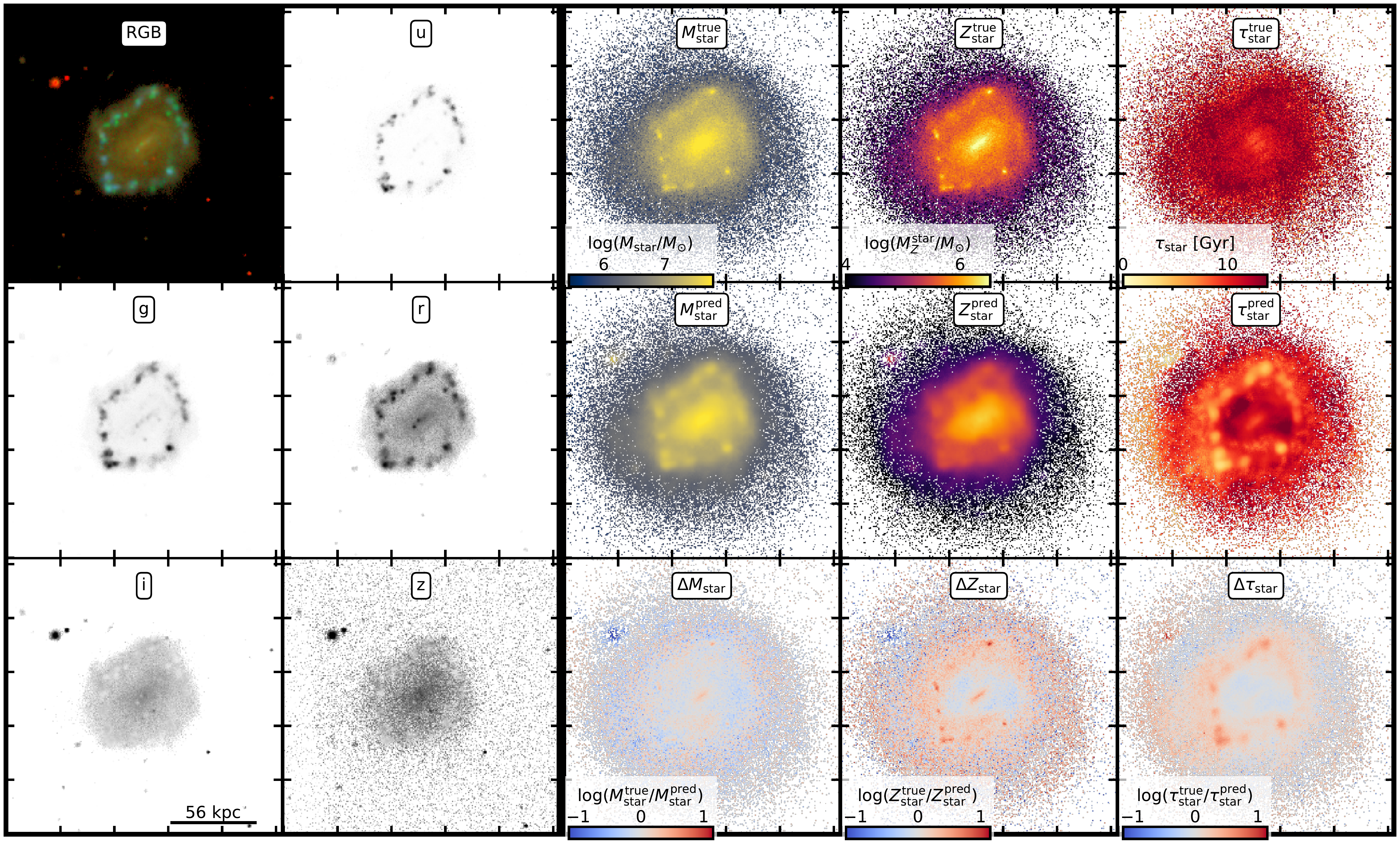}
\includegraphics[width=.49\textwidth]{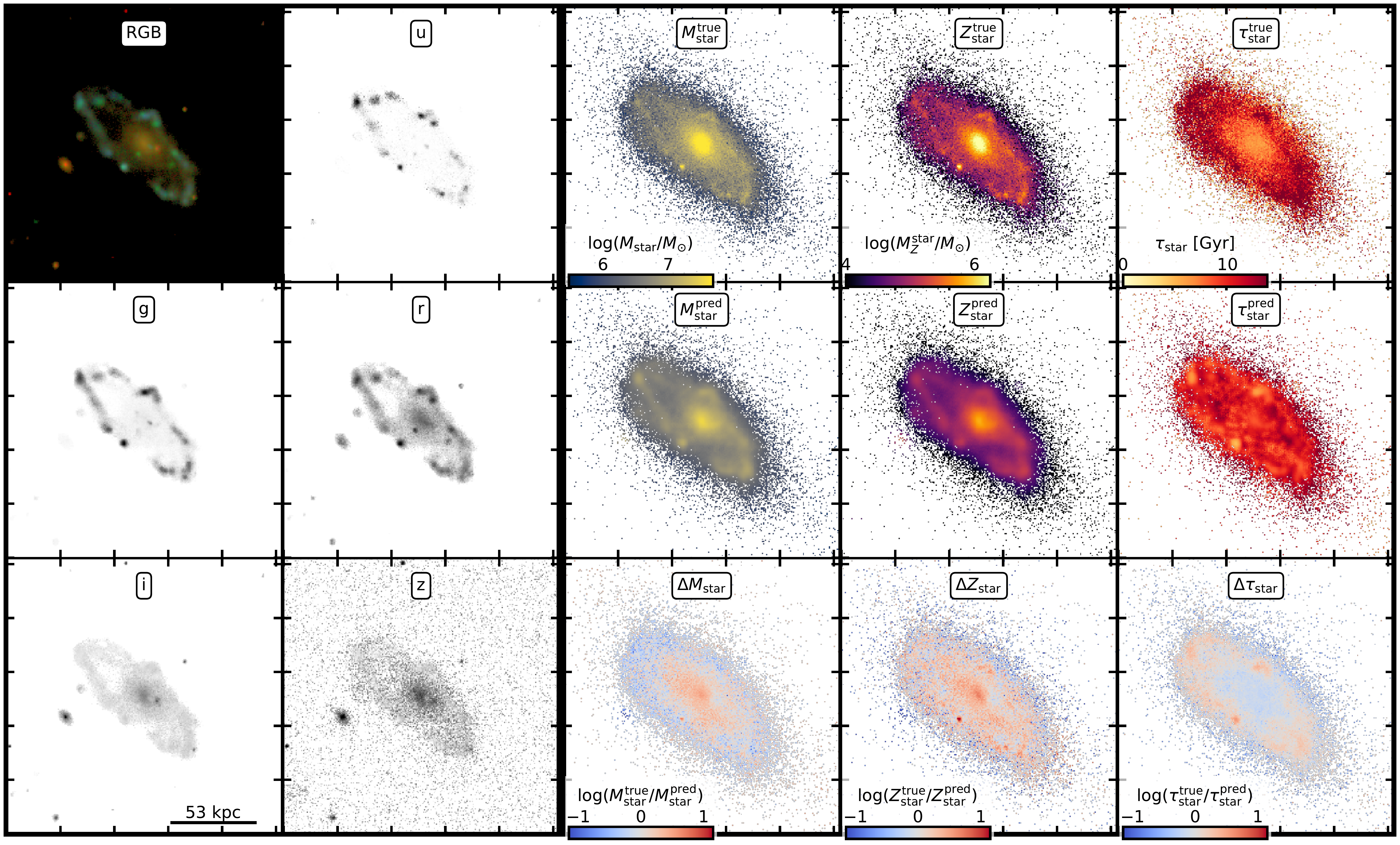}
\includegraphics[width=.49\textwidth]{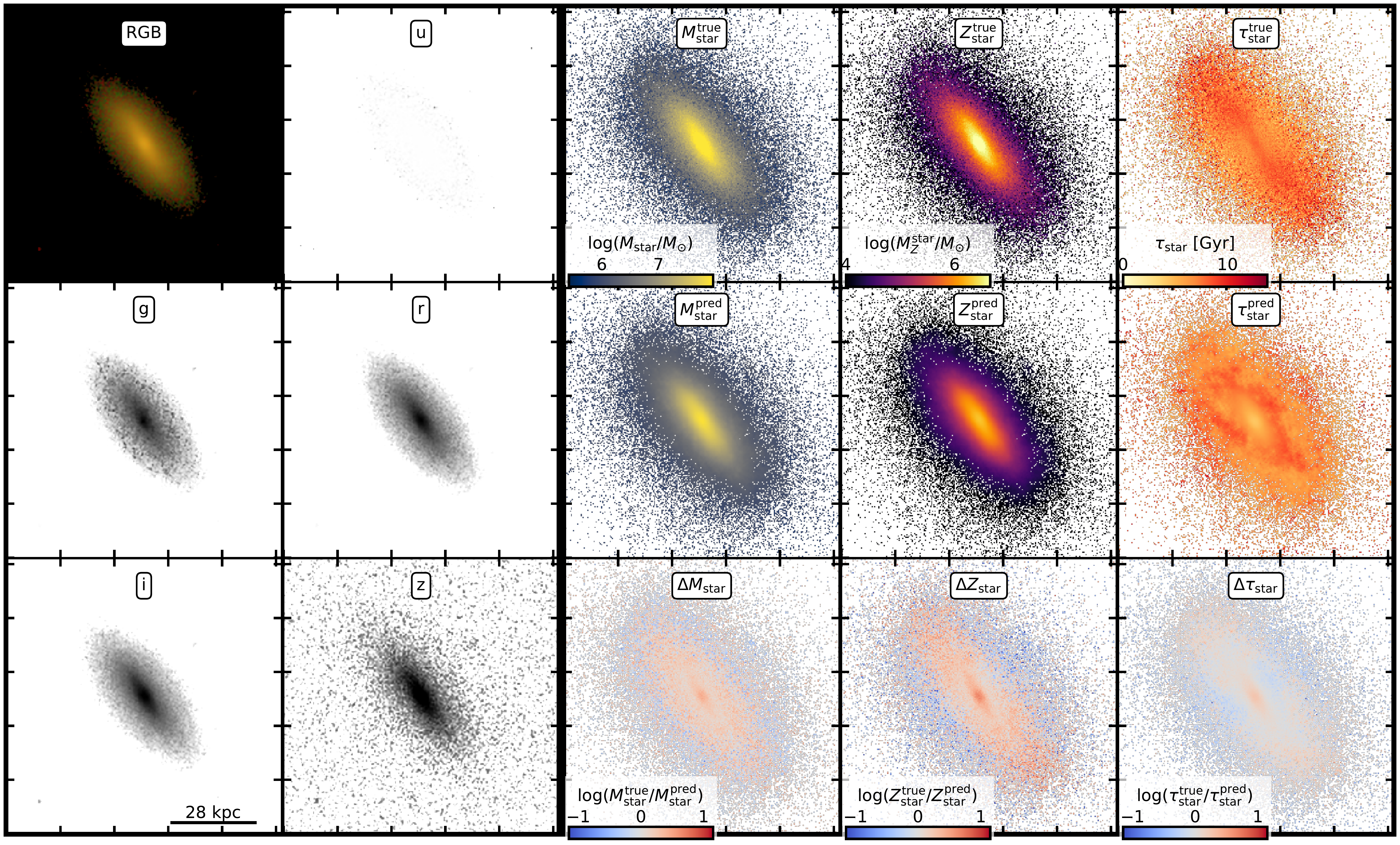}
\includegraphics[width=.49\textwidth]{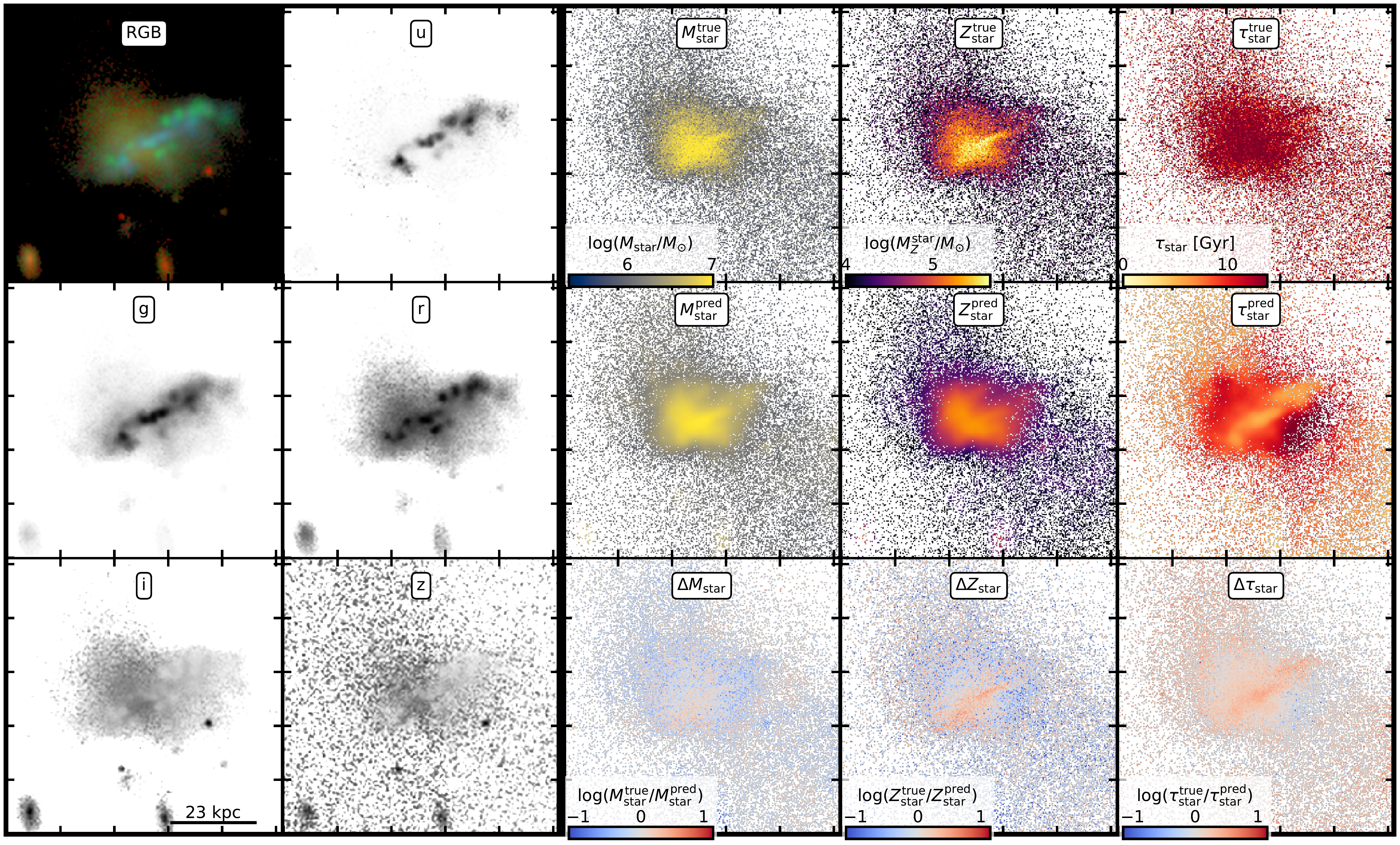}
\end{center}
\vspace{-.35cm}
\oscaptionpy{maps}{Randomly selected examples galaxies and their ML reconstruction. In each sub-panel, we show on the left an RGB image plus all five SDSS bands. To the right of that we show in the upper row the true stellar mass, metallicity and age maps. The middle row shows the ML prediction and the bottom row shows the logarithmic residual between true and predicted maps. Note, the SDSS fibre size is $3$ arcsecond corresponding to $\sim3$ kpc at a redshift of $z\sim0.05$ which for most galaxies in SDSS only covers the central parts. In contrast, our ML approach here is able to get information on physical properties for the entire galaxy out to several half-mass radii. This includes even the companion galaxies in merging galaxy pairs shown in the upper right example and the top row of Fig. \ref{fig:images2}.
}
\label{fig:images}
\end{figure*}

\begin{figure*}
\begin{center}
\includegraphics[width=.49\textwidth]{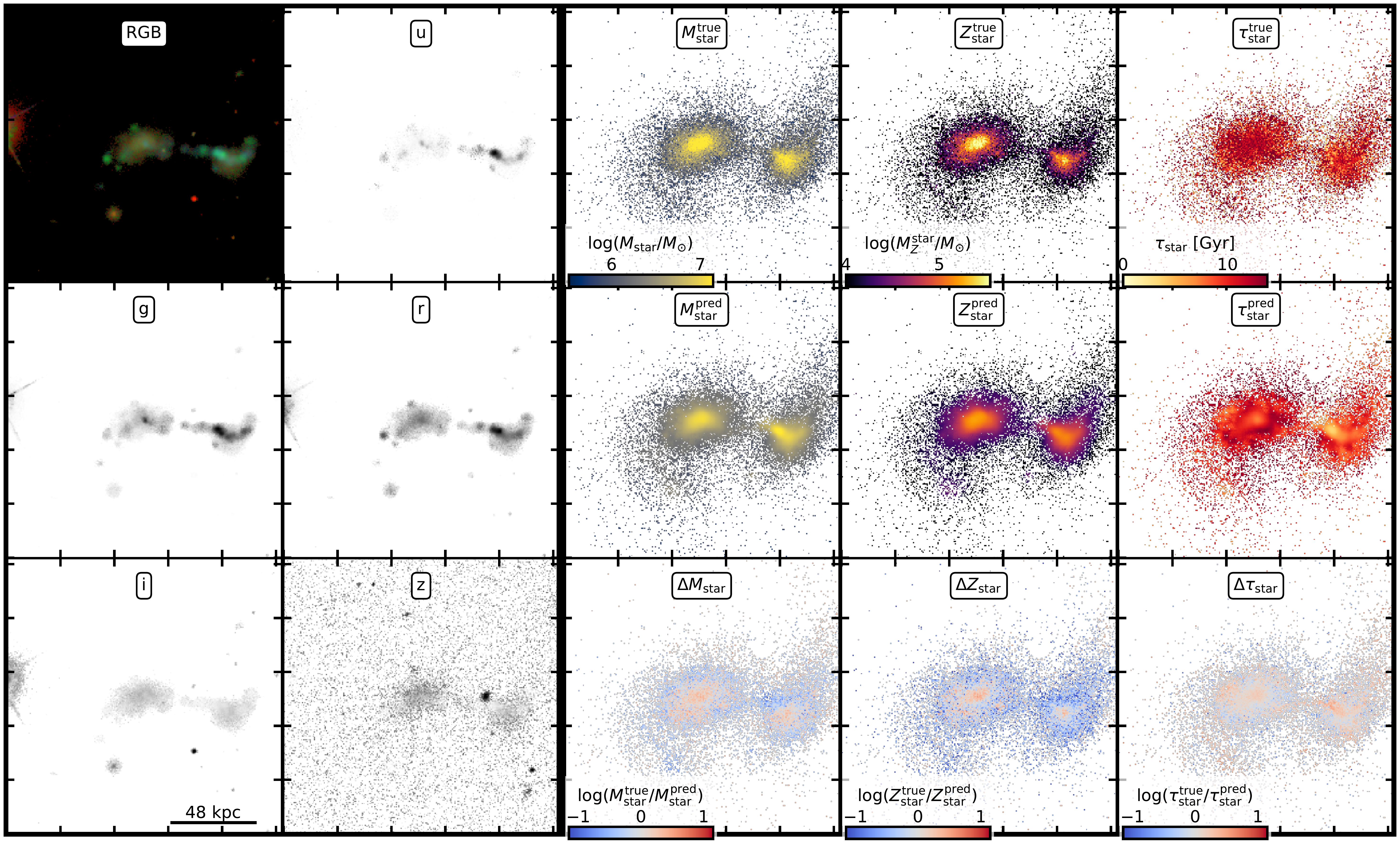}
\includegraphics[width=.49\textwidth]{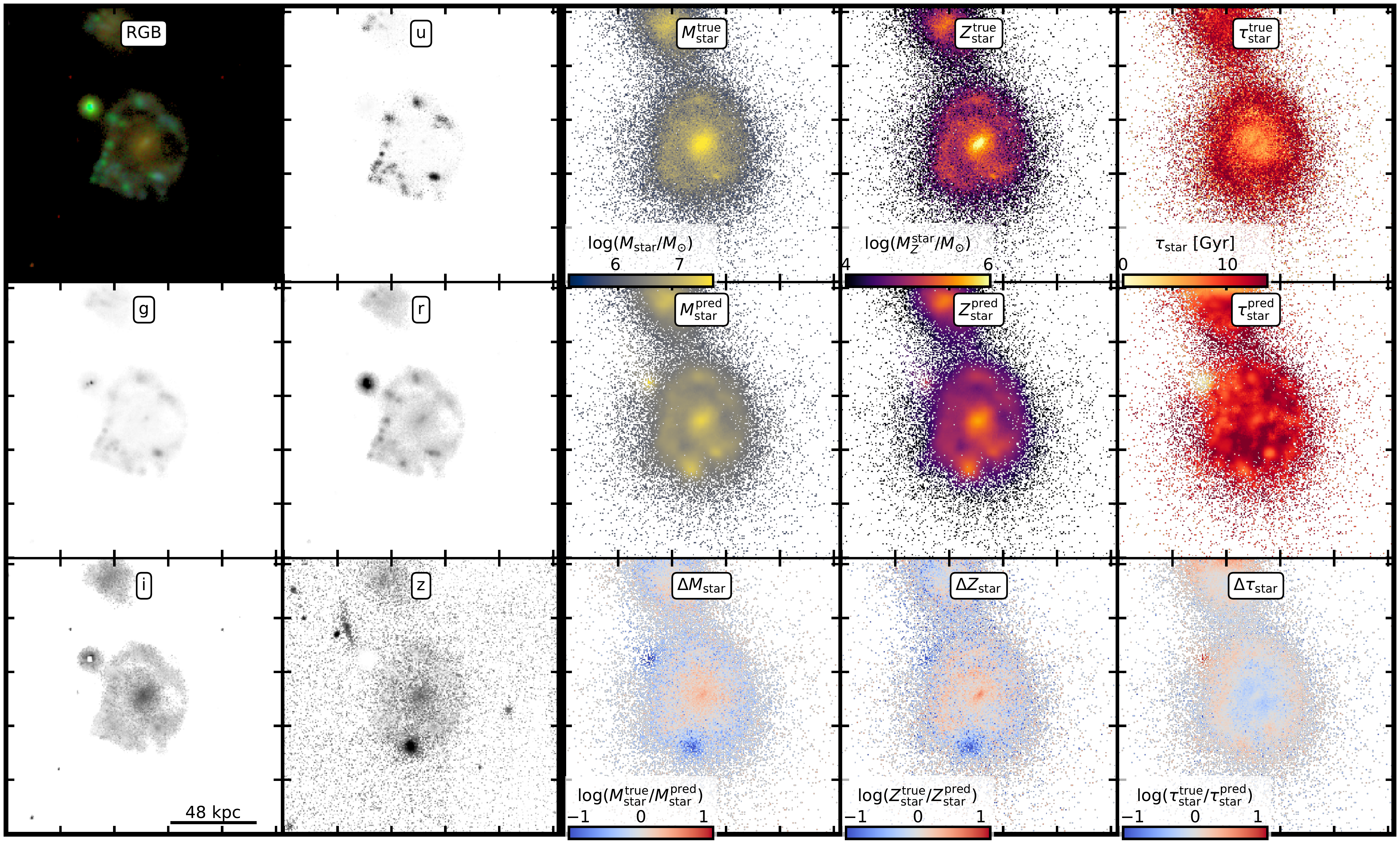}   
\includegraphics[width=.49\textwidth]{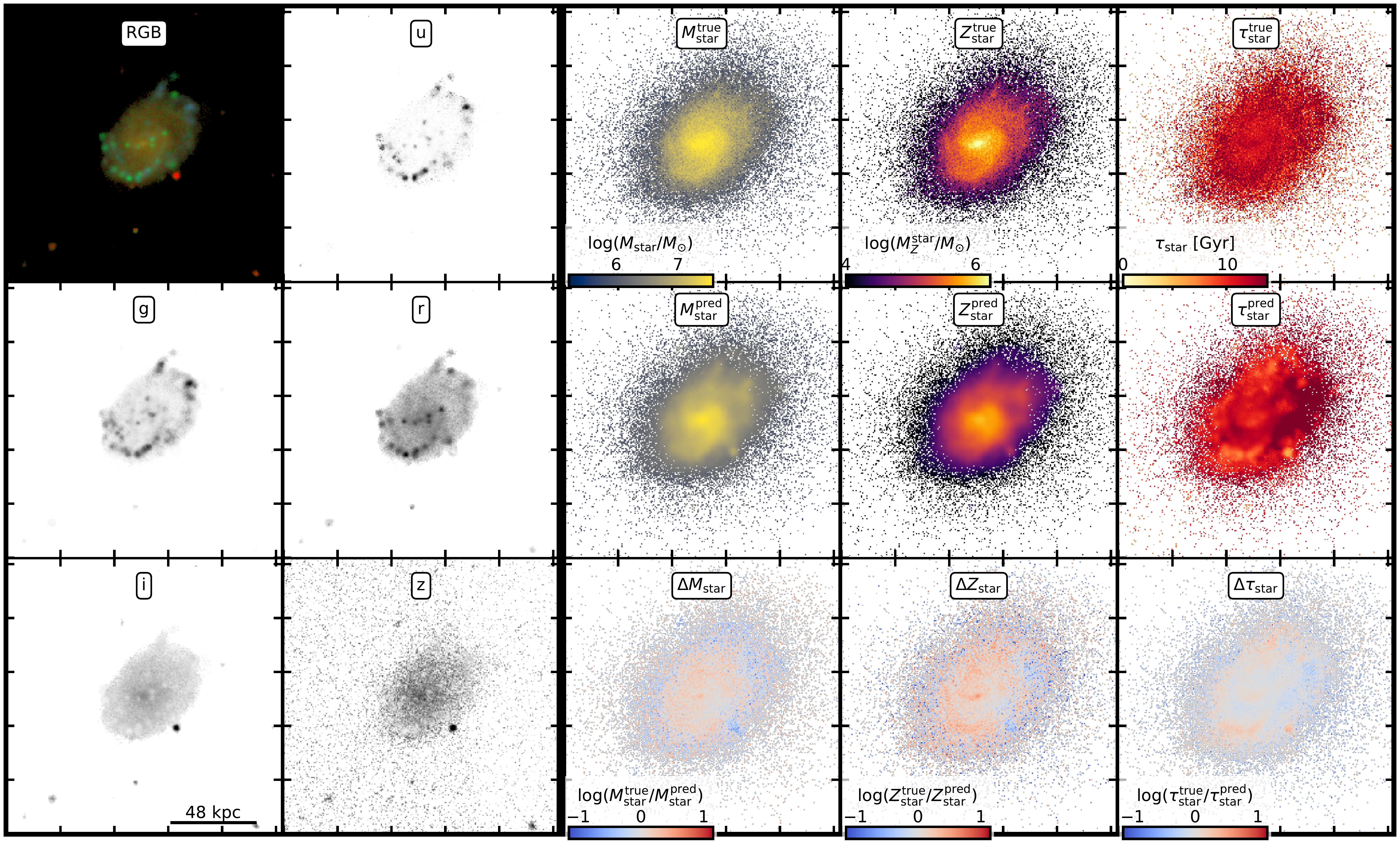}
\includegraphics[width=.49\textwidth]{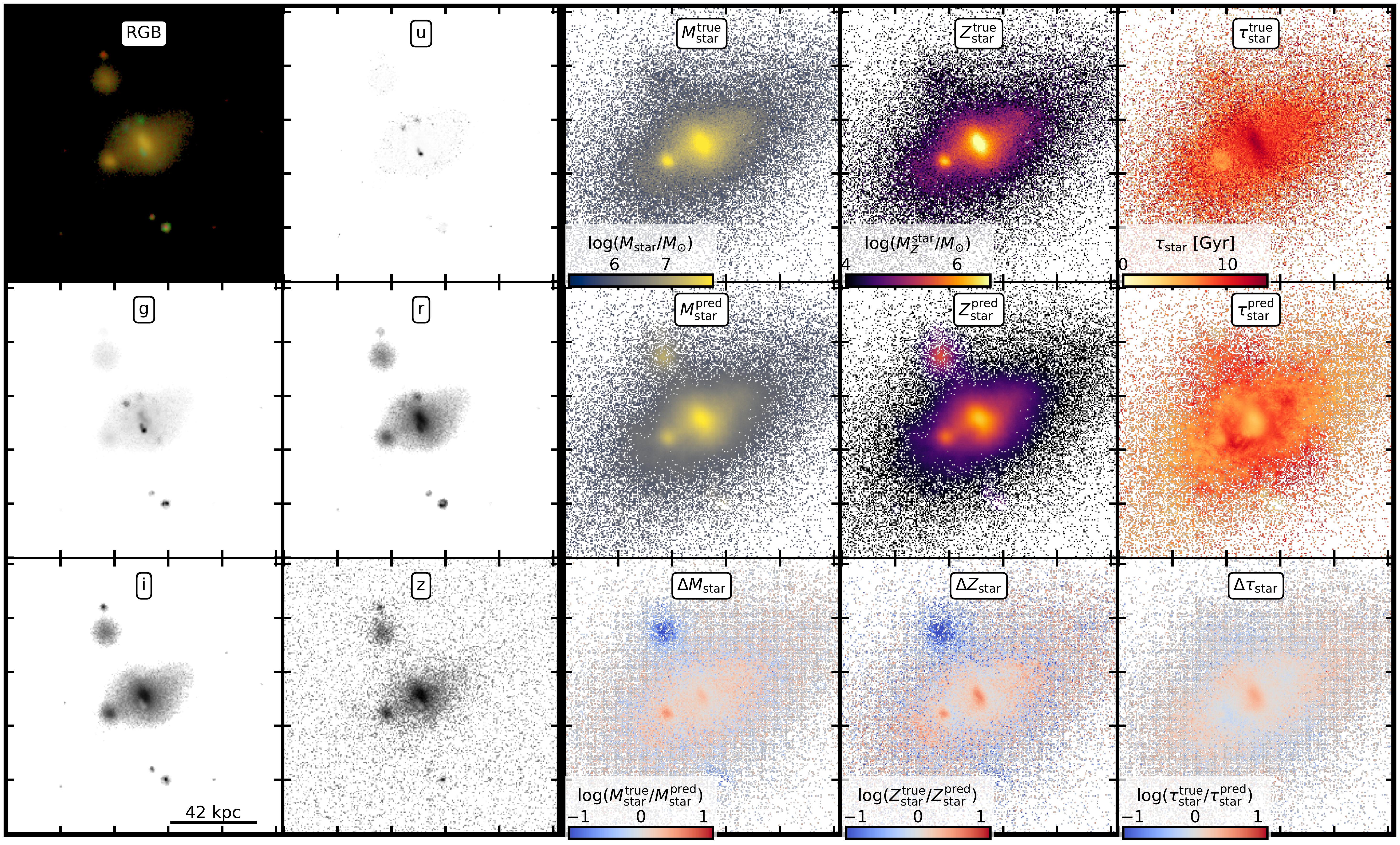}
\includegraphics[width=.49\textwidth]{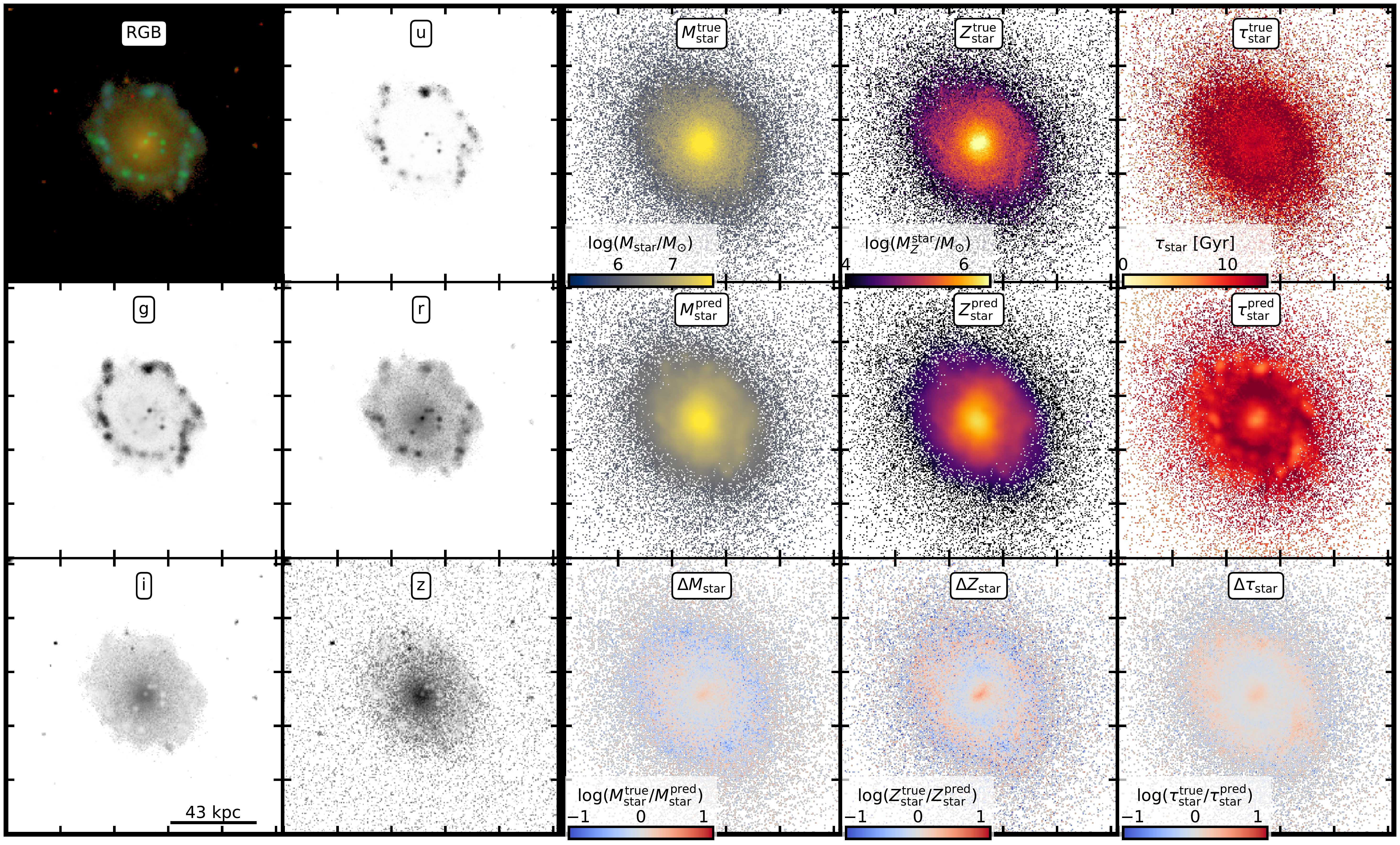}
\includegraphics[width=.49\textwidth]{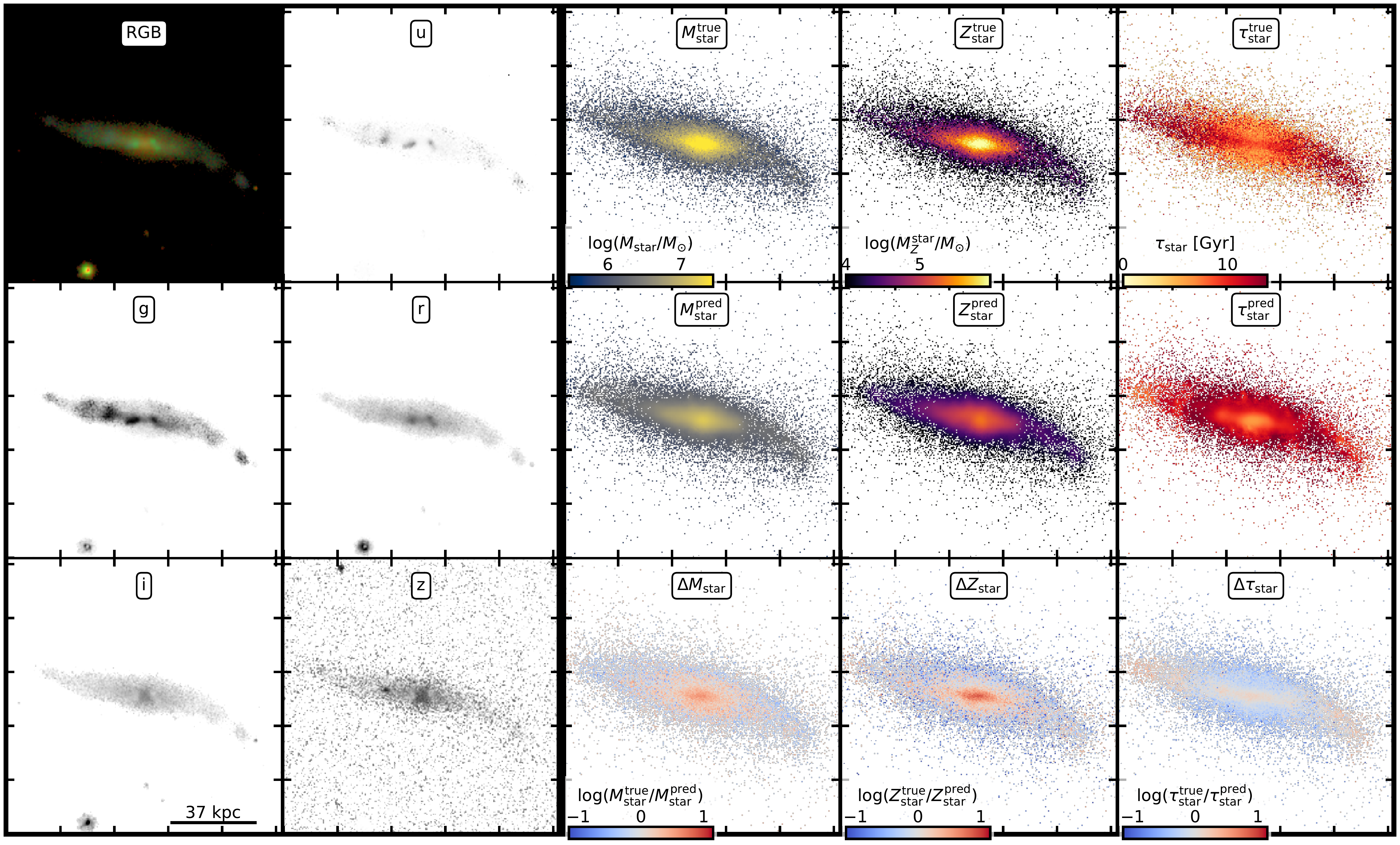}
\includegraphics[width=.49\textwidth]{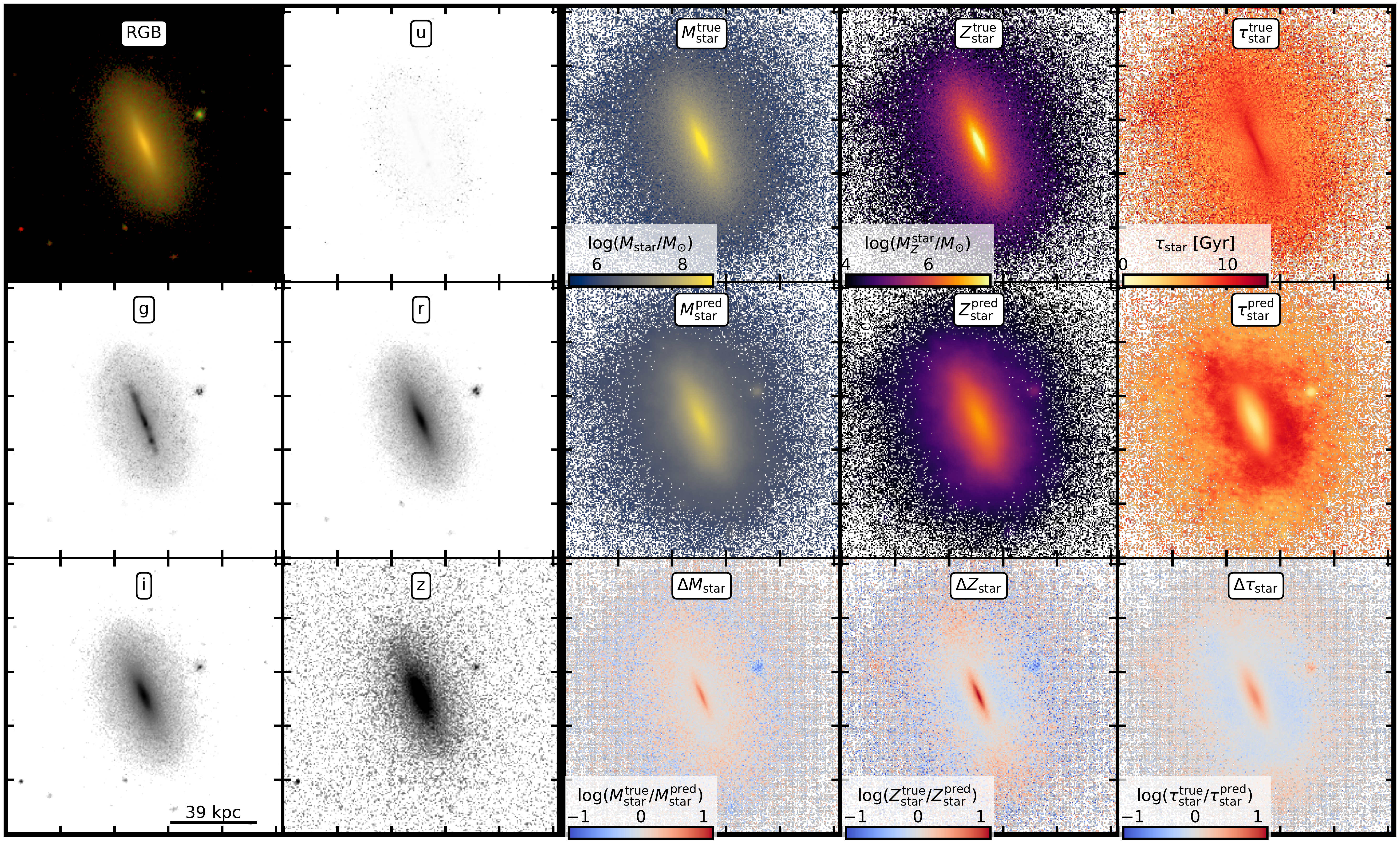}
\includegraphics[width=.49\textwidth]{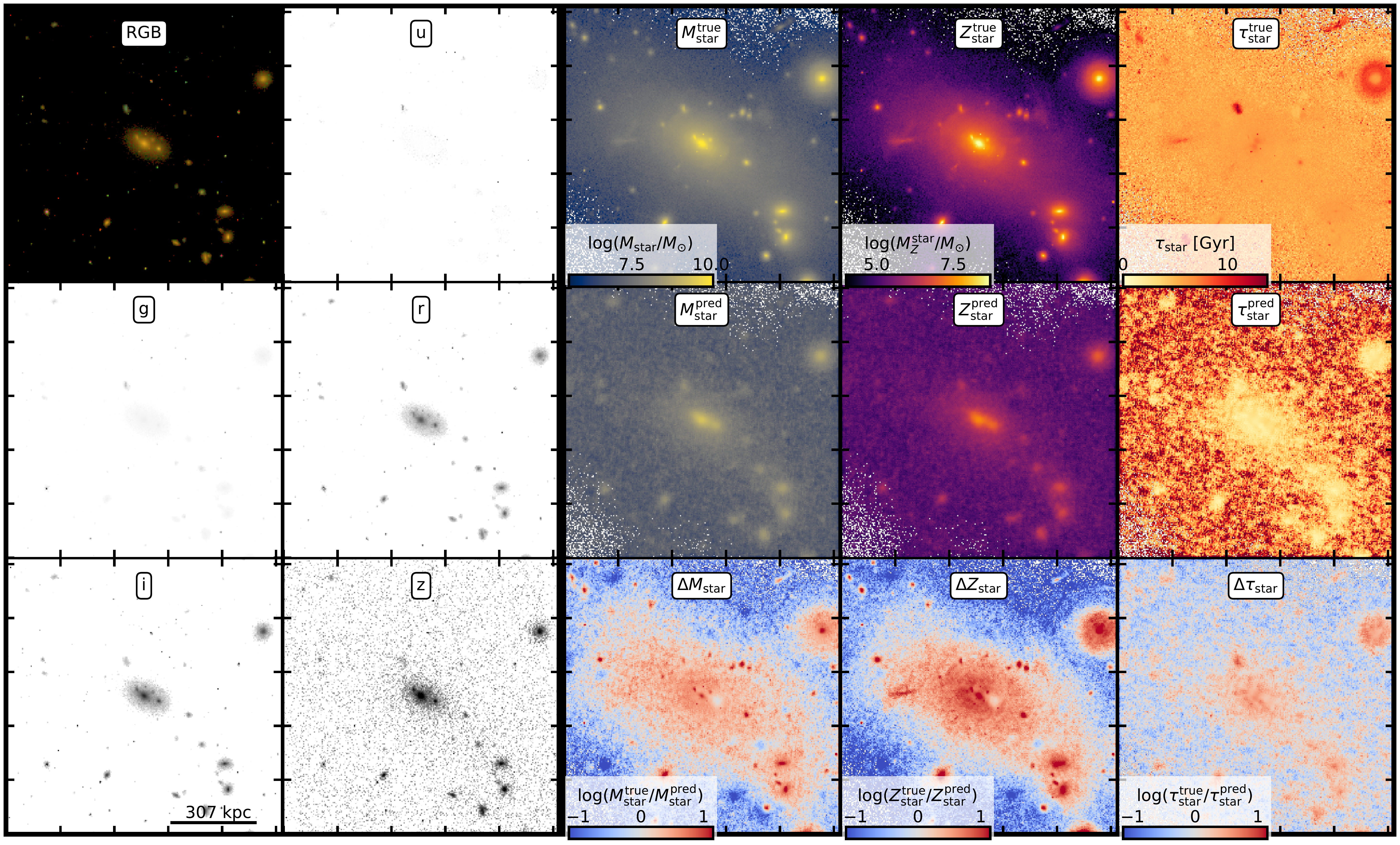}
\end{center}
\vspace{-.35cm}
\oscaptionpy{maps}{Same as Fig. \ref{fig:images}. Note the excellent reconstruction of merging galaxy pairs in the upper row and the ability of the network to reconstruct morphological features such as edge-on disks or cluster galaxies (bottom right) well.   
}
\label{fig:images2}
\end{figure*}

While our proof-of-concept work is based on SDSS mock images to verify our procedure and to quantify its accuracy, the CNN architecture can easily be applied to real photometric surveys or even IFU surveys such as SDSS-MANGA, SAMI or CALIFA leveraging the reliance on hydrodynamical simulations. In this section, we lay out the details of the ML workflow (including a description of the CNN architecture in Section \ref{subsec:ML}) and explain the dataset of synthetic photometric images (Section \ref{subsec:sdss}) as well as corresponding maps of fundamental galaxy properties (e.g. stellar mass, gas mass or SFR maps).
We have combined our training, prediction and analysis procedures into a publicly available python code named \textit{PICASSSO (Painting Intrinsic Attributes onto SDSS Objects)}\footnote{training part at \url{https://github.com/Steffen-Wolf/picasso_training} and analysis part at \url{https://github.com/TobiBu/picassso}}. 

\subsection{Synthetic observations} \label{subsec:sdss}

Modern cosmological simulations of galaxy formation have mastered the task of calculating model galaxies that resemble statistical properties of the observed galaxy population \citep[e.g.][]{Vogelsberger2014,Schaye2015,Wang2015,Dolag2016,Hopkins2018,Pillepich2018,Buck2019d,Buck2021}. Therefore, such model galaxies are ideally suited to build the training data for machine learning applications. Here, we study the information content of $\sim27,000$ photometric images of galaxies at $z \sim 0$ taken from the publicly available Illustris Project \citep[\url{www.illustris-project.org/data}][]{Vogelsberger2014} which is a suite of hydrodynamical simulations of galaxy formation in a periodic volume of size $(106.5$ Mpc$)^3$. The Illustris Project uses the Arepo code for gravity and gas dynamics \citep{Springel2010} which simultaneously evolves the gas, star, and dark matter components from cosmological initial conditions. The galaxy formation physics models include primordial and metal line gas cooling, star formation, gas recycling, metal enrichment, supermassive black hole (SMBH) growth, and gas heating by feedback from supernovae and SMBHs \citep{Vogelsberger2013}.
Here we present results from the highest resolution simulation, which has a baryonic mass resolution of $1.26\times10^6\Msun$. 

\subsubsection{Radiative transfer post-processing of model galaxies}

For each model galaxy we create images for four viewing directions (face-on, edge-on and two random directions) in the five SDSS filters ($u,g,r,i,z$) following the procedures described in \citet{Torrey2015} and \citet{Snyder2015}. These authors generated noiseless high-resolution images with the {\sc Sunrise}\footnote{Sunrise is freely available at \href{https://bitbucket.org/lutorm/sunrise}{https://bitbucket.org/lutorm/sunrise}} code \citep{Jonsson2006}. 

{\sc Sunrise} assigns to each star particle a spectral energy distribution (SED) based on its mass, age, and metallicity using stellar population models by \citet{Bruzual2003} with a \citet{Chabrier2003} initial mass function. Each star particle radiates from a region with a size directly proportional to the radius that encloses its 16 nearest neighbours which makes the emission region smaller (larger) in areas with high (low) stellar density. This adaptive smoothing technique leads to much more realistic surface brightness maps in regions of low stellar density \citep[see][for full details]{Torrey2015}. The publicly available images were created without the dust radiative transfer (RT) option enabled in {\sc Sunrise} in order to avoid over-modelling galaxies with ISM resolution insufficient to gain substantial accuracy by performing the RT \citep[see][for more details]{Snyder2015}. Since the subject of this paper is a proof-of-concept of the ML technique and dust attenuation will not strongly change the morphology of low redshift galaxies \citep[but see e.g.][for the effect of dust on high redshift star forming galaxies]{Buck2017}. Thus, we are confident that this omission of realism is no major restriction to our study.

\subsubsection{SDSS mock images}
The output of the radiative transfer (RT) step in {\sc Sunrise} is the SED at each position of $512$x$512$ pixels in each of four cameras chosen to view the galaxy face-on, edge-on and from two random directions. From this, Sunrise creates raw mock images by integrating the (optionally redshifted) SED in each pixel over a set of common astronomical filters, from the UV through IR. In this paper, we aim to create SDSS mock images and thus perform this filter synthesis assuming each galaxy resides at redshift $z=0.05$ by adopting the SDSS $u,g,r,i,z$ filters. The spatial extent of each image is set to ten times the 3D stellar half-mass radius, and therefore the physical pixel scale varies. This results in pixel sizes of roughly $100-300$ pc which is sufficiently small to simulate SDSS images of sources at $z \sim 0.05$ regardless of the limited spatial resolution of the simulation.

We apply several layers of image realism to the photometric images in order to account for foreground and background sources, point-spread function (PSF) blurring and observational photon noise. Following the procedure explained in \citet{Snyder2015}, we convolve each high-resolution image with a Gaussian (PSF) with full-width at half-maximum (FWHM) of 1.0 kpc. Next, the images are re-binned to a constant pixel scale of $0.24$ kpc (approximately $1/3$ of the Illustris stellar gravitational softening). These parameters are chosen to roughly correspond to $1$ arcsecond seeing for observations of a source located at $z = 0.05$, where the physical scale is roughly 1 kpc/arcsec. This choice was made such that the image properties roughly correspond to many sources in the Sloan Digital Sky Survey main galaxy sample \citep{Strauss2002} or to an HST WFC3 survey of sources at $z \sim 0.5$.

In a next step, we add sky shot noise assuming Gaussian random noise independently applied to each of the pixels such that the final average signal-to-noise ratio of each pixel is $25$. Further we add real SDSS background images in the appropriate photometric band to create fully synthetic $u,g,r,i,z$ galaxy images (see e.g. \ref{fig:images} and \ref{fig:images2}). For this, we first download mosaics from the SDSS DR10 \citep{Ahn2014} Science Archive Server with the mosaic web tool (\href{data.sdss3.org/mosaics}{data.sdss3.org/mosaics}) and then randomly select a region of an appropriate size for each synthetic image (assuming the galaxies are at $z = 0.05$) and add it to the simulated galaxy image. More details on this procedure can be found in \citep[][Section 2.]{Snyder2015}. Finally, for computational reasons we create our target images with a fixed pixel count of (256x256) for our fiducial images and successively lower pixel counts for the resolution study in Section \ref{sec:resolution}. Thus, at each resolution level we have a total of $\sim2.7 \times 10^5$ synthetic images. 

The Figures \ref{fig:images} and \ref{fig:images2} show each eight example galaxies from our validation set. In the two left columns of each sub-panels we show a three color mock image created from the SDSS $g,r,i$ bands as well as all five $u,g,r,i,z$ single band images used as input to the network. Additionally, we show in the right three columns the resulting reconstructed physical maps (top panels) for stellar mass, stellar metal mass and average stellar age predicted by the ML algorithm from the photometric images. In comparison to this we also show the \textit{true} counterparts (middle panels) extracted from the simulation. The bottom row shows the pixel residual maps between predicted and true maps.

Note, for spectroscopic observations, the SDSS fibre size is $3$ arcsecond corresponding to $\sim3$ kpc at a redshift of $z\sim0.05$ which for most galaxies in SDSS only covers the very central parts of each galaxy. In contrast, our ML approach here is able to get information on physical properties for the entire galaxy out to several half-mass radii and even for merging systems (see lower right panel of Fig. \ref{fig:images}).

\begin{figure*}
\begin{center}
\includegraphics[trim={0 475 0 0 cm},clip,width=\textwidth]{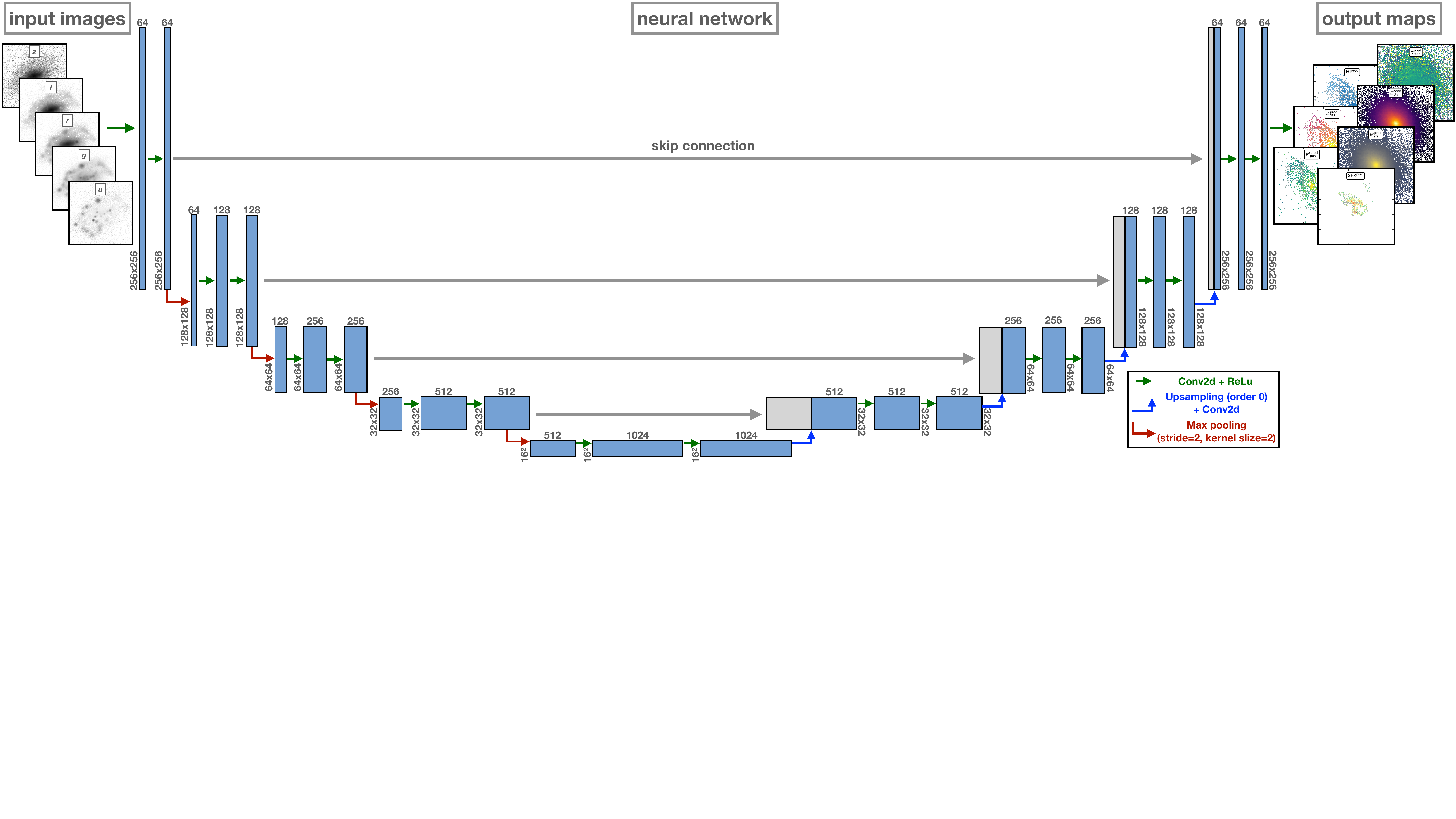}
\end{center}
\vspace{-.35cm}
\oscaptionnet{Neural network architecture: We use a U-net architecture with skip connections. Each blue box corresponds to a multi-channel feature map with the number of channels denoted on top of each box. The x-y image size is provided at the lower left edge of each box. Grey boxes represent copied feature maps. Red, blue and green arrows denote the different operations of down-sampling by max pooling, up-sampling and 2d-convolutions.}
\label{fig:network}
\end{figure*}

\subsection{Maps of physical quantities}

In addition to the mock photometric images we create surface density maps of $7$ physical galaxy properties with the same pixel sizes as the final photometric images (e.g. 256x256 in the fiducial case). In particular we consider the properties stellar mass, $M_{\rm star}$, stellar metallicity, $Z_{\rm star}$, stellar age, $t_{\rm star}$, star formation rate, SFR, gas mass, $M_{\rm gas}$, gas metallicity, $Z_{\rm gas}$, and neutral hydrogen mass, $M_{\rm HI}$. This amounts to another $\sim 1.9\times 10^5$ images in our dataset. No image realism is applied to the true physical images. The reasoning behind this is that we seek to create mock photometric images closely resembling the data from real galaxy surveys while the target physical maps in real surveys might already have accounted for contamination and applied some form of de-noising/de-blurring, modelling the effects of the point spread function (PSF). Thus, we rely on the ability of the networks to learn some kind of automatic de-noising simultaneously to reconstructing the physical properties. However, since we are not interested in the exact de-noising/de-blurring of images and thus in future work we will improve upon this and model de-blurring/de-noising in a hierarchical manner similar to the approach taken by \citep{Lanusse2019}.

\subsubsection{Analysis pipeline}
\label{subsec:analysis}

In the course of this paper we will compare galaxy properties derived from the predicted images and the true underlying physical maps. Since galaxy images are scaled to encompass an area of $\pm5$ half mass radii around each galaxy and galaxies can be randomly orientated (e.g. not only edge-on or face-on) we apply an ellipse fitting method\footnote{We use the ellipse fitting technique from the publicly available python package photutils (\href{https://photutils.readthedocs.io/en/stable/}{https://photutils.readthedocs.io/en/stable/})} to the true stellar mass map in order to define which pixels belong to the galaxy. For stability and better comparison, the ellipse found from the true stellar mass image is kept fixed during analysis of all other properties both for true and predicted properties. Fiducial analysis is performed on all pixels inside an ellipse of semi-major axis size of two half mass radii, $2R_{\rm half}$. Using ellipses with different semi-major axis values we are further able to assess any radial dependence of our image reconstruction. 

Note, the true maps derived from the Illustris galaxies may contain empty pixels, especially in the outskirt of the galaxies where the stellar density or the gas density decreases strongly. Therefore, when we compare predicted vs. true properties, we restrict ourselves to only take into account those pixels, that have non-zero values in the true properties although the network is able to also predict values for originally empty pixels owing to its convolutional nature.  

\section{Machine Learning setup} \label{subsec:ML}

Thanks to the widespread availability of affordable graphics processing units (GPUs) capable of performing general purpose, highly parallel computing, neural networks have become the tool of choice for image classification and regression problems. Connecting the imaging data to the underlying physical properties derived from spectroscopy is essentially an image regression problem. Deriving spatial relations in images is most readily done using multiple layers of image convolutions and adaptively constraining the filters in use \citep[see e.g.][]{Krizhevsky2012} -- concepts which have been in use in modern astronomy since its beginning. 
In the field of image recognition and analysis, these concepts have been formalised and combined which resulted in algorithms such as convolutional neural networks \citep[CNN][]{CNN2015} and/or generative adversarial neural networks \citep{Goodfellow2014}. Such networks efficiently adapt to (learn) relations in the images which are of the order of the convolution filters (or kernels). Like in classical regression the parameters of a model function, here the filters themselves, are the subject of the regression. These filters will be adapted/learned through training the network, i.e. minimising a predefined loss function on a set of images where the property to be learned is known. The more layers of convolution are included in a network the better its capability to adapt to more and more abstract features \citep{Zeiler2014}. If the number of layers becomes large enough, the neural network is called \textit{deep}.

\begin{figure*}
\begin{center}
\includegraphics[width=\textwidth]{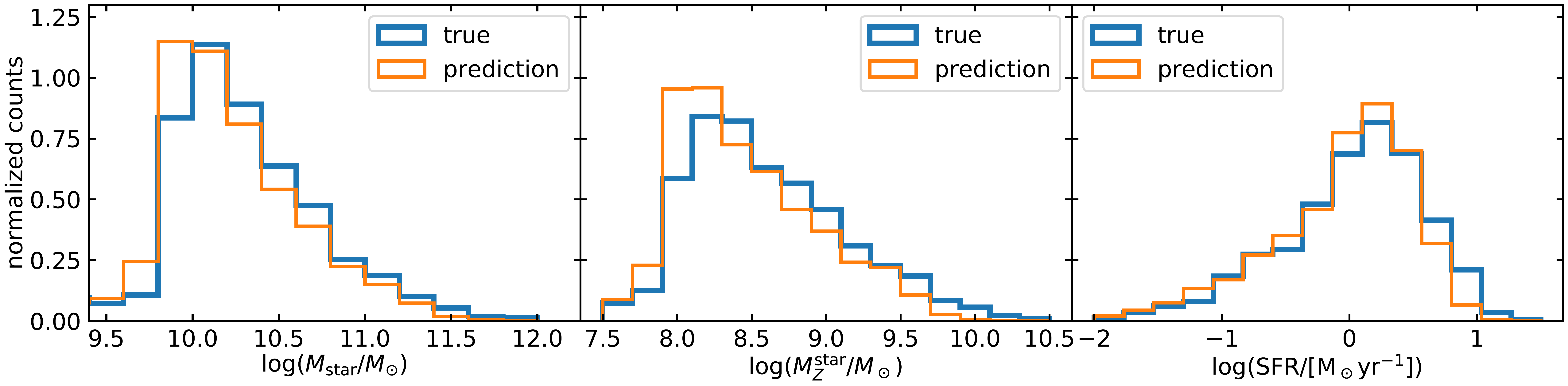}
\includegraphics[width=\textwidth]{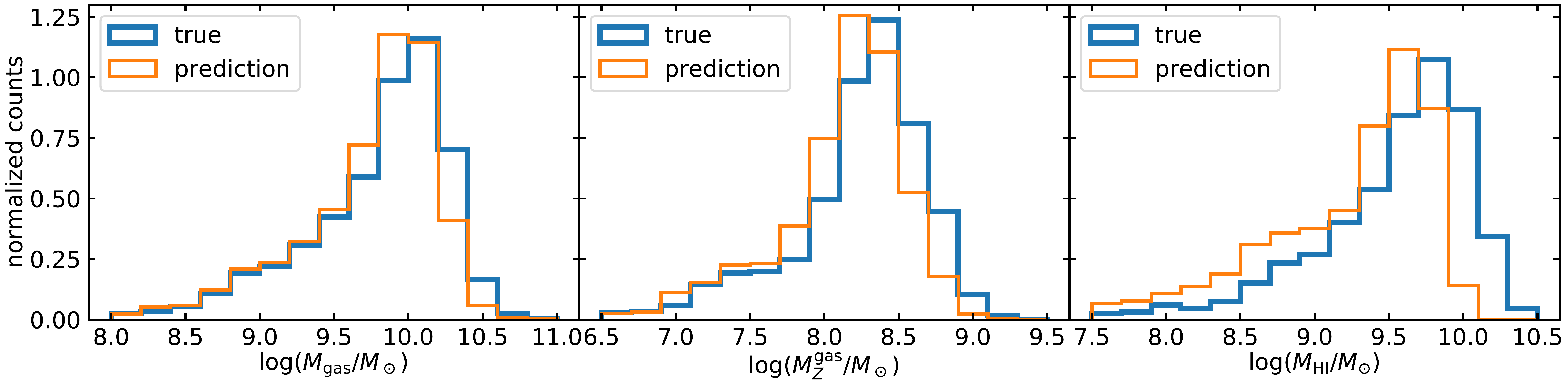}
\end{center}
\vspace{-.35cm}
\oscaption{Analysis_ugriz}{Distributions of the true (blue) and predicted (orange) galaxy properties. Upper panels show stellar properties, from left to right we display stellar mass, metallicity and SFR. Lower panels show gaseous properties displaying from left to right the gas mass, metallicity and neutral hydrogen abundance.}
\label{fig:hist}
\end{figure*}

\subsection{Network Architecture}

In this work, we use a convolutional neural network~(CNN) to learn the connection between the 5 band photometry from SDSS \textit{ugriz} and the underlying physical galaxy properties (on a pixel-by-pixel basis). Both input and output of this network are images and we therefore utilize an encoder-decoder style architecture with skip connections, commonly used in these tasks. This architecture type was first developed for biological-image  processing\citep{Ronneberger2015} and is referred to as a U-net due to its use of downsampling and subsequent upsampling layers. Similar architectures are also referred to as Hourglass networks \cite{newell2016stacked} or pix2pix networks \cite{isola2017image} for image-to-image translation.

This enables us to not only connect the imaging data to scalar values of total mass, metallicity, stellar age or SFR but to further perform full 2D analysis on the image data itself such as calculation of mass, metal or age gradients or spatial variations of metallicity, SFR or stellar age.

The specific network architecture used for this work is shown in Fig. \ref{fig:network}. We employ a U-Net with 4 down-/up-sampling layers each of which has 2 convolutional layers and skip connections. We follow \cite{Ronneberger2015} and employ max-pooling to decreases the image resolution in every down-sampling layer and use a ReLU activation function.

\subsubsection{Dataset Splits and Augmentation}

Our analysis is based on 27,558 pairs $(x, y)$ of SDSS mock images ($x \in \mathbb{R}^{5 \times W \times H}$) with the corresponding physical properties ($y \in \mathbb{R}^{7 \times W \times H}$). We use 20\% of all images for our analysis (test set) and the rest for network training. From the training set, we again reserve 20\% of all image pairs for validation and selecting the best network during the training run (lowest loss on the validation data set).

The training set size can be artificially increased through data augmentation. We utilize the symmetry of our mock images and augment the images with random 90-degree rotations, flips and image transpositions. Random noise is added to the images during the rendering of the mock images but fixed during training.

\subsection{Network Training}

For the training purposes we consider the network as a function $f_\theta$ with adjustable parameters $\theta$. It maps any given mock image $x$ to a prediction of physical properties $\hat{y} = f_\theta(x)$. During the training we aim to find network parameters that minimize the regression loss for all training image pairs (x, y):

\begin{align}
    \mathcal{L}(x,y) =  \sum_{i=0}^W \sum_{j=0}^H \sum_{c=0}^7 w_{ij} || \mathop{\log_{10}}(y_{cij}) - f_\theta(x)_{cij} ||_2
\end{align}

The loss decomposes over all pixels (index $i$ and $j$) and the 7 physical properties. Since the physical properties span multiple orders of magnitude, we chose to train the network to predict the physical properties in log-space. After training, we remap to the true physical properties in our analysis using $10^{\hat{y}}$. We use the weight factor $w_{ij}$ to mask pixels that correspond to bins without any simulated particles. For all pixels with known physical properties we set $w_{ij}=1$ and for all other pixels we use $w_{ij} = 10^-6$ and $y_{ijc} = 10^{-100}$ as a regularization factor.

We train the network for 13 epochs with a batch size of 16 using an Adam optimizer~\cite{kingma2014adam}. We start with a learning rate of $10^{-5}$ and reduce it by a factor of $10$ after epoch 8 and 10. One training run takes 10 hours on a single GeForce RTX 2080 Ti.

\section{Results: The information content of photometric images}
\label{sec:results}

\begin{figure*}
\vspace{-.4cm}
\begin{center}
\includegraphics[height=.925\textheight]{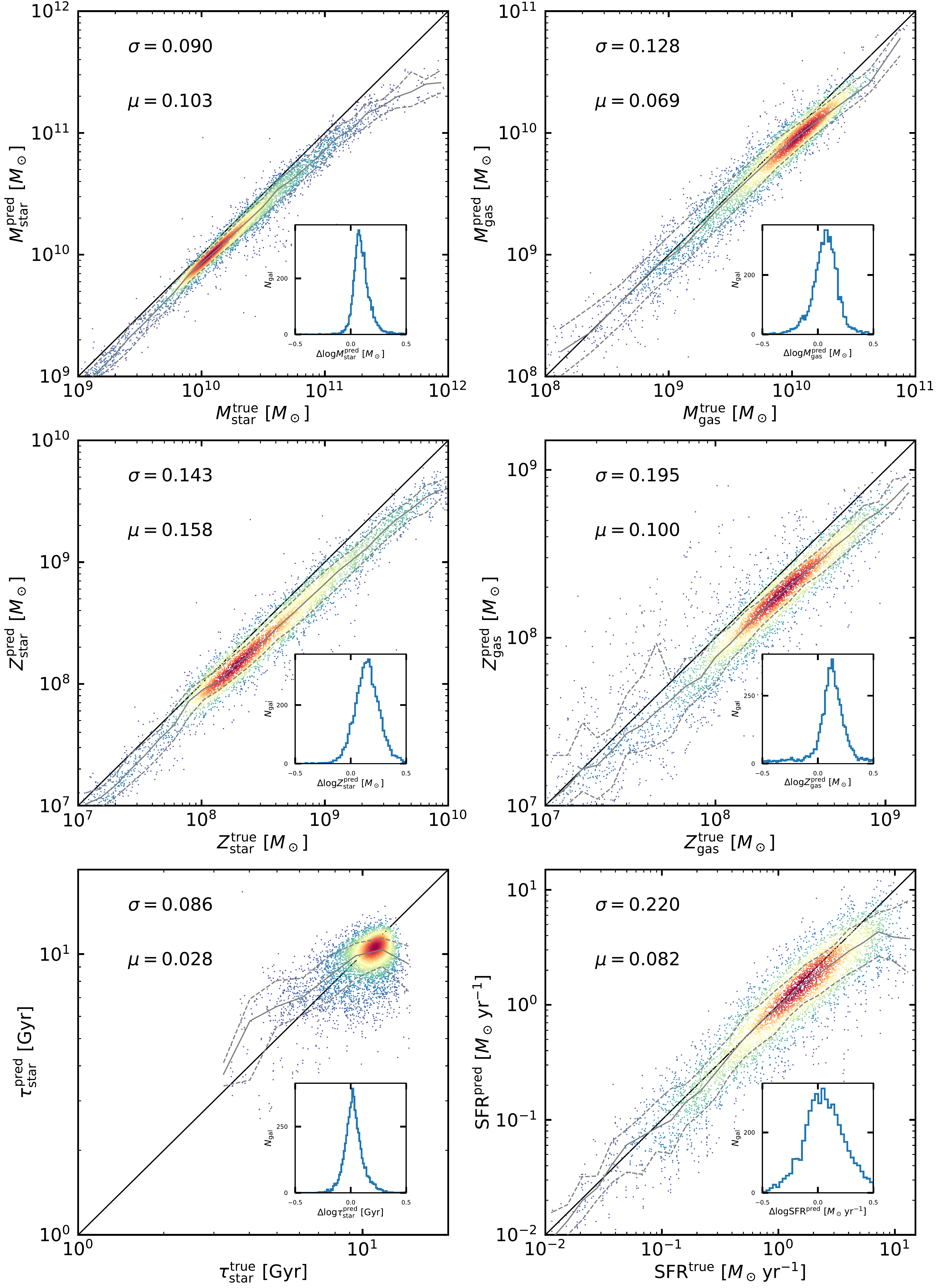}
\end{center}
\vspace{-.5cm}
\oscaption{Analysis_ugriz}{True vs. predicted galaxy properties as measured from all pixels inside two half mass radii, $2\,R_{\rm half}$. The top row shows stellar and gas mass, the middle row shows stellar and gas metallicities and the bottom row shows the average stellar age within $2\,R_{\rm half}$ and the SFR. The diagonal black line shows the 1:1 relation and the color-coding highlights a kernel-density estimation of the galaxy density. The gray solid lines shows the running average with dashed lines highlighting the one sigma scatter around it. The inset panel shows the logarithmic difference between true and predicted galaxy property. The values for mean, $\mu$, and scatter, $\sigma$, of this distribution are printed in the upper left corner of each panel. In general, galaxy properties are reconstructed with high precision although we find that all properties are slightly under-predicted.}
\label{fig:true_vs_pred}
\end{figure*}

Properly evaluating the quality of the reconstruction of physical properties from photometric data is a difficult problem. A traditional metric measuring the per-pixel mean-squared error does not assess the joint statistics of the result. Therefore, this approach does not measure the very structure encoded in the image that we aim to reproduce with structured losses. Furthermore, we are dealing with galaxy images for which specific quantities are of interest, such as e.g. the total mass within a given radius or the radial gradients of the properties of interest.
To more holistically evaluate the quality of our results, we employ two strategies: We evaluate how well we recover the total sum of pixel values in a given area of the image (e.g. total stellar/gas mass within $R_{\rm half}$) and how well we recover those properties as a function of radius. 

Another practical aspect to test our algorithm for is the sample size used for training and the image resolution requirements. While gathering (low resolution) photometric data is relatively cheap and therefore datasets are usually huge, gathering the initial training datasets consisting of photometric input images plus reconstructed physical maps (e.g. stellar mass or metallicity maps) might become a challenge. Therefore, in Section \ref{sec:bands},  \ref{sec:resolution} and \ref{sec:sample_size} we independently vary the number of input wavelength bands, image resolution and training sample size in order to evaluate how many different wavelength observations, image pairs or which image resolution is required to meet a given target accuracy. This effectively constrains the survey which can be used as training set. It gives a lower limit to the total number of available image pairs needed (e.g. SDSS MANGA has $\sim10.000$ galaxies observed) or similarly, for a given survey, it provides a limit on the redshift range that might be accessible by constraining the minimum number of pixel per galaxy needed. At the same time, reducing the number of wavelength bands to a single image at fixed resolution tests how important color information is over morphological information. Similarly, reducing the image resolution at fixed color information (e.g. all 5 SDSS bands) studies the importance of morphological features over color.  

\begin{figure*}
\begin{center}
\includegraphics[width=\textwidth]{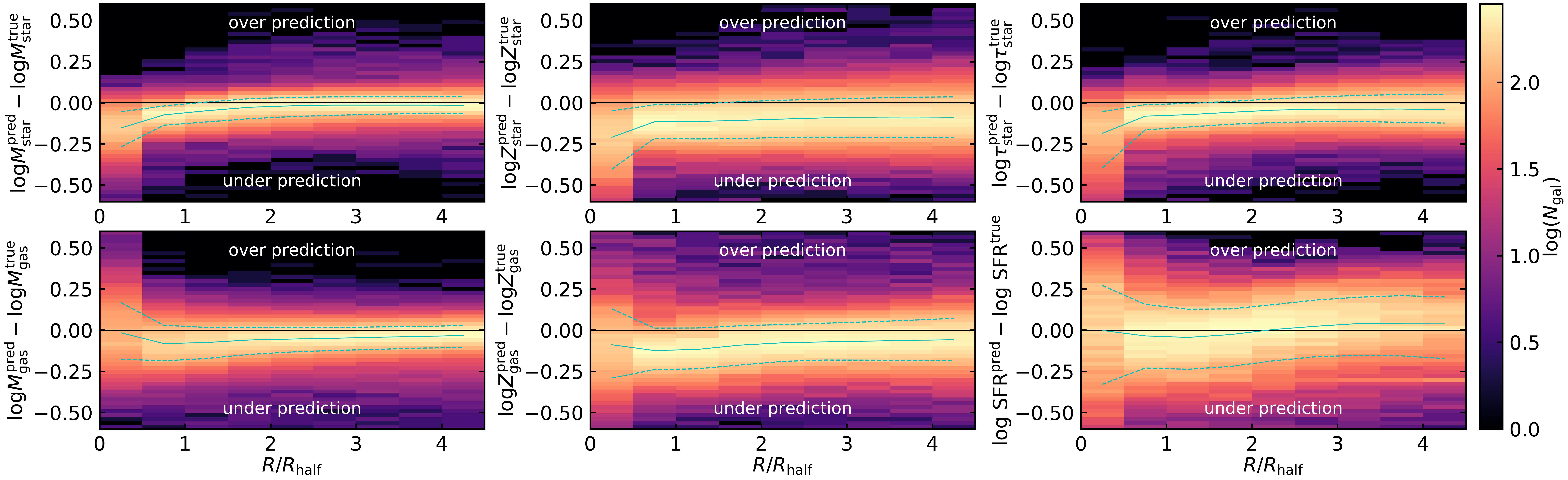}
\end{center}
\vspace{-.35cm}
\oscaption{Analysis_ugriz}{Accuracy of the predicted properties as a function of radius. For each radial bin we plot the distribution of logarithmic errors. Cyan solid lines show the median and dashed lines the $16^{\rm th}$ and $84^{\rm th}$ percentiles. Upper panels show stellar properties, from left to right we display stellar mass, metallicity and age. Lower panels show gaseous properties, from left to right we display gas mass, metallicity and SFR.}
\label{fig:true_vs_pred_rad}
\end{figure*}

\subsection{Galaxy population properties}

Looking at Figs. \ref{fig:images} and \ref{fig:images2}, we can already identify several successes but also failure modes of the model. In general, mass and metals are distributed following the light profile of the galaxy and residuals are small. Both, face-on and edge-on but also inclined galaxies are well reconstructed and internal structures such as rings and clumps or disk warps are recovered.
Similarly, ongoing galaxy mergers are well captured by the model and both galaxies, the central one and the merging galaxy, are simultaneously well modelled. 
However, from the lower right panel of Fig. \ref{fig:images2} we see that although the internal structure of cluster galaxies can be well recovered, their magnitude is under-predicted. The reason for this is, that our training set, the Illustris simulation, has only a small number of such massive clusters and thus, during training the network only encounters very few examples of these clusters. This makes learning their statistical properties difficult.

We perform our final assessment of the quality and accuracy of our method on the validation sample drawn from the whole SDSS mock Illustris data set by \emph{stratified sampling}, i.e. we sorted the images by mass and selected every 5$^{\rm{th}}$ image as a test image in order to represent the long tailed distribution (see Fig. \ref{fig:hist}).

The distribution of stellar (top panels) and gaseous properties (bottom panels) of the galaxies from the validation set is shown in Fig. \ref{fig:hist} by the blue histograms. Here we have defined the galaxy properties as the sum over all pixels within an ellipse of semi-major axis length of two projected half mass radii. Galaxies in the validation set span a stellar mass range from $M_{\rm star}\sim10^{9.5}-10^{12}\Msun$ and a metal mass range of $M_{Z}^{\rm star}\sim10^{7.5}-10^{10.5}\Msun$, both peaked at the low mass end with a tail towards larger masses. The overall SFR of those galaxies spans a range between $\sim10^{-2}-10^{1}\Msun\rm{yr}^{-1}$ with the peak of the distribution around SFR$\sim1-2\Msun\rm{yr}^{-1}$. Corresponding gas masses range from $M_{\rm gas}\sim10^{8}-10^{11}\Msun$ and gas metal masses from $M_{Z}^{\rm gas}\sim10^{7}-10^{9}\Msun$. Both are peaked at the higher mass end with a tail towards lower masses. The galaxy neutral hydrogen fractions show values between $M_{\rm HI}^{\rm gas}\sim10^{7.5}-10^{10.5}\Msun$ with the peak of the distribution again at the high mass end. 
With orange histograms we show the predicted galaxy properties from the ML algorithm whose statistical properties are in good agreement with the true galaxy properties. We find slight offsets towards lower values of gas metallicity and neutral hydrogen fraction for our predictions.

\subsection{Individual galaxy properties}

A more detailed comparison between predicted galaxy properties and their true values is shown in Fig. \ref{fig:true_vs_pred} where we plot the true value on the x-axis against the prediction on the y-axis for each galaxy in our sample. The left panels show stellar properties and the right hand side shows gaseous properties. From top to bottom we show mass, metal mass and stellar age or SFR. In this plot, each point represents one galaxy and compares the total sum of all pixels within two half mass radii, $2\,R_{\rm half}$, between the predicted and the true image. Color coding of the points highlights a kernel density estimate of the galaxy number density. In each panel the inset shows the distribution of the residuals between predicted and true value in logarithmic scale defined as $\Delta\phi=\log(\phi_{\rm true})-\log(\phi_{\rm pred})$. The corresponding mean residual, $\mu$, and the $1\sigma$ scatter are printed in the upper left corner of each panel. The mean residual can be interpreted as a measure of how well we can predict a spectroscopic property based on photometric images while the corresponding scatter is a measure of how accurate those predictions are on average.
	
Figure \ref{fig:true_vs_pred} shows that our image reconstruction is unbiased for most galaxy properties and does not show any strong dependence on galaxy property. For each property our predictions scatter along the 1:1 line with relatively small intrinsic scatter of $\sim 20\%$. We note some systematic deviations from the 1:1 relation at the highest masses, metal masses as well as at the highest SFRs/stellar ages. Stellar and gas masses are best reproduced by the ML algorithm with systematic residual scatter of $\sim0.09$ dex and $\sim0.13$ dex, respectively. Note, this is much smaller than traditional mass estimates from mass-to-light ratios or color images \citep[e.g.][]{Courteau2014,Roediger2015}. 
Correspondingly, metal masses are predicted to better than $\sim0.14$ dex for stellar metal mass and $\sim0.19$ dex for gaseous metal mass. Average stellar ages can be predicted to better than $\sim0.09$ dex while the SFR has intrinsic residual scatter of $\sim0.22$ dex. We further note that all properties are slightly under-predicted, at most for the stellar metal mass with $\sim0.16$ dex (see also next section for more details).

\begin{figure*}
\begin{center}
\includegraphics[width=.485\textwidth]{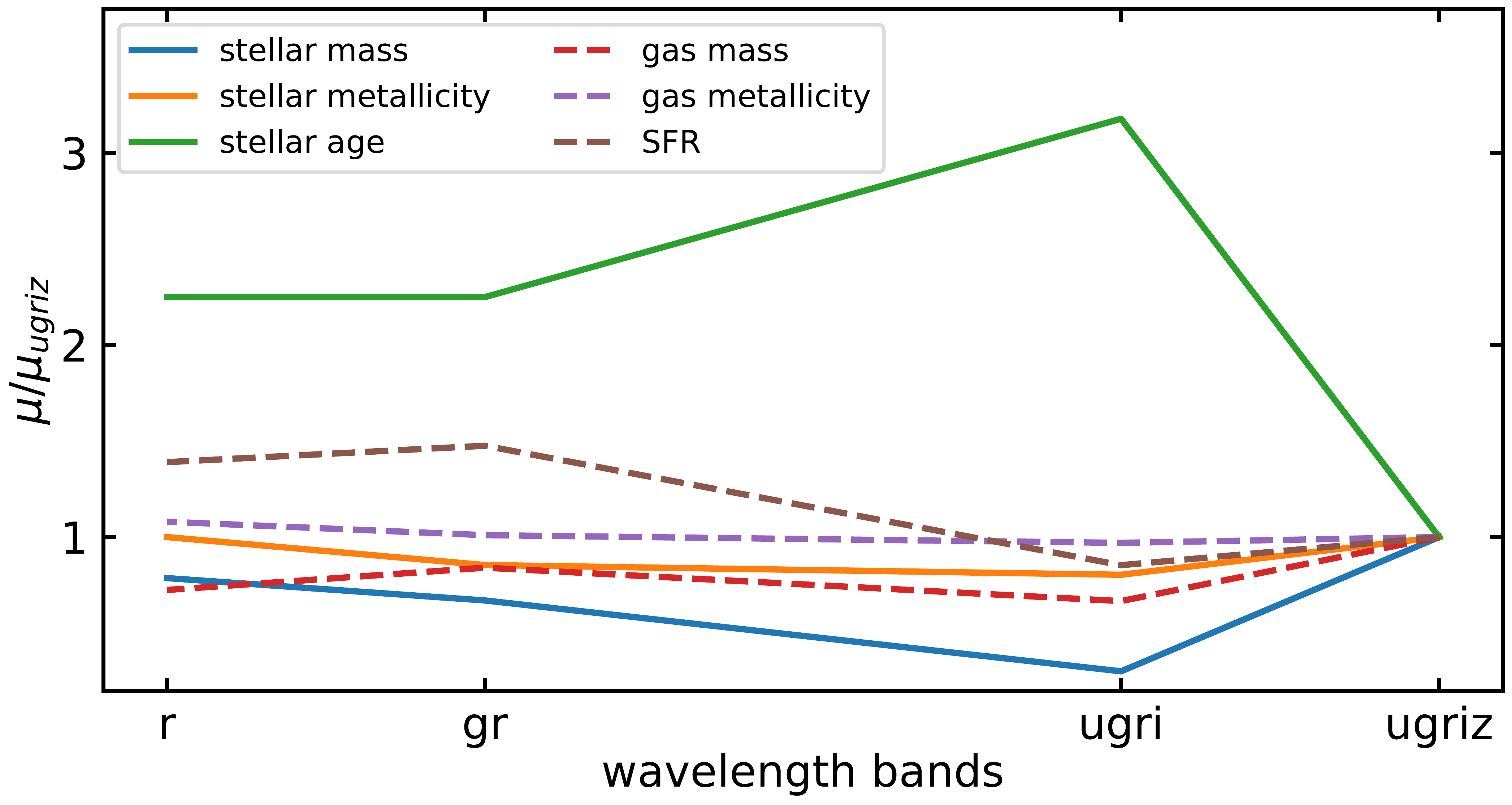}
\includegraphics[width=.5\textwidth]{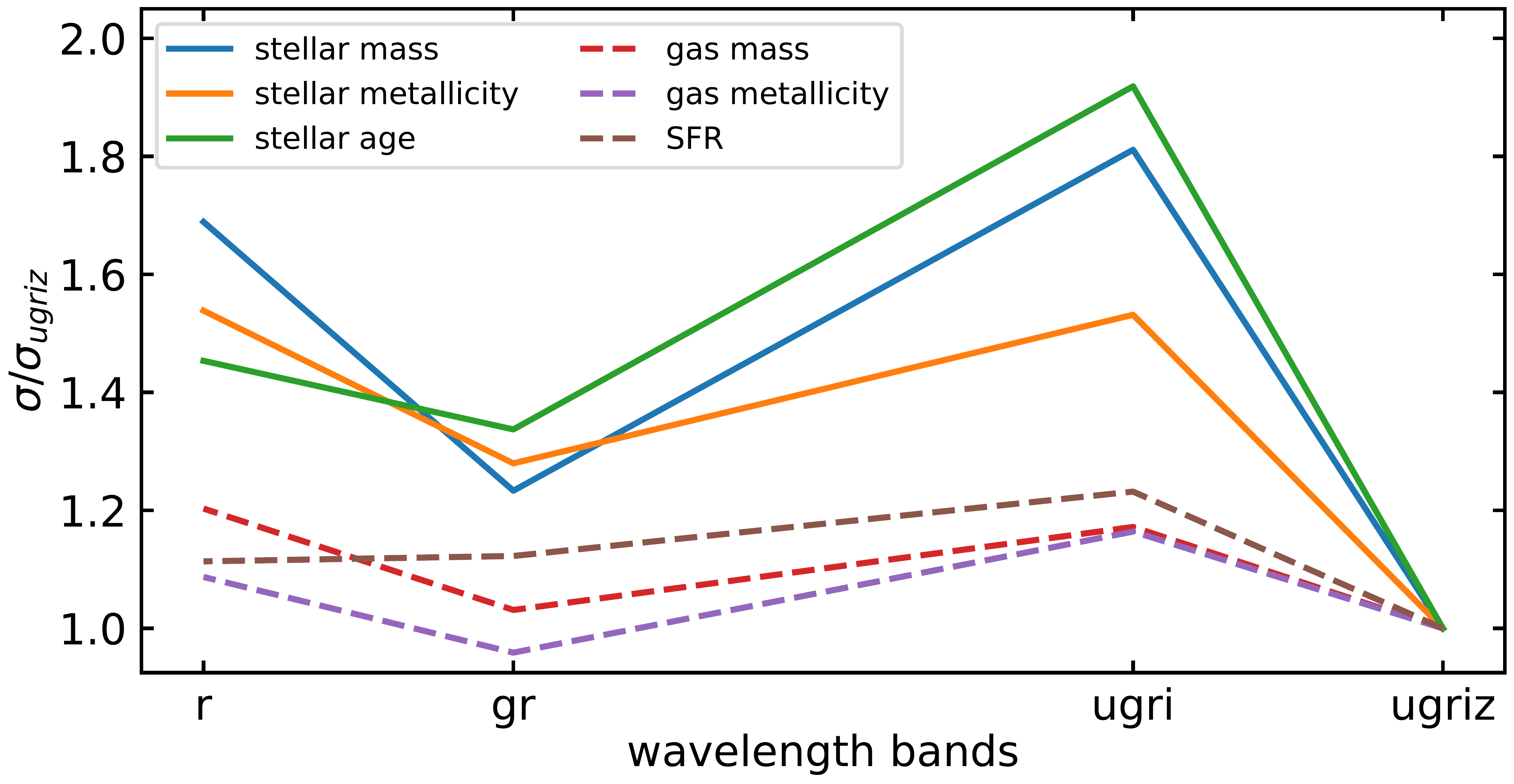}
\end{center}
\vspace{-.35cm}
\oscaption{channel_and_resolution_experiments}{Dependence of the reconstruction accuracy and precision as a function of wavelength bands used. The left panel shows the accuracy to which galaxy properties within $2\,R_{\rm half}$ can be reconstructed and the right panel shows how the precision of the reconstruction changes with wavelength bands used.}
\label{fig:band_comp}
\end{figure*}

\subsection{Testing for systematics with galaxy radius}
\label{sec:radial}

In order to exploit the full information content of the two dimensional images it is important to quantify how well we can predict galaxy properties at different galactic radii. For example, we have seen in Fig. \ref{fig:true_vs_pred} that galaxy properties within twice the projected stellar half mass radius are reasonably well reconstructed. However, a priori it is not clear if the ML algorithm achieves this by simply predicting an average pixel value or indeed by reproducing the actual two dimensional structure of the galaxy image. However, Fig. \ref{fig:images} shows the potential of the CNN to actually reconstruct small morphological features of the galaxies as well as merging galaxy pairs. 

Thus, in Fig. \ref{fig:true_vs_pred_rad} we investigate how well different properties are reproduced as a function of galactic radius. To this extent, we compare the residual between true and predicted properties in different elliptical annuli ranging from $0.5-5.0\, R_{\rm half}$ in semi-major axis length in bins of $0.5\, R_{\rm half}$. The upper panels in Fig. \ref{fig:true_vs_pred_rad} show stellar properties and lower panels the gaseous properties. From left to right we show mass, metal mass and stellar age or SFR. Overall we do not find a strong radial dependence of the reconstructed galaxy properties but we note some interesting features.

While stellar mass is equally well reproduced at each radius with a residual scatter of less than $0.1$ dex, the gas mass reconstruction shows increased scatter inside one half mass radius. Looking at the metal mass we find that the stellar metal mass is under-predicted by $\sim0.15$ dex at all radii and the scatter is $\sim0.2$ dex. Only in the very galaxy center we find a slight increase of under-prediction. Gas metal mass (and similarly average stellar age) does not show much radial dependence and is under-predicted by $\sim0.1$ dex at each radial bin.  
The SFR is surprisingly well predicted with no systematic bias but shows the largest scatter of $\sim0.25$ dex as we have already seen from Fig. \ref{fig:true_vs_pred}.

\subsection{Evaluating the ML prediction as a function of available wavelength bands}
\label{sec:bands}

\begin{figure*}
\begin{center}
\includegraphics[width=.5\textwidth]{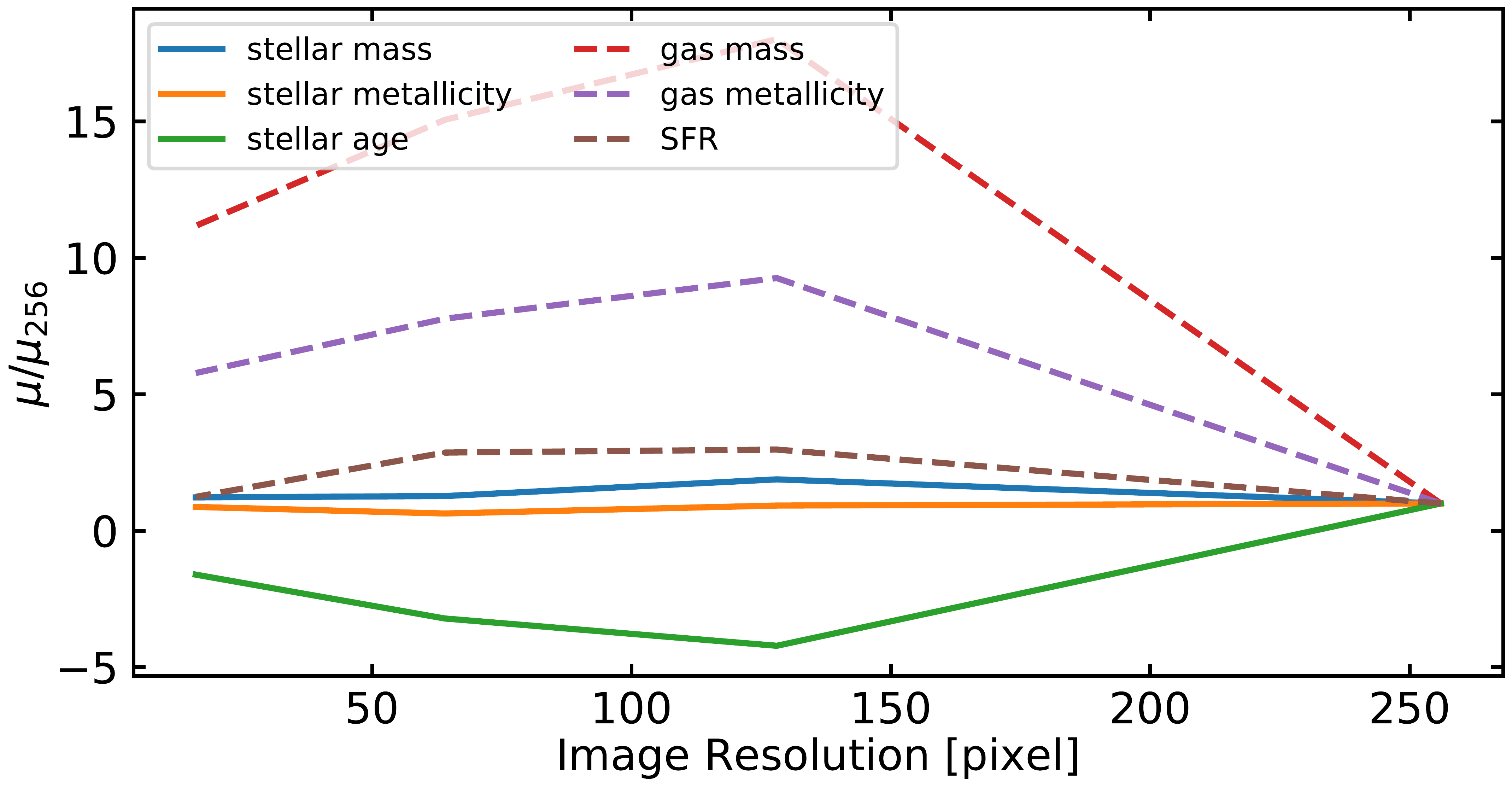}
\includegraphics[width=.4875\textwidth]{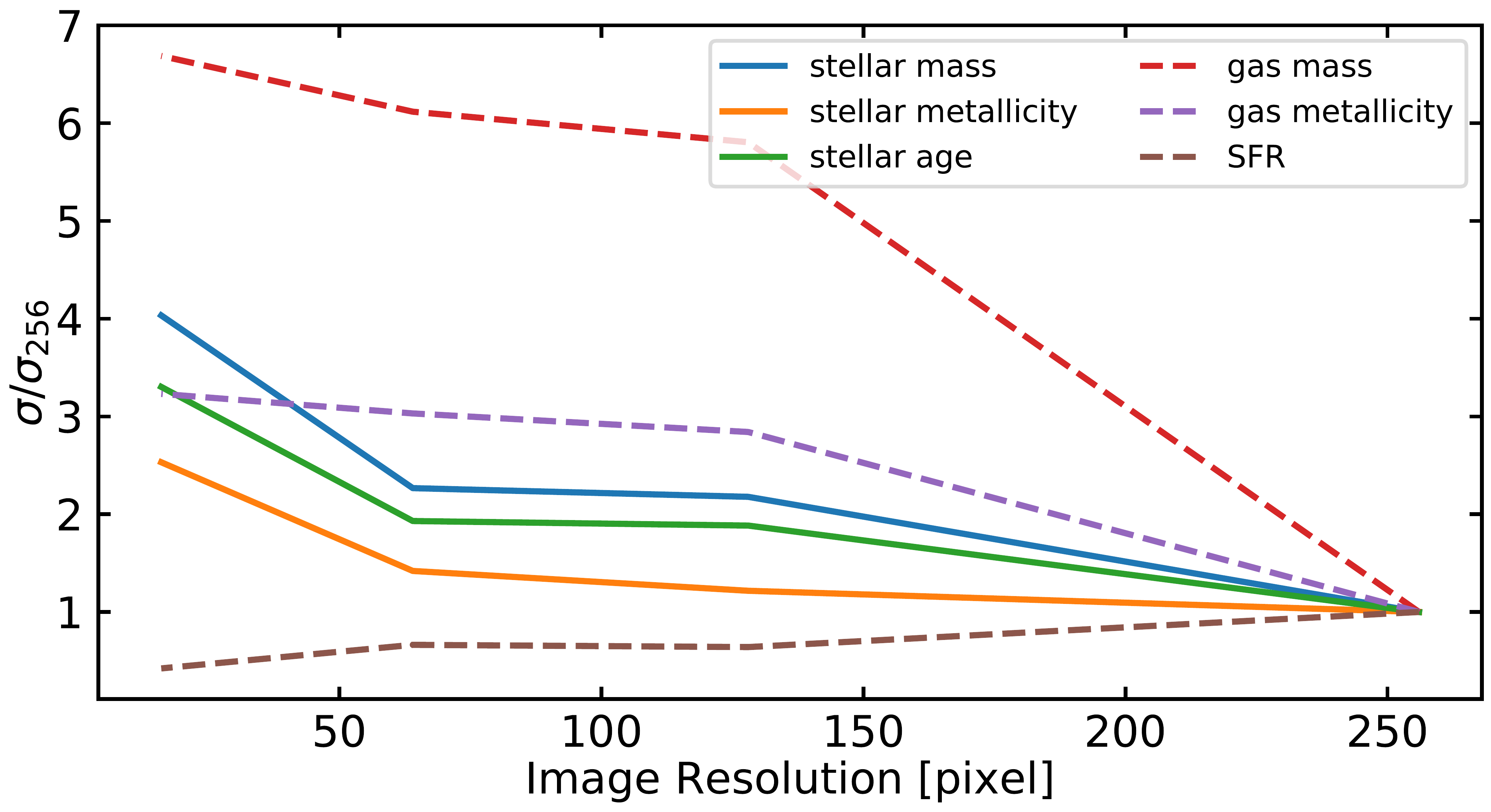}
\end{center}
\vspace{-.35cm}
\oscaption{channel_and_resolution_experiments}{Dependence of the reconstruction accuracy and precision as a function of image resolution. The left panel shows the accuracy to which galaxy properties within $2\,R_{\rm half}$ can be reconstructed and the right panel shows how the precision of the reconstruction changes with image resolution.}
\label{fig:res_comp}
\end{figure*}

The accuracy of the galaxy reconstruction might depend on the number of available wavelength bands. This simply highlights the fact that stellar populations of different metallicity and age obey a different spectral energy distribution. With more available bands and therefore better wavelength coverage, a better representation of the entire spectrum can be achieved. Therefore, leaving out specific bands during training reduces the amount of information passed to the network. We have explored a few representative combinations of bands to study this effect. We have repeated the analysis of the previous section using only the \emph{ugri}, \emph{gr} and \emph{r} band images during training and prediction. This aims at probing the effect of leaving out the infrared part of the spectrum (\emph{z} and/or \emph{i}-band), the blue part of the spectrum (\emph{u}-band) or only using a single band without any color information. 

In Fig. \ref{fig:band_comp} we show how the accuracy and precision of the galaxy reconstruction changes with available wavelength bands. To this end, we compare how the values of $\mu$ and $\sigma$ as calculated for Fig. \ref{fig:true_vs_pred} change. For completeness, for each combination of bands we show a similar figure to Fig. \ref{fig:true_vs_pred} in the appendix \ref{sec:ap_bands} (Figs. \ref{fig:true_vs_pred2}, \ref{fig:true_vs_pred3}, \ref{fig:true_vs_pred4}). The left panel of Fig. \ref{fig:band_comp} shows the change of $\mu$ normalized to $\mu_{ ugriz}$ for the \emph{ugriz} combination of bands. The right panel shows $\sigma$ normalized to $\sigma_{ ugriz}$, respectively.
This figure shows that, except for stellar age, $\mu$ does not change much as a function of available bands. Stellar/gas mass and metallicity for the whole population are roughly equally well reconstructed even when using only \emph{r}-band information. This might suggest that total luminosity in combination with morphology seems to be enough to predict these properties. SFR on the other hand is less accurately reconstructed as $\mu$ increases with decreasing number of bands.  

The scatter, $\sigma$, for stellar properties generally increases with decreasing number of bands while the one for gas properties only increase mildly by $\lesssim20\%$. Interestingly, the increase in stellar scatter is largest ($\sim60\%$) when leaving out only the \emph{z}-band and stays roughly constant when reducing the number of bands further. A similar result is further obtained for the radial dependence of the reconstruction. 
It is yet unclear if this means that the \emph{z}-band carries a lot information or if this would also happen for a combination of \emph{griz} bands due to the reduced color information. Pin-pointing this is currently outside the scope of this study.

In summary, the analysis done for Fig. \ref{fig:band_comp} shows that on average the accuracy of the prediction is only little affected by color information but the precision decreases with decreasing number of wavelength bands. 

\subsection{Evaluating the ML prediction as a function of image size}
\label{sec:resolution}

Similar to color information from using different wavelength bands, an important factor for the reconstruction of galaxy properties might come from the morphological information in the image. 
Here we test, how well spatially resolved a galaxy image must be to achieve good accuracy in the reconstruction of galaxy properties. Therefore we reduce the spatial resolution of our fiducial images (256x256 pixel) by factors of two and retrain the network. Thus we study how well galaxy properties can be reconstructed when images of 128, 64 and 16 pixels are used. 

Since image resolution is affected by either the resolving power of the telescope or by the distance of the object, this either tests at fixed telescope resolution how far away a galaxy sample can be to achieve a good reconstruction or at fixed distance it tests for the minimum resolution of a telescope.
We summarise our results with Fig. \ref{fig:res_comp} which in the left panel again shows the change of $\mu$ normalized to $\mu_{256}$ for the fiducial 256x256 \emph{ugriz} combination of bands and the right panel shows $\sigma$ normalized to $\sigma_{256}$, respectively. 

We find that stellar properties such as stellar mass, metallicity or age are less affected by image resolution than gas properties. We find that stellar mass and metallicity, as well as SFR, are equally accurately predicted across all image resolutions. Stellar age on the other hand is strongly over-predicted. Only the precision for stellar mass and metallicity decreases constantly up to a factor of 3 increased scatter at the lowest image resolution compared to the fiducial resolution of 256x256 pixels. The precision for SFR on the other hand stays roughly constant across all resolutions. Gas mass and metallicity are heavily under-predicted (up to a factor of $\gsim10$ compared to our fiducial resolution) when the image resolution is changed and similarly their precision decreases with decreasing image resolution. The precision of gas metallicity decreases similarly to the stellar metallicity and gas mass precision degrades strongly with decreasing image resolution, up to a factor of 7 compared to our fiducial resolution.

\begin{figure*}
\begin{center}
\includegraphics[width=.5\textwidth]{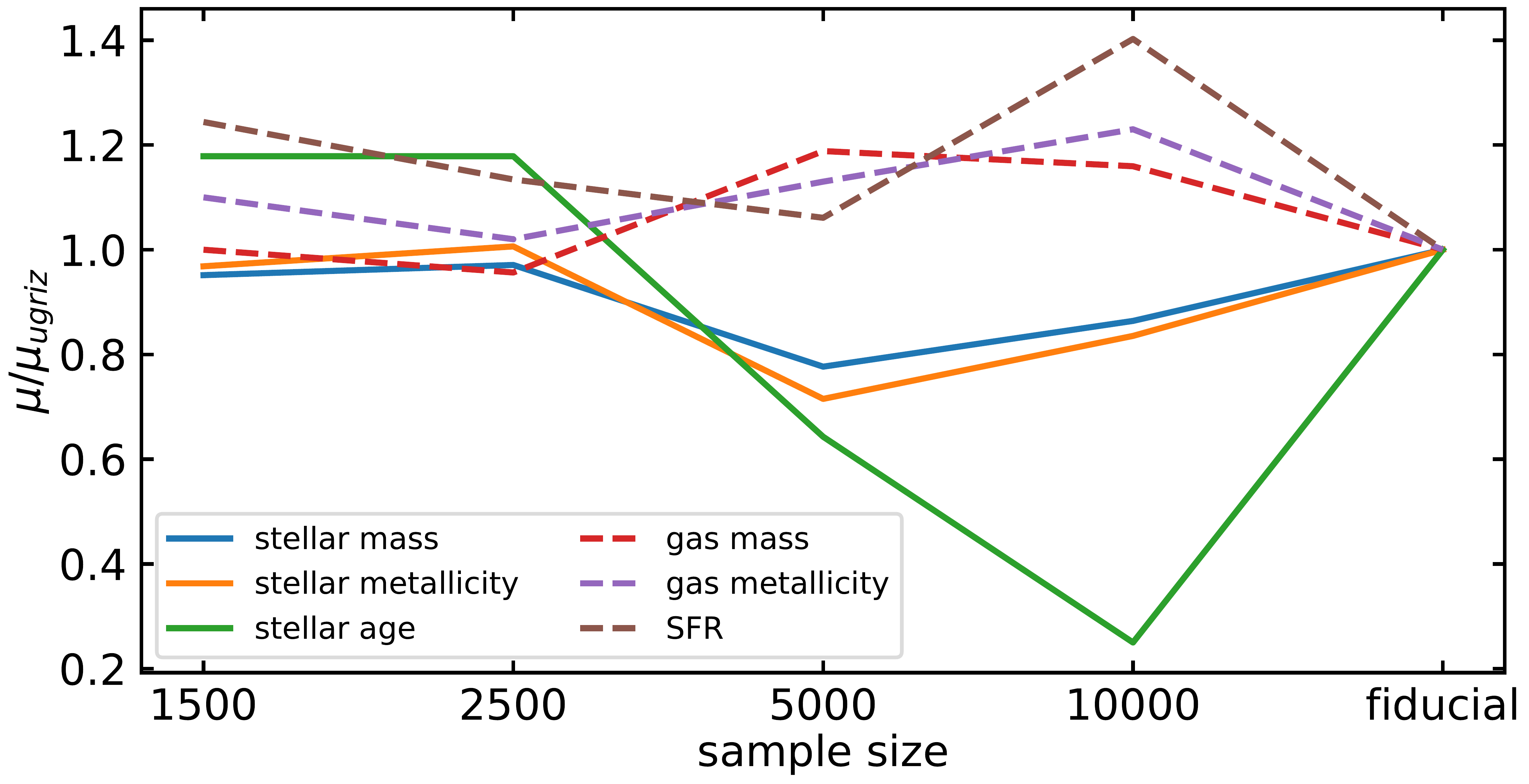}
\includegraphics[width=.4875\textwidth]{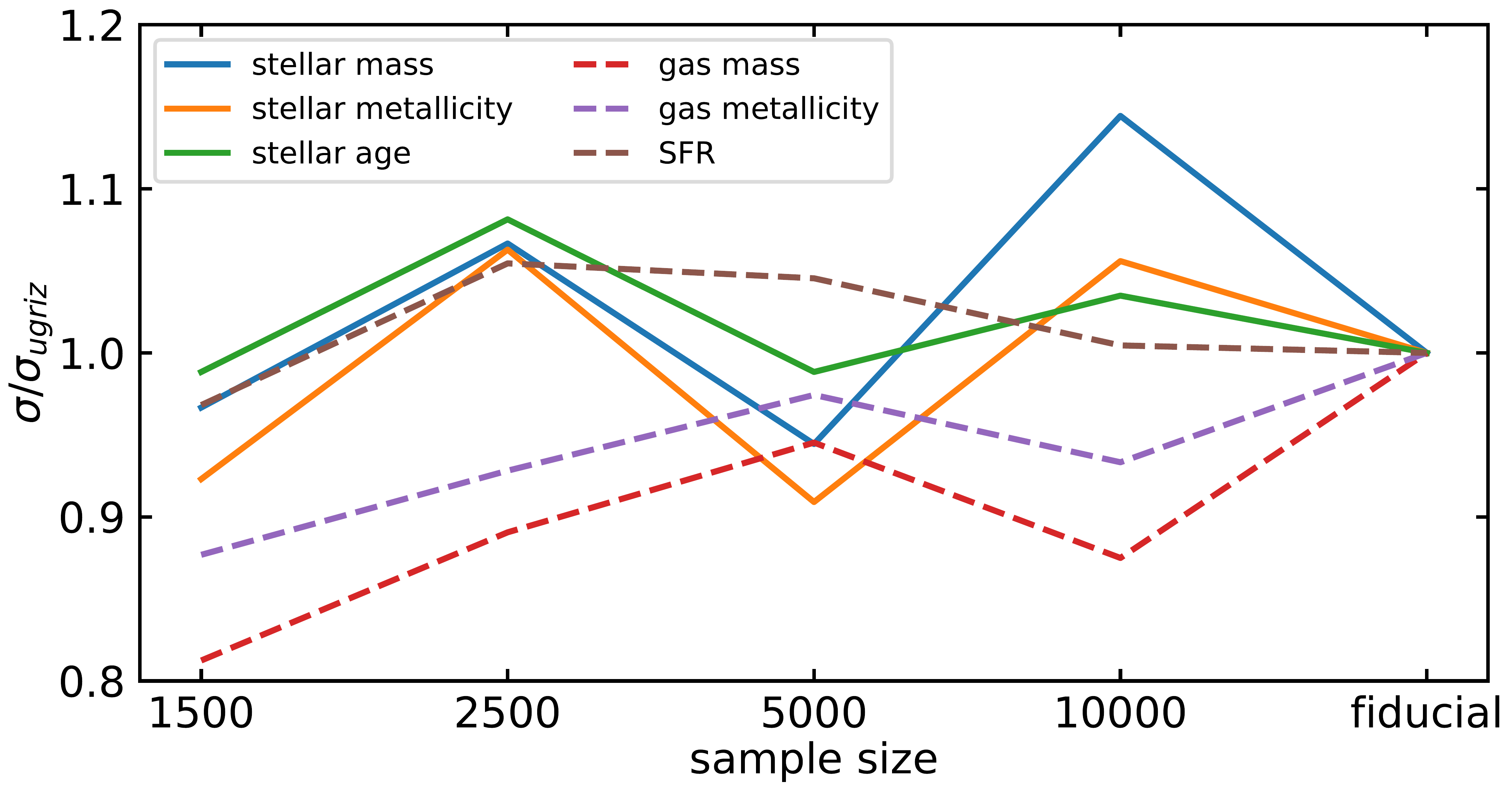}
\end{center}
\vspace{-.35cm}
\oscaption{channel_and_resolution_experiments}{Dependence of the reconstruction accuracy and precision as a function of training sample size. The left panel shows the accuracy to which galaxy properties within $2\,R_{\rm half}$ can be reconstructed and the right panel shows how the precision of the reconstruction changes with the number of galaxy images in the training sample.}
\label{fig:sample_comp}
\end{figure*}

We attribute this to the fact that the gas morphology shows more complexity than the stellar body.
This can readily be seen from the example galaxy shown in Figs. \ref{fig:app_res_comp} and \ref{fig:app_res_comp_gas} in the appendix \ref{sec:ap_res}. Furthermore, we train on stellar light from \emph{ugriz} SDSS bands and naturally, the main contribution to these images comes from the stellar body of the galaxy. The gas of the galaxy will only marginally affect the SED via absorption and emission lines. Thus, it is not unexpected that gas properties are less well reproduced when the image resolution is changing. In fact, it is actually surprising that gas properties can be reproduced at all (even in our fiducial setup) given that galaxy images are mainly tracing stellar light.

The tests performed in this sub-section make us conclude that galaxy morphology contains a significant amount of information about galactic properties. Foremost gaseous properties such as gas mass or metallicity are mainly learned from morphological features. This becomes obvious when comparing to the results from the previous sub-section where we tested the information content of galaxy colors at fixed image resolution and thus at a fixed amount of morphological features resolved. Note, the total luminosity of the galaxies stays the same when resolution is changed. Thus, the only variable for this test is the amount of resolved morphology. This makes us conclude that galaxy morphology carries much information about galaxy properties, not only in a global sense for total stellar mass but also as a resolved map-like feature \citep[see also][for the importance of morphology in predicting galaxy SFR]{Yesuf2021}. 

These findings complement the textbook knowledge of galaxy luminosity and color being a good predictor for its total (stellar) mass \citep[e.g.][]{Kauffmann2003} and similarly galaxy color being a good predictor for its metallicity (modulo stellar age effects) \citep[e.g.][]{Tremonti2004,Gallazzi2005}. Furthermore, in contrast to this established knowledge, adding and employing galaxy morphology through modern image recognition tools does not only enable to reconstruct galaxy properties from images alone, it further allows an estimate of the 2d maps of those properties. This presents a major advantage with respect to previous analysis where only a global measurement could be obtained.

\subsection{Evaluating the ML prediction as a function of training sample size}
\label{sec:sample_size}

Finally, in order to apply the concept presented in this work to real galaxy images we need to retrain the network on a sample of real galaxy images and their reconstructed physical properties, e.g. from IFU observations. To this extent is necessary to know the sensitivity of the algorithm to the number of galaxy images in the training sample, i.e. we want to know if a sample of galaxies of the size of SAMI (3000 galaxies) or merely of the size of MANGA (10,000 galaxies) is needed to achieve a good reconstruction accuracy.

We have retrained our neural network varying the training sample size from 1 500 up to the fiducial sample size of 17 637 galaxy image pairs and present the results in Fig.~\ref{fig:sample_comp}. While the accuracy and precision vary by about $10-20\%$ when changing the sample size we do not find any clear trend with decreasing sample size. We attribute the change in reconstruction accuracy and precision to the fact that retraining and determining the optimal network weights is a stochastic process. Thus, testing the statistical significance of the $\sim10\%$ changes would require retraining a statistical sample of a few ten networks which is outside the scope of this work.
We conclude from Fig.~\ref{fig:sample_comp} that within the reach of current IFU sample sizes the accuracy and precision of the neural network reconstruction does not change significantly. However, when evaluating the accuracy and precision in a real application it will be necessary to train an ensemble of networks to carefully pin-down the reconstruction uncertainties.

\section{Discussion: Application to observations}
\label{sec:disc}

Before we conclude, we will discuss in this section how the reconstruction of galaxy properties via ML applications in a real observational setup could work.
In this work, we have presented a proof-of-concept how modern CNNs are able to reconstruct physical galaxy properties from multi-band galaxy images by making use of the combined spatial color, luminosity and morphological information contained in the images. There are several advantages of employing ML techniques for this task: (i) morphological features can be taken into account via the CNN, (ii) a generally faster analysis of image data but also that (iii) image analysis can be \textbf{automated} easier and \textbf{generalizes} better as no manual, human biased parameter choices have to be made (except for hyper-parameter choices of the network). Furthermore, \textbf{transferablity} to other datasets is easier as a simple retraining on new data is needed. Sometimes, only the higher layers of the network which take care of the survey specific features need to be retrained and deeper layers that cope with the more general features can simply be ported (transfer learning).

In terms of \textbf{uncertainty}, the approach presented here relies on the fact, that enough training data to learn the statistical relations of galaxies exists and that the quality of this data is good enough in terms of sampling a fair representation of the objects. As a proof-of-concept, we have used mock SDSS \emph{ugriz} images from the Illustris dataset which has enough objects sampling a range of galactic morphologies. additionally, the true galaxy properties can easily be created from the raw simulation data. This in turn means, the network effectively learns the statistical properties of the Illustris Universe which obviously differs from the real Universe. Therefore, it is not recommended to train on mock images of any simulated model and then apply the network to real galaxy images.

There are two options to proceed from here: (i) use a training set, where \emph{ground truth} data is self-consistently derived from observational data, i.e. IFU observations where resolved galaxy properties are derived e.g. via spectral energy distribution fitting. (ii) Modify the supervised ML network to not blindly learn statistical relations of the data but be able to either include prior knowledge about galaxies or to move to unsupervised ML techniques to alleviate the pure reliance on \emph{ground truth} data.
 
Point (i) is the most straightforward way of applying the proposed reconstruction network to real observational data such SAMI, MANGA or CALIFA. Those IFU studies have enabled the classical reconstruction of galaxy properties such as stellar mass and metallicity but also gas mass and gas metallicity from the spectral information included in the  IFU data of several thousands of galaxies. Combining these datasets with images of photometric surveys of the same objects will then results in image-physical property pairs that can be fed to the CNN to train it. Thus, one would be able to transfer the knowledge gained from the "small" sample of IFU observed galaxies onto the larger set of galaxies from photometric images (millions of galaxies already in SDSS). Especially with existing large photometric surveys such as e.g. the Dark Energy Survey \citep[DES][]{DES,DES2005}, the Dark Energy Spectroscopic Instrument (DESI) Legacy Imaging Surveys \citep{DESI} or upcoming large photometric surveys from the next-generation of 30m class ground-based telescopes or space-based observations like EUCLID \citep{euclid}, JWST \citep{jwst}, the Rubin Observatory Legacy Survey of Space and Time \citep[LSST][]{LSST,LSST2019} or the Nancy Grace Roman Telescope in combination with photometric redshift estimates \citep[e.g.][]{Henghes2021} this approach might enable to use their imaging data in an unforeseen way.  


\section{Summary \& Conclusion} 
\label{sec:concl}

In this work we set out to study the information content of galaxy images. We are specifically interested to evaluate how well physical properties of galaxies such as stellar mass or metallicity maps, stellar age maps or gas mass and metallicity maps can be predicted from multi-band photometric images. To tackle this question we build upon modern ML techniques for image recognition such as CNNs. As a proof-of-concept we have studied more than 27,000 SDSS \emph{ugriz} mock images from the Illustris simulation suite. We have extracted maps of physical quantities (stellar/gas mass and metallicity, stellar age and SFR and neutral HI gas mass) on the same scale and resolution as the corresponding SDSS mock images. The network architecture and all code is publicly available on github (network at \url{https://github.com/Steffen-Wolf/picasso_training} analysis and plotting routines at \url{https://github.com/TobiBu/picassso}).

In our fiducial setup we have used galaxy images in all 5 SDSS bands (\emph{ugriz}) with a resolution of 256x256 pixels as input to the network in order to predict maps of all seven physical quantities (stellar and gas mass/metallicity, HI mass, stellar age and SFR) at the same time. Additionally, we test for the importance of color information in the prediction by reducing the number of input bands from 5 to 1 as well as the importance of image resolution by reducing the number of pixels from 256 to 16 per side. Further, we test how many galaxy images in the training sample are needed to achieve a decent prediction accuracy by varying the number of training images from the fiducial 17 637 down to only 1 500. 

Our results are summarized as follows:
\begin{itemize}
\item The ML network tested in this work (see Fig. \ref{fig:network} for a sketch of the architecture) is able to predict global stellar and gaseous galaxy properties from broad band photometric images for galaxies spanning a range of more than 3 orders of magnitude in mass and thus luminosity (see Fig. \ref{fig:hist}). Remarkably, this includes not only the reconstruction of the central galaxy in an image but also merging or overlapping galaxies (see e.g. Fig. \ref{fig:images}.  
\item In our fiducial setup, using all five \emph{ugriz} SDSS bands, we recover the true stellar properties on a pixel by pixel basis with only little scatter, $\lesssim10\%$ compared to $50\%$ statistical uncertainty from traditional mass-to-light-ratio based methods. Gaseous properties such as gas mass and metallicity are slightly less well predicted with a scatter of around $\sim20\%$. Similarly, stellar age and SFR are reconstructed with a population wide scatter of $\sim20-50\%$ (c.f. Fig. \ref{fig:true_vs_pred}).
\item Since our algorithm reconstructs full 2d maps of galaxies from the corresponding images, we are able to evaluate how well each property is reconstructed at any given radius (see Fig. \ref{fig:true_vs_pred_rad}). We find no systematic dependence of the reconstruction accuracy with radius except for the very central parts ($R\sim0.5\,R_{\rm half}$) of galaxies where stellar properties are slightly under-predicted.   
\item In order to examine how important color information is in comparison to galaxy morphology, we vary the number of input bands at fixed image resolution (see Fig. \ref{fig:band_comp}) as well as the image resolution at fixed number of input bands (Fig. \ref{fig:res_comp}). We find that reducing the number of input bands causes a $\sim40-60\%$ increase in scatter for stellar properties and $\sim20\%$ for gaseous properties. On the other hand, down grading the resolution blows up the scatter by up to a factor of 3-4 for stellar properties and $\sim7$ for gas mass compared to the fiducial resolution. This makes us conclude that galaxy morphology in addition to color carries significant amount of information for the reconstruction of galaxy properties.
\item Finally, we explore how large the training sample size needs to be to enable a decent reconstruction accuracy and find no significant variations of accuracy with training sample size. Within the numbers of current IFU surveys ($\sim1,500$ - 10,000 galaxies) the reconstruction accuracy does not vary much.  
\end{itemize}

\section*{Data Availability}
The network architecture is publicly available on github at \url{https://github.com/Steffen-Wolf/picasso_training} with an example notebook to download the pre-trained network weights available under \url{https://github.com/Steffen-Wolf/picasso_training/blob/main/notebooks/example_load_and_predict.ipynb}. All analysis code is available at \url{https://github.com/TobiBu/picassso} including Jupyter notebooks containing all plotting routines. 

\section*{Acknowledgments}
We like to thank Paul Torrey for helping us to access the Illustris SDSS mock images. We further thank Nik Arora, Sven Buder, Aura Obreja and Christoph Pfrommer for a careful reading of an earlier version of this draft. 
Analysis has been performed on the ISAAC cluster of the Max-Planck-Institut für Astronomie and the HYDRA and DRACO clusters at the Rechenzentrum in Garching.
Steffen Wolf is supported by the Medical Research Council, as part of United Kingdom Research and Innovation (also known as UK Research and Innovation) [MCUP1201/23].
This research made use of the {\sc{matplotlib}} \citep{matplotlib}, {\sc{SciPy}} \citep{scipy}, {\sc{astropy}} \citep{astropy2018} and {\sc{NumPy, IPython and Jupyter}} \citep{numpy,ipython,jupyter} {\sc{python}} packages.
This research made use of Photutils, an Astropy package for detection and photometry of astronomical sources \citep{Bradley2020}.
Hyperlink figures to code access are inspired by Sven Buder and Rodrigo Luger's \href{https://github.com/rodluger/showyourwork}{showyourwork package}.



\bibliography{astro-ph.bib}

\begin{thebibliography}{}
\makeatletter
\relax
\def\mn@urlcharsother{\let\do\@makeother \do\$\do\&\do\#\do\^\do\_\do\%\do\~}
\def\mn@doi{\begingroup\mn@urlcharsother \@ifnextchar [ {\mn@doi@}
  {\mn@doi@[]}}
\def\mn@doi@[#1]#2{\def\@tempa{#1}\ifx\@tempa\@empty \href
  {http://dx.doi.org/#2} {doi:#2}\else \href {http://dx.doi.org/#2} {#1}\fi
  \endgroup}
\def\mn@eprint#1#2{\mn@eprint@#1:#2::\@nil}
\def\mn@eprint@arXiv#1{\href {http://arxiv.org/abs/#1} {{\tt arXiv:#1}}}
\def\mn@eprint@dblp#1{\href {http://dblp.uni-trier.de/rec/bibtex/#1.xml}
  {dblp:#1}}
\def\mn@eprint@#1:#2:#3:#4\@nil{\def\@tempa {#1}\def\@tempb {#2}\def\@tempc
  {#3}\ifx \@tempc \@empty \let \@tempc \@tempb \let \@tempb \@tempa \fi \ifx
  \@tempb \@empty \def\@tempb {arXiv}\fi \@ifundefined
  {mn@eprint@\@tempb}{\@tempb:\@tempc}{\expandafter \expandafter \csname
  mn@eprint@\@tempb\endcsname \expandafter{\@tempc}}}

\bibitem[\protect\citeauthoryear{{Abazajian} et~al.,}{{Abazajian}
  et~al.}{2009}]{SDSS}
{Abazajian} K.~N.,  et~al., 2009, \mn@doi [\apjs]
  {10.1088/0067-0049/182/2/543}, \href
  {http://adsabs.harvard.edu/abs/2009ApJS..182..543A} {182, 543}

\bibitem[\protect\citeauthoryear{{Abbott} et~al.,}{{Abbott} et~al.}{2018}]{DES}
{Abbott} T.~M.~C.,  et~al., 2018, \mn@doi [\apjs] {10.3847/1538-4365/aae9f0},
  \href {https://ui.adsabs.harvard.edu/abs/2018ApJS..239...18A} {239, 18}

\bibitem[\protect\citeauthoryear{{Agarwal}, {Dav{\'e}}  \& {Bassett}}{{Agarwal}
  et~al.}{2018}]{Agarwal2018}
{Agarwal} S.,  {Dav{\'e}} R.,   {Bassett} B.~A.,  2018, \mn@doi [\mnras]
  {10.1093/mnras/sty1169}, \href
  {https://ui.adsabs.harvard.edu/abs/2018MNRAS.478.3410A} {478, 3410}

\bibitem[\protect\citeauthoryear{{Ahn} et~al.,}{{Ahn} et~al.}{2014}]{Ahn2014}
{Ahn} C.~P.,  et~al., 2014, \mn@doi [\apjs] {10.1088/0067-0049/211/2/17}, \href
  {https://ui.adsabs.harvard.edu/abs/2014ApJS..211...17A} {211, 17}

\bibitem[\protect\citeauthoryear{{Arora} et~al.,}{{Arora}
  et~al.}{2021}]{Arora2021}
{Arora} N.,  et~al., 2021, arXiv e-prints, \href
  {https://ui.adsabs.harvard.edu/abs/2021arXiv210907487A} {p. arXiv:2109.07487}

\bibitem[\protect\citeauthoryear{{Bai}, {Liu}, {Wang}  \& {Yang}}{{Bai}
  et~al.}{2019}]{Bai2019}
{Bai} Y.,  {Liu} J.,  {Wang} S.,   {Yang} F.,  2019, \mn@doi [\aj]
  {10.3847/1538-3881/aaf009}, \href
  {https://ui.adsabs.harvard.edu/abs/2019AJ....157....9B} {157, 9}

\bibitem[\protect\citeauthoryear{{Baldry}, {Balogh}, {Bower}, {Glazebrook},
  {Nichol}, {Bamford}  \& {Budavari}}{{Baldry} et~al.}{2006}]{Baldry2006}
{Baldry} I.~K.,  {Balogh} M.~L.,  {Bower} R.~G.,  {Glazebrook} K.,  {Nichol}
  R.~C.,  {Bamford} S.~P.,   {Budavari} T.,  2006, \mn@doi [\mnras]
  {10.1111/j.1365-2966.2006.11081.x}, \href
  {https://ui.adsabs.harvard.edu/abs/2006MNRAS.373..469B} {373, 469}

\bibitem[\protect\citeauthoryear{{Beck} et~al.,}{{Beck}
  et~al.}{2018}]{Beck2018}
{Beck} M.~R.,  et~al., 2018, \mn@doi [\mnras] {10.1093/mnras/sty503}, \href
  {https://ui.adsabs.harvard.edu/abs/2018MNRAS.476.5516B} {476, 5516}

\bibitem[\protect\citeauthoryear{{Bell}, {McIntosh}, {Katz}  \&
  {Weinberg}}{{Bell} et~al.}{2003}]{Bell2003}
{Bell} E.~F.,  {McIntosh} D.~H.,  {Katz} N.,   {Weinberg} M.~D.,  2003, \mn@doi
  [\apjs] {10.1086/378847}, \href
  {https://ui.adsabs.harvard.edu/abs/2003ApJS..149..289B} {149, 289}

\bibitem[\protect\citeauthoryear{{Bluck}, {Maiolino}, {S{\'a}nchez}, {Ellison},
  {Thorp}, {Piotrowska}, {Teimoorinia}  \& {Bundy}}{{Bluck}
  et~al.}{2020}]{Bluck2020}
{Bluck} A. F.~L.,  {Maiolino} R.,  {S{\'a}nchez} S.~F.,  {Ellison} S.~L.,
  {Thorp} M.~D.,  {Piotrowska} J.~M.,  {Teimoorinia} H.,   {Bundy} K.~A.,
  2020, \mn@doi [\mnras] {10.1093/mnras/stz3264}, \href
  {https://ui.adsabs.harvard.edu/abs/2020MNRAS.492...96B} {492, 96}

\bibitem[\protect\citeauthoryear{{Bonjean}, {Aghanim}, {Salom{\'e}}, {Beelen},
  {Douspis}  \& {Soubri{\'e}}}{{Bonjean} et~al.}{2019}]{Bonjean2019}
{Bonjean} V.,  {Aghanim} N.,  {Salom{\'e}} P.,  {Beelen} A.,  {Douspis} M.,
  {Soubri{\'e}} E.,  2019, \mn@doi [\aap] {10.1051/0004-6361/201833972}, \href
  {https://ui.adsabs.harvard.edu/abs/2019A&A...622A.137B} {622, A137}

\bibitem[\protect\citeauthoryear{{Boone}}{{Boone}}{2021}]{Boone2021}
{Boone} K.,  2021, arXiv e-prints, \href
  {https://ui.adsabs.harvard.edu/abs/2021arXiv210913999B} {p. arXiv:2109.13999}

\bibitem[\protect\citeauthoryear{{Bottrell} et~al.,}{{Bottrell}
  et~al.}{2019}]{Bottrell2019}
{Bottrell} C.,  et~al., 2019, \mn@doi [\mnras] {10.1093/mnras/stz2934}, \href
  {https://ui.adsabs.harvard.edu/abs/2019MNRAS.490.5390B} {490, 5390}

\bibitem[\protect\citeauthoryear{Bradley et~al.,}{Bradley
  et~al.}{2020}]{Bradley2020}
Bradley L.,  et~al., 2020, astropy/photutils: 1.0.0,
  \mn@doi{10.5281/zenodo.4044744}, \url
  {https://doi.org/10.5281/zenodo.4044744}

\bibitem[\protect\citeauthoryear{{Bruzual} \& {Charlot}}{{Bruzual} \&
  {Charlot}}{2003}]{Bruzual2003}
{Bruzual} G.,  {Charlot} S.,  2003, \mn@doi [\mnras]
  {10.1046/j.1365-8711.2003.06897.x}, \href
  {http://adsabs.harvard.edu/abs/2003MNRAS.344.1000B} {344, 1000}

\bibitem[\protect\citeauthoryear{{Bryant} et~al.,}{{Bryant}
  et~al.}{2015}]{sami2015}
{Bryant} J.~J.,  et~al., 2015, \mn@doi [\mnras] {10.1093/mnras/stu2635}, \href
  {https://ui.adsabs.harvard.edu/abs/2015MNRAS.447.2857B} {447, 2857}

\bibitem[\protect\citeauthoryear{{Buck}, {Macci{\`o}}, {Obreja}, {Dutton},
  {Dom{\'{\i}}nguez-Tenreiro}  \& {Granato}}{{Buck} et~al.}{2017}]{Buck2017}
{Buck} T.,  {Macci{\`o}} A.~V.,  {Obreja} A.,  {Dutton} A.~A.,
  {Dom{\'{\i}}nguez-Tenreiro} R.,   {Granato} G.~L.,  2017, \mn@doi [\mnras]
  {10.1093/mnras/stx685}, \href
  {http://adsabs.harvard.edu/abs/2017MNRAS.468.3628B} {468, 3628}

\bibitem[\protect\citeauthoryear{{Buck}, {Ness}, {Macci{\`o}}, {Obreja}  \&
  {Dutton}}{{Buck} et~al.}{2018}]{Buck2018}
{Buck} T.,  {Ness} M.~K.,  {Macci{\`o}} A.~V.,  {Obreja} A.,   {Dutton} A.~A.,
  2018, \mn@doi [\apj] {10.3847/1538-4357/aac890}, \href
  {http://adsabs.harvard.edu/abs/2018ApJ...861...88B} {861, 88}

\bibitem[\protect\citeauthoryear{{Buck}, {Dutton}  \& {Macci{\`o}}}{{Buck}
  et~al.}{2019a}]{Buck2019}
{Buck} T.,  {Dutton} A.~A.,   {Macci{\`o}} A.~V.,  2019a, \mn@doi [\mnras]
  {10.1093/mnras/stz969}, \href
  {https://ui.adsabs.harvard.edu/abs/2019MNRAS.486.1481B} {486, 1481}

\bibitem[\protect\citeauthoryear{{Buck}, {Ness}, {Obreja}, {Macci{\`o}}  \&
  {Dutton}}{{Buck} et~al.}{2019b}]{Buck2019b}
{Buck} T.,  {Ness} M.,  {Obreja} A.,  {Macci{\`o}} A.~V.,   {Dutton} A.~A.,
  2019b, \mn@doi [\apj] {10.3847/1538-4357/aaffd0}, \href
  {https://ui.adsabs.harvard.edu/abs/2019ApJ...874...67B} {874, 67}

\bibitem[\protect\citeauthoryear{{Buck}, {Obreja}, {Macci{\`o}}, {Minchev},
  {Dutton}  \& {Ostriker}}{{Buck} et~al.}{2020}]{Buck2019d}
{Buck} T.,  {Obreja} A.,  {Macci{\`o}} A.~V.,  {Minchev} I.,  {Dutton} A.~A.,
  {Ostriker} J.~P.,  2020, \mn@doi [\mnras] {10.1093/mnras/stz3241}, \href
  {https://ui.adsabs.harvard.edu/abs/2020MNRAS.491.3461B} {491, 3461}

\bibitem[\protect\citeauthoryear{{Buck}, {Rybizki}, {Buder}, {Obreja},
  {Macci{\`o}}, {Pfrommer}, {Steinmetz}  \& {Ness}}{{Buck}
  et~al.}{2021}]{Buck2021}
{Buck} T.,  {Rybizki} J.,  {Buder} S.,  {Obreja} A.,  {Macci{\`o}} A.~V.,
  {Pfrommer} C.,  {Steinmetz} M.,   {Ness} M.,  2021, \mn@doi [\mnras]
  {10.1093/mnras/stab2736}, \href
  {https://ui.adsabs.harvard.edu/abs/2021MNRAS.508.3365B} {508, 3365}

\bibitem[\protect\citeauthoryear{{Buder} et~al.,}{{Buder}
  et~al.}{2021a}]{Buder2021b}
{Buder} S.,  et~al., 2021a, arXiv e-prints, \href
  {https://ui.adsabs.harvard.edu/abs/2021arXiv210904059B} {p. arXiv:2109.04059}

\bibitem[\protect\citeauthoryear{{Buder} et~al.,}{{Buder}
  et~al.}{2021b}]{Buder2021}
{Buder} S.,  et~al., 2021b, \mn@doi [\mnras] {10.1093/mnras/stab1242}, \href
  {https://ui.adsabs.harvard.edu/abs/2021MNRAS.506..150B} {506, 150}

\bibitem[\protect\citeauthoryear{{Bundy} et~al.,}{{Bundy}
  et~al.}{2015}]{manga2015}
{Bundy} K.,  et~al., 2015, \mn@doi [\apj] {10.1088/0004-637X/798/1/7}, \href
  {https://ui.adsabs.harvard.edu/abs/2015ApJ...798....7B} {798, 7}

\bibitem[\protect\citeauthoryear{{Campagne}}{{Campagne}}{2020}]{Campagne2020}
{Campagne} J.-E.,  2020, arXiv e-prints, \href
  {https://ui.adsabs.harvard.edu/abs/2020arXiv200210154C} {p. arXiv:2002.10154}

\bibitem[\protect\citeauthoryear{{Chabrier}}{{Chabrier}}{2003}]{Chabrier2003}
{Chabrier} G.,  2003, \mn@doi [\pasp] {10.1086/376392}, \href
  {http://adsabs.harvard.edu/abs/2003PASP..115..763C} {115, 763}

\bibitem[\protect\citeauthoryear{{Charnock}, {Lavaux}, {Wandelt}, {Boruah},
  {Jasche}  \& {Hudson}}{{Charnock} et~al.}{2020}]{Charnock2020}
{Charnock} T.,  {Lavaux} G.,  {Wandelt} B.~D.,  {Boruah} S.~S.,  {Jasche} J.,
  {Hudson} M.~J.,  2020, \mn@doi [\mnras] {10.1093/mnras/staa682}, \href
  {https://ui.adsabs.harvard.edu/abs/2020MNRAS.tmp..639C} {}

\bibitem[\protect\citeauthoryear{{Courteau} et~al.,}{{Courteau}
  et~al.}{2014}]{Courteau2014}
{Courteau} S.,  et~al., 2014, \mn@doi [Reviews of Modern Physics]
  {10.1103/RevModPhys.86.47}, \href
  {https://ui.adsabs.harvard.edu/abs/2014RvMP...86...47C} {86, 47}

\bibitem[\protect\citeauthoryear{{Dawson}, {Davis}, {Gomez}, {Schock}, {Zabel}
  \& {Williams}}{{Dawson} et~al.}{2020}]{Dawson2020}
{Dawson} J.~M.,  {Davis} T.~A.,  {Gomez} E.~L.,  {Schock} J.,  {Zabel} N.,
  {Williams} T.~G.,  2020, \mn@doi [\mnras] {10.1093/mnras/stz3097}, \href
  {https://ui.adsabs.harvard.edu/abs/2020MNRAS.491.2506D} {491, 2506}

\bibitem[\protect\citeauthoryear{{Dey} et~al.,}{{Dey} et~al.}{2019}]{DESI}
{Dey} A.,  et~al., 2019, \mn@doi [\aj] {10.3847/1538-3881/ab089d}, \href
  {https://ui.adsabs.harvard.edu/abs/2019AJ....157..168D} {157, 168}

\bibitem[\protect\citeauthoryear{{Dieleman}, {Willett}  \& {Dambre}}{{Dieleman}
  et~al.}{2015}]{Dieleman2015}
{Dieleman} S.,  {Willett} K.~W.,   {Dambre} J.,  2015, \mn@doi [\mnras]
  {10.1093/mnras/stv632}, \href
  {https://ui.adsabs.harvard.edu/abs/2015MNRAS.450.1441D} {450, 1441}

\bibitem[\protect\citeauthoryear{{Dobbels}, {Krier}, {Pirson}, {Viaene}, {De
  Geyter}, {Salim}  \& {Baes}}{{Dobbels} et~al.}{2019}]{Dobbels2019}
{Dobbels} W.,  {Krier} S.,  {Pirson} S.,  {Viaene} S.,  {De Geyter} G.,
  {Salim} S.,   {Baes} M.,  2019, \mn@doi [\aap] {10.1051/0004-6361/201834575},
  \href {https://ui.adsabs.harvard.edu/abs/2019A&A...624A.102D} {624, A102}

\bibitem[\protect\citeauthoryear{{Dolag}, {Komatsu}  \& {Sunyaev}}{{Dolag}
  et~al.}{2016}]{Dolag2016}
{Dolag} K.,  {Komatsu} E.,   {Sunyaev} R.,  2016, \mn@doi [\mnras]
  {10.1093/mnras/stw2035}, \href
  {http://adsabs.harvard.edu/abs/2016MNRAS.463.1797D} {463, 1797}

\bibitem[\protect\citeauthoryear{{Dom{\'e}nech-Moral}, {Mart{\'\i}nez-Serrano},
  {Dom{\'\i}nguez-Tenreiro}  \& {Serna}}{{Dom{\'e}nech-Moral}
  et~al.}{2012}]{Domenech-Moral2012}
{Dom{\'e}nech-Moral} M.,  {Mart{\'\i}nez-Serrano} F.~J.,
  {Dom{\'\i}nguez-Tenreiro} R.,   {Serna} A.,  2012, \mn@doi [\mnras]
  {10.1111/j.1365-2966.2012.20534.x}, \href
  {https://ui.adsabs.harvard.edu/abs/2012MNRAS.421.2510D} {421, 2510}

\bibitem[\protect\citeauthoryear{{Dom{\'\i}nguez S{\'a}nchez},
  {Huertas-Company}, {Bernardi}, {Tuccillo}  \& {Fischer}}{{Dom{\'\i}nguez
  S{\'a}nchez} et~al.}{2018}]{Dominguez2018}
{Dom{\'\i}nguez S{\'a}nchez} H.,  {Huertas-Company} M.,  {Bernardi} M.,
  {Tuccillo} D.,   {Fischer} J.~L.,  2018, \mn@doi [\mnras]
  {10.1093/mnras/sty338}, \href
  {https://ui.adsabs.harvard.edu/abs/2018MNRAS.476.3661D} {476, 3661}

\bibitem[\protect\citeauthoryear{{Dom{\'\i}nguez S{\'a}nchez}
  et~al.,}{{Dom{\'\i}nguez S{\'a}nchez} et~al.}{2019}]{Dominguez2019}
{Dom{\'\i}nguez S{\'a}nchez} H.,  et~al., 2019, \mn@doi [\mnras]
  {10.1093/mnras/sty3497}, \href
  {https://ui.adsabs.harvard.edu/abs/2019MNRAS.484...93D} {484, 93}

\bibitem[\protect\citeauthoryear{{Escamilla-Rivera}, {Carvajal Quintero}  \&
  {Capozziello}}{{Escamilla-Rivera} et~al.}{2020}]{Escamilla2020}
{Escamilla-Rivera} C.,  {Carvajal Quintero} M.~A.,   {Capozziello} S.,  2020,
  \mn@doi [\jcap] {10.1088/1475-7516/2020/03/008}, \href
  {https://ui.adsabs.harvard.edu/abs/2020JCAP...03..008E} {2020, 008}

\bibitem[\protect\citeauthoryear{{Falahkheirkhah}, {Yeh}, {Mittal}, {Pfister}
  \& {Bhargava}}{{Falahkheirkhah} et~al.}{2019}]{Falahkheirkhah2019}
{Falahkheirkhah} K.,  {Yeh} K.,  {Mittal} S.,  {Pfister} L.,   {Bhargava} R.,
  2019, arXiv e-prints, \href
  {https://ui.adsabs.harvard.edu/abs/2019arXiv191104410F} {p. arXiv:1911.04410}

\bibitem[\protect\citeauthoryear{{Gallazzi}, {Charlot}, {Brinchmann}, {White}
  \& {Tremonti}}{{Gallazzi} et~al.}{2005}]{Gallazzi2005}
{Gallazzi} A.,  {Charlot} S.,  {Brinchmann} J.,  {White} S. D.~M.,   {Tremonti}
  C.~A.,  2005, \mn@doi [\mnras] {10.1111/j.1365-2966.2005.09321.x}, \href
  {https://ui.adsabs.harvard.edu/abs/2005MNRAS.362...41G} {362, 41}

\bibitem[\protect\citeauthoryear{{Galligan}, {Katz}, {Kimm}, {Rosdahl},
  {Blaizot}, {Devriendt}  \& {Slyz}}{{Galligan} et~al.}{2019}]{Galligan2019}
{Galligan} T.~P.,  {Katz} H.,  {Kimm} T.,  {Rosdahl} J.,  {Blaizot} J.,
  {Devriendt} J.,   {Slyz} A.,  2019, arXiv e-prints, \href
  {https://ui.adsabs.harvard.edu/abs/2019arXiv190101264G} {p. arXiv:1901.01264}

\bibitem[\protect\citeauthoryear{{Gardner} et~al.,}{{Gardner}
  et~al.}{2006}]{jwst}
{Gardner} J.~P.,  et~al., 2006, \mn@doi [\ssr] {10.1007/s11214-006-8315-7},
  \href {https://ui.adsabs.harvard.edu/abs/2006SSRv..123..485G} {123, 485}

\bibitem[\protect\citeauthoryear{{Goodfellow}, {Pouget-Abadie}, {Mirza}, {Xu},
  {Warde-Farley}, {Ozair}, {Courville}  \& {Bengio}}{{Goodfellow}
  et~al.}{2014}]{Goodfellow2014}
{Goodfellow} I.~J.,  {Pouget-Abadie} J.,  {Mirza} M.,  {Xu} B.,  {Warde-Farley}
  D.,  {Ozair} S.,  {Courville} A.,   {Bengio} Y.,  2014, arXiv e-prints, \href
  {https://ui.adsabs.harvard.edu/abs/2014arXiv1406.2661G} {p. arXiv:1406.2661}

\bibitem[\protect\citeauthoryear{{Henghes}, {Pettitt}, {Thiyagalingam}, {Hey}
  \& {Lahav}}{{Henghes} et~al.}{2021}]{Henghes2021}
{Henghes} B.,  {Pettitt} C.,  {Thiyagalingam} J.,  {Hey} T.,   {Lahav} O.,
  2021, arXiv e-prints, \href
  {https://ui.adsabs.harvard.edu/abs/2021arXiv210902503H} {p. arXiv:2109.02503}

\bibitem[\protect\citeauthoryear{{Hezaveh}, {Perreault Levasseur}  \&
  {Marshall}}{{Hezaveh} et~al.}{2017}]{Hezaveh2017}
{Hezaveh} Y.~D.,  {Perreault Levasseur} L.,   {Marshall} P.~J.,  2017, \mn@doi
  [\nat] {10.1038/nature23463}, \href
  {https://ui.adsabs.harvard.edu/abs/2017Natur.548..555H} {548, 555}

\bibitem[\protect\citeauthoryear{{Ho}, {Rau}, {Ntampaka}, {Farahi}, {Trac}  \&
  {P{\'o}czos}}{{Ho} et~al.}{2019}]{Ho2019}
{Ho} M.,  {Rau} M.~M.,  {Ntampaka} M.,  {Farahi} A.,  {Trac} H.,   {P{\'o}czos}
  B.,  2019, \mn@doi [\apj] {10.3847/1538-4357/ab4f82}, \href
  {https://ui.adsabs.harvard.edu/abs/2019ApJ...887...25H} {887, 25}

\bibitem[\protect\citeauthoryear{{Hocking}, {Geach}, {Sun}  \&
  {Davey}}{{Hocking} et~al.}{2018}]{Hocking2018}
{Hocking} A.,  {Geach} J.~E.,  {Sun} Y.,   {Davey} N.,  2018, \mn@doi [\mnras]
  {10.1093/mnras/stx2351}, \href
  {https://ui.adsabs.harvard.edu/abs/2018MNRAS.473.1108H} {473, 1108}

\bibitem[\protect\citeauthoryear{{Hopkins} et~al.,}{{Hopkins}
  et~al.}{2018}]{Hopkins2018}
{Hopkins} P.~F.,  et~al., 2018, \mn@doi [\mnras] {10.1093/mnras/sty1690}, \href
  {http://adsabs.harvard.edu/abs/2018MNRAS.480..800H} {480, 800}

\bibitem[\protect\citeauthoryear{{Huertas-Company} et~al.,}{{Huertas-Company}
  et~al.}{2015}]{Huertas-Company2015}
{Huertas-Company} M.,  et~al., 2015, \mn@doi [\apjs]
  {10.1088/0067-0049/221/1/8}, \href
  {https://ui.adsabs.harvard.edu/abs/2015ApJS..221....8H} {221, 8}

\bibitem[\protect\citeauthoryear{Hunter}{Hunter}{2007}]{matplotlib}
Hunter J.~D.,  2007, \mn@doi [Computing In Science \& Engineering]
  {10.1109/MCSE.2007.55}, 9, 90

\bibitem[\protect\citeauthoryear{Isola, Zhu, Zhou  \& Efros}{Isola
  et~al.}{2017}]{isola2017image}
Isola P.,  Zhu J.-Y.,  Zhou T.,   Efros A.~A.,  2017, in Proceedings of the
  IEEE conference on computer vision and pattern recognition. pp 1125--1134

\bibitem[\protect\citeauthoryear{{Ivezi{\'c}} et~al.,}{{Ivezi{\'c}}
  et~al.}{2019}]{LSST2019}
{Ivezi{\'c}} {\v{Z}}.,  et~al., 2019, \mn@doi [\apj]
  {10.3847/1538-4357/ab042c}, \href
  {https://ui.adsabs.harvard.edu/abs/2019ApJ...873..111I} {873, 111}

\bibitem[\protect\citeauthoryear{{Jeffrey}, {Lanusse}, {Lahav}  \&
  {Starck}}{{Jeffrey} et~al.}{2020}]{Jeffrey2020}
{Jeffrey} N.,  {Lanusse} F.,  {Lahav} O.,   {Starck} J.-L.,  2020, \mn@doi
  [\mnras] {10.1093/mnras/staa127}, \href
  {https://ui.adsabs.harvard.edu/abs/2020MNRAS.492.5023J} {492, 5023}

\bibitem[\protect\citeauthoryear{{Jo} \& {Kim}}{{Jo} \& {Kim}}{2019}]{Jo2019}
{Jo} Y.,  {Kim} J.-h.,  2019, \mn@doi [\mnras] {10.1093/mnras/stz2304}, \href
  {https://ui.adsabs.harvard.edu/abs/2019MNRAS.489.3565J} {489, 3565}

\bibitem[\protect\citeauthoryear{Jones, Oliphant, Peterson  et~al.}{Jones
  et~al.}{01  }]{scipy}
Jones E.,  Oliphant T.,  Peterson P.,   et~al., 2001--, {SciPy}: Open source
  scientific tools for {Python}, \url {http://www.scipy.org/}

\bibitem[\protect\citeauthoryear{{Jonsson}}{{Jonsson}}{2006}]{Jonsson2006}
{Jonsson} P.,  2006, \mn@doi [\mnras] {10.1111/j.1365-2966.2006.10884.x}, \href
  {https://ui.adsabs.harvard.edu/abs/2006MNRAS.372....2J} {372, 2}

\bibitem[\protect\citeauthoryear{{Kamdar}, {Turk}  \& {Brunner}}{{Kamdar}
  et~al.}{2016}]{Kamdar2016}
{Kamdar} H.~M.,  {Turk} M.~J.,   {Brunner} R.~J.,  2016, \mn@doi [\mnras]
  {10.1093/mnras/stv2981}, \href
  {https://ui.adsabs.harvard.edu/abs/2016MNRAS.457.1162K} {457, 1162}

\bibitem[\protect\citeauthoryear{{Kauffmann} et~al.,}{{Kauffmann}
  et~al.}{2003}]{Kauffmann2003}
{Kauffmann} G.,  et~al., 2003, \mn@doi [\mnras]
  {10.1046/j.1365-8711.2003.06291.x}, \href
  {https://ui.adsabs.harvard.edu/abs/2003MNRAS.341...33K} {341, 33}

\bibitem[\protect\citeauthoryear{{Kim} \& {Brunner}}{{Kim} \&
  {Brunner}}{2017}]{Kim2017}
{Kim} E.~J.,  {Brunner} R.~J.,  2017, \mn@doi [\mnras] {10.1093/mnras/stw2672},
  \href {https://ui.adsabs.harvard.edu/abs/2017MNRAS.464.4463K} {464, 4463}

\bibitem[\protect\citeauthoryear{Kingma \& Ba}{Kingma \&
  Ba}{2014}]{kingma2014adam}
Kingma D.~P.,  Ba J.,  2014, arXiv preprint arXiv:1412.6980

\bibitem[\protect\citeauthoryear{Kluyver et~al.,}{Kluyver
  et~al.}{2016}]{jupyter}
Kluyver T.,  et~al., 2016, in Loizides F.,  Schmidt B.,  eds, Positioning and
  Power in Academic Publishing: Players, Agents and Agendas. pp 87 -- 90

\bibitem[\protect\citeauthoryear{Krizhevsky, Sutskever  \& Hinton}{Krizhevsky
  et~al.}{2012}]{Krizhevsky2012}
Krizhevsky A.,  Sutskever I.,   Hinton G.~E.,  2012, in Pereira F.,  Burges C.
  J.~C.,  Bottou L.,   Weinberger K.~Q.,  eds, , Advances in Neural Information
  Processing Systems 25.
Curran Associates, Inc., pp 1097--1105, \url
  {http://papers.nips.cc/paper/4824-imagenet-classification-with-deep-convolutional-neural-networks.pdf}

\bibitem[\protect\citeauthoryear{{LSST Science Collaboration} et~al.,}{{LSST
  Science Collaboration} et~al.}{2009}]{LSST}
{LSST Science Collaboration} et~al., 2009, arXiv e-prints, \href
  {https://ui.adsabs.harvard.edu/abs/2009arXiv0912.0201L} {p. arXiv:0912.0201}

\bibitem[\protect\citeauthoryear{{Lanusse}, {Ma}, {Li}, {Collett}, {Li},
  {Ravanbakhsh}, {Mandelbaum}  \& {P{\'o}czos}}{{Lanusse}
  et~al.}{2018}]{Lanusse2018}
{Lanusse} F.,  {Ma} Q.,  {Li} N.,  {Collett} T.~E.,  {Li} C.-L.,  {Ravanbakhsh}
  S.,  {Mandelbaum} R.,   {P{\'o}czos} B.,  2018, \mn@doi [\mnras]
  {10.1093/mnras/stx1665}, \href
  {https://ui.adsabs.harvard.edu/abs/2018MNRAS.473.3895L} {473, 3895}

\bibitem[\protect\citeauthoryear{{Lanusse}, {Melchior}  \&
  {Moolekamp}}{{Lanusse} et~al.}{2019}]{Lanusse2019}
{Lanusse} F.,  {Melchior} P.,   {Moolekamp} F.,  2019, arXiv e-prints, \href
  {https://ui.adsabs.harvard.edu/abs/2019arXiv191203980L} {p. arXiv:1912.03980}

\bibitem[\protect\citeauthoryear{{Laureijs} et~al.,}{{Laureijs}
  et~al.}{2011}]{euclid}
{Laureijs} R.,  et~al., 2011, arXiv e-prints, \href
  {https://ui.adsabs.harvard.edu/abs/2011arXiv1110.3193L} {p. arXiv:1110.3193}

\bibitem[\protect\citeauthoryear{{Lintott} et~al.,}{{Lintott}
  et~al.}{2008}]{galaxyzoo2008}
{Lintott} C.~J.,  et~al., 2008, \mn@doi [\mnras]
  {10.1111/j.1365-2966.2008.13689.x}, \href
  {https://ui.adsabs.harvard.edu/abs/2008MNRAS.389.1179L} {389, 1179}

\bibitem[\protect\citeauthoryear{{Lovell}, {Acquaviva}, {Thomas}, {Iyer},
  {Gawiser}  \& {Wilkins}}{{Lovell} et~al.}{2019}]{Lovell2019}
{Lovell} C.~C.,  {Acquaviva} V.,  {Thomas} P.~A.,  {Iyer} K.~G.,  {Gawiser} E.,
    {Wilkins} S.~M.,  2019, \mn@doi [\mnras] {10.1093/mnras/stz2851}, \href
  {https://ui.adsabs.harvard.edu/abs/2019MNRAS.490.5503L} {490, 5503}

\bibitem[\protect\citeauthoryear{{Lucie-Smith}, {Peiris}  \&
  {Pontzen}}{{Lucie-Smith} et~al.}{2019}]{Lucie-Smith2019}
{Lucie-Smith} L.,  {Peiris} H.~V.,   {Pontzen} A.,  2019, \mn@doi [\mnras]
  {10.1093/mnras/stz2599}, \href
  {https://ui.adsabs.harvard.edu/abs/2019MNRAS.490..331L} {490, 331}

\bibitem[\protect\citeauthoryear{{Margalef-Bentabol}, {Huertas-Company},
  {Charnock}, {Margalef-Bentabol}, {Bernardi}, {Dubois}, {Storey-Fisher}  \&
  {Zanis}}{{Margalef-Bentabol} et~al.}{2020}]{Margalef2020}
{Margalef-Bentabol} B.,  {Huertas-Company} M.,  {Charnock} T.,
  {Margalef-Bentabol} C.,  {Bernardi} M.,  {Dubois} Y.,  {Storey-Fisher} K.,
  {Zanis} L.,  2020, arXiv e-prints, \href
  {https://ui.adsabs.harvard.edu/abs/2020arXiv200308263M} {p. arXiv:2003.08263}

\bibitem[\protect\citeauthoryear{{Menou}}{{Menou}}{2019}]{Menou2019}
{Menou} K.,  2019, \mn@doi [\mnras] {10.1093/mnras/stz2477}, \href
  {https://ui.adsabs.harvard.edu/abs/2019MNRAS.489.4802M} {489, 4802}

\bibitem[\protect\citeauthoryear{Newell, Yang  \& Deng}{Newell
  et~al.}{2016}]{newell2016stacked}
Newell A.,  Yang K.,   Deng J.,  2016, in European conference on computer
  vision. pp 483--499

\bibitem[\protect\citeauthoryear{{Ntampaka}, {Trac}, {Sutherland}, {Battaglia},
  {P{\'o}czos}  \& {Schneider}}{{Ntampaka} et~al.}{2015}]{Ntampaka2015}
{Ntampaka} M.,  {Trac} H.,  {Sutherland} D.~J.,  {Battaglia} N.,  {P{\'o}czos}
  B.,   {Schneider} J.,  2015, \mn@doi [\apj] {10.1088/0004-637X/803/2/50},
  \href {https://ui.adsabs.harvard.edu/abs/2015ApJ...803...50N} {803, 50}

\bibitem[\protect\citeauthoryear{{Ntampaka}, {Trac}, {Sutherland},
  {Fromenteau}, {Poczos}  \& {Schneider}}{{Ntampaka}
  et~al.}{2018}]{Ntampaka2018}
{Ntampaka} M.,  {Trac} H.,  {Sutherland} D.,  {Fromenteau} S.,  {Poczos} B.,
  {Schneider} J.,  2018, in American Astronomical Society Meeting Abstracts
  \#231. p. 225.04

\bibitem[\protect\citeauthoryear{{Ntampaka} et~al.,}{{Ntampaka}
  et~al.}{2019}]{Ntampaka2019}
{Ntampaka} M.,  et~al., 2019, \mn@doi [\apj] {10.3847/1538-4357/ab14eb}, \href
  {https://ui.adsabs.harvard.edu/abs/2019ApJ...876...82N} {876, 82}

\bibitem[\protect\citeauthoryear{{Ntampaka}, {Eisenstein}, {Yuan}  \&
  {Garrison}}{{Ntampaka} et~al.}{2020}]{Ntampaka2020}
{Ntampaka} M.,  {Eisenstein} D.~J.,  {Yuan} S.,   {Garrison} L.~H.,  2020,
  \mn@doi [\apj] {10.3847/1538-4357/ab5f5e}, \href
  {https://ui.adsabs.harvard.edu/abs/2020ApJ...889..151N} {889, 151}

\bibitem[\protect\citeauthoryear{{O'Shea} \& {Nash}}{{O'Shea} \&
  {Nash}}{2015}]{CNN2015}
{O'Shea} K.,  {Nash} R.,  2015, arXiv e-prints, \href
  {https://ui.adsabs.harvard.edu/abs/2015arXiv151108458O} {p. arXiv:1511.08458}

\bibitem[\protect\citeauthoryear{{Obreja}, {Macci{\`o}}, {Moster}, {Dutton},
  {Buck}, {Stinson}  \& {Wang}}{{Obreja} et~al.}{2018}]{Obreja2018}
{Obreja} A.,  {Macci{\`o}} A.~V.,  {Moster} B.,  {Dutton} A.~A.,  {Buck} T.,
  {Stinson} G.~S.,   {Wang} L.,  2018, \mn@doi [\mnras]
  {10.1093/mnras/sty1022}, \href
  {https://ui.adsabs.harvard.edu/abs/2018MNRAS.477.4915O} {477, 4915}

\bibitem[\protect\citeauthoryear{{Obreja} et~al.,}{{Obreja}
  et~al.}{2019}]{Obreja2019}
{Obreja} A.,  et~al., 2019, \mn@doi [\mnras] {10.1093/mnras/stz1563}, \href
  {https://ui.adsabs.harvard.edu/abs/2019MNRAS.487.4424O} {487, 4424}

\bibitem[\protect\citeauthoryear{{Pasquato} \& {Chung}}{{Pasquato} \&
  {Chung}}{2016}]{Pasquato2016}
{Pasquato} M.,  {Chung} C.,  2016, \mn@doi [\aap]
  {10.1051/0004-6361/201425181}, \href
  {https://ui.adsabs.harvard.edu/abs/2016A&A...589A..95P} {589, A95}

\bibitem[\protect\citeauthoryear{{Pasquato} \& {Chung}}{{Pasquato} \&
  {Chung}}{2019}]{Pasquato2019}
{Pasquato} M.,  {Chung} C.,  2019, \mn@doi [\mnras] {10.1093/mnras/stz2766},
  \href {https://ui.adsabs.harvard.edu/abs/2019MNRAS.490.3392P} {490, 3392}

\bibitem[\protect\citeauthoryear{{Peek} \& {Burkhart}}{{Peek} \&
  {Burkhart}}{2019}]{Peek2019}
{Peek} J.~E.~G.,  {Burkhart} B.,  2019, \mn@doi [\apjl]
  {10.3847/2041-8213/ab3a9e}, \href
  {https://ui.adsabs.harvard.edu/abs/2019ApJ...882L..12P} {882, L12}

\bibitem[\protect\citeauthoryear{{Peng} et~al.,}{{Peng}
  et~al.}{2010}]{Peng2010}
{Peng} Y.-j.,  et~al., 2010, \mn@doi [\apj] {10.1088/0004-637X/721/1/193},
  \href {https://ui.adsabs.harvard.edu/abs/2010ApJ...721..193P} {721, 193}

\bibitem[\protect\citeauthoryear{{Peng}, {Lilly}, {Renzini}  \&
  {Carollo}}{{Peng} et~al.}{2012}]{Peng2012}
{Peng} Y.-j.,  {Lilly} S.~J.,  {Renzini} A.,   {Carollo} M.,  2012, \mn@doi
  [\apj] {10.1088/0004-637X/757/1/4}, \href
  {https://ui.adsabs.harvard.edu/abs/2012ApJ...757....4P} {757, 4}

\bibitem[\protect\citeauthoryear{P\'erez \& Granger}{P\'erez \&
  Granger}{2007}]{ipython}
P\'erez F.,  Granger B.~E.,  2007, \mn@doi [Computing in Science and
  Engineering] {10.1109/MCSE.2007.53}, 9, 21

\bibitem[\protect\citeauthoryear{{P{\'e}rez-Carrasco}, {Cabrera-Vives},
  {Martinez-Marin}, {Cerulo}, {Demarco}, {Protopapas}, {Godoy}  \&
  {Huertas-Company}}{{P{\'e}rez-Carrasco} et~al.}{2019}]{Perez2019}
{P{\'e}rez-Carrasco} M.,  {Cabrera-Vives} G.,  {Martinez-Marin} M.,  {Cerulo}
  P.,  {Demarco} R.,  {Protopapas} P.,  {Godoy} J.,   {Huertas-Company} M.,
  2019, \mn@doi [\pasp] {10.1088/1538-3873/aaeeb4}, \href
  {https://ui.adsabs.harvard.edu/abs/2019PASP..131j8002P} {131, 108002}

\bibitem[\protect\citeauthoryear{{Petrillo} et~al.,}{{Petrillo}
  et~al.}{2017}]{Petrillo2017}
{Petrillo} C.~E.,  et~al., 2017, \mn@doi [\mnras] {10.1093/mnras/stx2052},
  \href {https://ui.adsabs.harvard.edu/abs/2017MNRAS.472.1129P} {472, 1129}

\bibitem[\protect\citeauthoryear{{Petrillo} et~al.,}{{Petrillo}
  et~al.}{2019}]{Petrillo2019}
{Petrillo} C.~E.,  et~al., 2019, \mn@doi [\mnras] {10.1093/mnras/stz189}, \href
  {https://ui.adsabs.harvard.edu/abs/2019MNRAS.484.3879P} {484, 3879}

\bibitem[\protect\citeauthoryear{{Pillepich} et~al.,}{{Pillepich}
  et~al.}{2018}]{Pillepich2018}
{Pillepich} A.,  et~al., 2018, \mn@doi [\mnras] {10.1093/mnras/stx2656}, \href
  {http://adsabs.harvard.edu/abs/2018MNRAS.473.4077P} {473, 4077}

\bibitem[\protect\citeauthoryear{{Price-Whelan} et~al.,}{{Price-Whelan}
  et~al.}{2018}]{astropy2018}
{Price-Whelan} A.~M.,  et~al., 2018, \mn@doi [\aj] {10.3847/1538-3881/aabc4f},
  \href {https://ui.adsabs.harvard.edu/#abs/2018AJ....156..123T} {156, 123}

\bibitem[\protect\citeauthoryear{{Ramanah}, {Lavaux}, {Jasche}  \& {Wand
  elt}}{{Ramanah} et~al.}{2019}]{Ramanah2019}
{Ramanah} D.~K.,  {Lavaux} G.,  {Jasche} J.,   {Wand elt} B.~D.,  2019, \mn@doi
  [\aap] {10.1051/0004-6361/201834117}, \href
  {https://ui.adsabs.harvard.edu/abs/2019A&A...621A..69R} {621, A69}

\bibitem[\protect\citeauthoryear{{Ratcliffe}, {Ness}, {Johnston}  \&
  {Sen}}{{Ratcliffe} et~al.}{2020}]{Ratcliffe2020}
{Ratcliffe} B.~L.,  {Ness} M.~K.,  {Johnston} K.~V.,   {Sen} B.,  2020, \mn@doi
  [\apj] {10.3847/1538-4357/abac61}, \href
  {https://ui.adsabs.harvard.edu/abs/2020ApJ...900..165R} {900, 165}

\bibitem[\protect\citeauthoryear{{Ratcliffe}, {Ness}, {Buck}, {Johnston},
  {Sen}, {Beraldo e Silva}  \& {Debattista}}{{Ratcliffe}
  et~al.}{2021}]{Ratcliffe2021}
{Ratcliffe} B.~L.,  {Ness} M.~K.,  {Buck} T.,  {Johnston} K.~V.,  {Sen} B.,
  {Beraldo e Silva} L.,   {Debattista} V.~P.,  2021, arXiv e-prints, \href
  {https://ui.adsabs.harvard.edu/abs/2021arXiv210708088R} {p. arXiv:2107.08088}

\bibitem[\protect\citeauthoryear{{Roediger} \& {Courteau}}{{Roediger} \&
  {Courteau}}{2015}]{Roediger2015}
{Roediger} J.~C.,  {Courteau} S.,  2015, \mn@doi [\mnras]
  {10.1093/mnras/stv1499}, \href
  {https://ui.adsabs.harvard.edu/abs/2015MNRAS.452.3209R} {452, 3209}

\bibitem[\protect\citeauthoryear{{Ronneberger}, {Fischer}  \&
  {Brox}}{{Ronneberger} et~al.}{2015}]{Ronneberger2015}
{Ronneberger} O.,  {Fischer} P.,   {Brox} T.,  2015, arXiv e-prints, \href
  {https://ui.adsabs.harvard.edu/abs/2015arXiv150504597R} {p. arXiv:1505.04597}

\bibitem[\protect\citeauthoryear{{S{\'a}nchez} et~al.,}{{S{\'a}nchez}
  et~al.}{2012}]{califa2012}
{S{\'a}nchez} S.~F.,  et~al., 2012, \mn@doi [\aap]
  {10.1051/0004-6361/201117353}, \href
  {https://ui.adsabs.harvard.edu/abs/2012A&A...538A...8S} {538, A8}

\bibitem[\protect\citeauthoryear{{Schaye} et~al.,}{{Schaye}
  et~al.}{2015}]{Schaye2015}
{Schaye} J.,  et~al., 2015, \mn@doi [\mnras] {10.1093/mnras/stu2058}, \href
  {http://adsabs.harvard.edu/abs/2015MNRAS.446..521S} {446, 521}

\bibitem[\protect\citeauthoryear{{Smirnov} \& {Markov}}{{Smirnov} \&
  {Markov}}{2017}]{Smirnov2017}
{Smirnov} E.~A.,  {Markov} A.~B.,  2017, \mn@doi [\mnras]
  {10.1093/mnras/stx999}, \href
  {https://ui.adsabs.harvard.edu/abs/2017MNRAS.469.2024S} {469, 2024}

\bibitem[\protect\citeauthoryear{{Snyder} et~al.,}{{Snyder}
  et~al.}{2015}]{Snyder2015}
{Snyder} G.~F.,  et~al., 2015, \mn@doi [\mnras] {10.1093/mnras/stv2078}, \href
  {https://ui.adsabs.harvard.edu/abs/2015MNRAS.454.1886S} {454, 1886}

\bibitem[\protect\citeauthoryear{{Soo} et~al.,}{{Soo} et~al.}{2018}]{Soo2018}
{Soo} J. Y.~H.,  et~al., 2018, \mn@doi [\mnras] {10.1093/mnras/stx3201}, \href
  {https://ui.adsabs.harvard.edu/abs/2018MNRAS.475.3613S} {475, 3613}

\bibitem[\protect\citeauthoryear{{Springel}}{{Springel}}{2010}]{Springel2010}
{Springel} V.,  2010, \mn@doi [\mnras] {10.1111/j.1365-2966.2009.15715.x},
  \href {http://adsabs.harvard.edu/abs/2010MNRAS.401..791S} {401, 791}

\bibitem[\protect\citeauthoryear{{Strateva} et~al.,}{{Strateva}
  et~al.}{2001}]{Strateva2001}
{Strateva} I.,  et~al., 2001, \mn@doi [\aj] {10.1086/323301}, \href
  {https://ui.adsabs.harvard.edu/abs/2001AJ....122.1861S} {122, 1861}

\bibitem[\protect\citeauthoryear{{Strauss} et~al.,}{{Strauss}
  et~al.}{2002}]{Strauss2002}
{Strauss} M.~A.,  et~al., 2002, \mn@doi [\aj] {10.1086/342343}, \href
  {https://ui.adsabs.harvard.edu/abs/2002AJ....124.1810S} {124, 1810}

\bibitem[\protect\citeauthoryear{{The Dark Energy Survey Collaboration}}{{The
  Dark Energy Survey Collaboration}}{2005}]{DES2005}
{The Dark Energy Survey Collaboration} 2005, arXiv e-prints, \href
  {https://ui.adsabs.harvard.edu/abs/2005astro.ph.10346T} {pp
  astro--ph/0510346}

\bibitem[\protect\citeauthoryear{{Torrey} et~al.,}{{Torrey}
  et~al.}{2015}]{Torrey2015}
{Torrey} P.,  et~al., 2015, \mn@doi [\mnras] {10.1093/mnras/stu2592}, \href
  {https://ui.adsabs.harvard.edu/abs/2015MNRAS.447.2753T} {447, 2753}

\bibitem[\protect\citeauthoryear{{Tremonti} et~al.,}{{Tremonti}
  et~al.}{2004}]{Tremonti2004}
{Tremonti} C.~A.,  et~al., 2004, \mn@doi [\apj] {10.1086/423264}, \href
  {https://ui.adsabs.harvard.edu/abs/2004ApJ...613..898T} {613, 898}

\bibitem[\protect\citeauthoryear{{Van Oort}, {Xu}, {Offner}  \&
  {Gutermuth}}{{Van Oort} et~al.}{2019}]{vanOort2019}
{Van Oort} C.~M.,  {Xu} D.,  {Offner} S. S.~R.,   {Gutermuth} R.~A.,  2019,
  \mn@doi [\apj] {10.3847/1538-4357/ab275e}, \href
  {https://ui.adsabs.harvard.edu/abs/2019ApJ...880...83V} {880, 83}

\bibitem[\protect\citeauthoryear{{Villaescusa-Navarro}
  et~al.,}{{Villaescusa-Navarro} et~al.}{2021}]{Villaescusa-Navarro2021}
{Villaescusa-Navarro} F.,  et~al., 2021, arXiv e-prints, \href
  {https://ui.adsabs.harvard.edu/abs/2021arXiv210909747V} {p. arXiv:2109.09747}

\bibitem[\protect\citeauthoryear{{Vogelsberger}, {Genel}, {Sijacki}, {Torrey},
  {Springel}  \& {Hernquist}}{{Vogelsberger} et~al.}{2013}]{Vogelsberger2013}
{Vogelsberger} M.,  {Genel} S.,  {Sijacki} D.,  {Torrey} P.,  {Springel} V.,
  {Hernquist} L.,  2013, \mn@doi [\mnras] {10.1093/mnras/stt1789}, \href
  {https://ui.adsabs.harvard.edu/abs/2013MNRAS.436.3031V} {436, 3031}

\bibitem[\protect\citeauthoryear{{Vogelsberger} et~al.,}{{Vogelsberger}
  et~al.}{2014}]{Vogelsberger2014}
{Vogelsberger} M.,  et~al., 2014, \mn@doi [\mnras] {10.1093/mnras/stu1536},
  \href {http://adsabs.harvard.edu/abs/2014MNRAS.444.1518V} {444, 1518}

\bibitem[\protect\citeauthoryear{Walt, Colbert  \& Varoquaux}{Walt
  et~al.}{2011}]{numpy}
Walt S. v.~d.,  Colbert S.~C.,   Varoquaux G.,  2011, \mn@doi [Computing in
  Science and Engg.] {10.1109/MCSE.2011.37}, 13, 22

\bibitem[\protect\citeauthoryear{{Wang}, {Dutton}, {Stinson}, {Macci{\`o}},
  {Penzo}, {Kang}, {Keller}  \& {Wadsley}}{{Wang} et~al.}{2015}]{Wang2015}
{Wang} L.,  {Dutton} A.~A.,  {Stinson} G.~S.,  {Macci{\`o}} A.~V.,  {Penzo} C.,
   {Kang} X.,  {Keller} B.~W.,   {Wadsley} J.,  2015, \mn@doi [\mnras]
  {10.1093/mnras/stv1937}, \href
  {http://adsabs.harvard.edu/abs/2015MNRAS.454...83W} {454, 83}

\bibitem[\protect\citeauthoryear{{Wang}, {Ma}, {Li}  \& {Xia}}{{Wang}
  et~al.}{2020}]{Wang2020}
{Wang} G.-J.,  {Ma} X.-J.,  {Li} S.-Y.,   {Xia} J.-Q.,  2020, \mn@doi [\apjs]
  {10.3847/1538-4365/ab620b}, \href
  {https://ui.adsabs.harvard.edu/abs/2020ApJS..246...13W} {246, 13}

\bibitem[\protect\citeauthoryear{{Wilman}, {Zibetti}  \&
  {Budav{\'a}ri}}{{Wilman} et~al.}{2010}]{Wilman2010}
{Wilman} D.~J.,  {Zibetti} S.,   {Budav{\'a}ri} T.,  2010, \mn@doi [\mnras]
  {10.1111/j.1365-2966.2010.16845.x}, \href
  {https://ui.adsabs.harvard.edu/abs/2010MNRAS.406.1701W} {406, 1701}

\bibitem[\protect\citeauthoryear{{Wilson}, {Nayyeri}, {Cooray}  \&
  {H{\"a}u{\ss}ler}}{{Wilson} et~al.}{2020}]{Wilson2020}
{Wilson} D.,  {Nayyeri} H.,  {Cooray} A.,   {H{\"a}u{\ss}ler} B.,  2020,
  \mn@doi [\apj] {10.3847/1538-4357/ab5a79}, \href
  {https://ui.adsabs.harvard.edu/abs/2020ApJ...888...83W} {888, 83}

\bibitem[\protect\citeauthoryear{{Wu} \& {Boada}}{{Wu} \&
  {Boada}}{2019}]{Wu2019}
{Wu} J.~F.,  {Boada} S.,  2019, \mn@doi [\mnras] {10.1093/mnras/stz333}, \href
  {https://ui.adsabs.harvard.edu/abs/2019MNRAS.484.4683W} {484, 4683}

\bibitem[\protect\citeauthoryear{{Xu}, {Ho}, {Trac}, {Schneider}, {Poczos}  \&
  {Ntampaka}}{{Xu} et~al.}{2013}]{Xu2013}
{Xu} X.,  {Ho} S.,  {Trac} H.,  {Schneider} J.,  {Poczos} B.,   {Ntampaka} M.,
  2013, \mn@doi [\apj] {10.1088/0004-637X/772/2/147}, \href
  {https://ui.adsabs.harvard.edu/abs/2013ApJ...772..147X} {772, 147}

\bibitem[\protect\citeauthoryear{{Yesuf}, {Ho}  \& {Faber}}{{Yesuf}
  et~al.}{2021}]{Yesuf2021}
{Yesuf} H.~M.,  {Ho} L.~C.,   {Faber} S.~M.,  2021, arXiv e-prints, \href
  {https://ui.adsabs.harvard.edu/abs/2021arXiv210908882Y} {p. arXiv:2109.08882}

\bibitem[\protect\citeauthoryear{Zeiler \& Fergus}{Zeiler \&
  Fergus}{2014}]{Zeiler2014}
Zeiler M.,  Fergus R.,  2014, in Computer Vision, ECCV 2014 - 13th European
  Conference, Proceedings. Springer Verlag, pp 818--833,
  \mn@doi{10.1007/978-3-319-10590-1_53}

\bibitem[\protect\citeauthoryear{{Zibetti}, {Charlot}  \& {Rix}}{{Zibetti}
  et~al.}{2009}]{Zibetti2009}
{Zibetti} S.,  {Charlot} S.,   {Rix} H.-W.,  2009, \mn@doi [\mnras]
  {10.1111/j.1365-2966.2009.15528.x}, \href
  {https://ui.adsabs.harvard.edu/abs/2009MNRAS.400.1181Z} {400, 1181}

\bibitem[\protect\citeauthoryear{{de Diego} et~al.,}{{de Diego}
  et~al.}{2021}]{Diego2021}
{de Diego} J.~A.,  et~al., 2021, arXiv e-prints, \href
  {https://ui.adsabs.harvard.edu/abs/2021arXiv210809415D} {p. arXiv:2108.09415}

\makeatother
\end{thebibliography}

\appendix
\section{Reconstruction with fewer wavelength bands}
\label{sec:ap_bands}

The accuracy of the galaxy reconstruction depends on the number of available wavelength bands. 

\begin{figure}
\vspace{-.4cm}
\begin{center}
\includegraphics[height=.5\textheight]{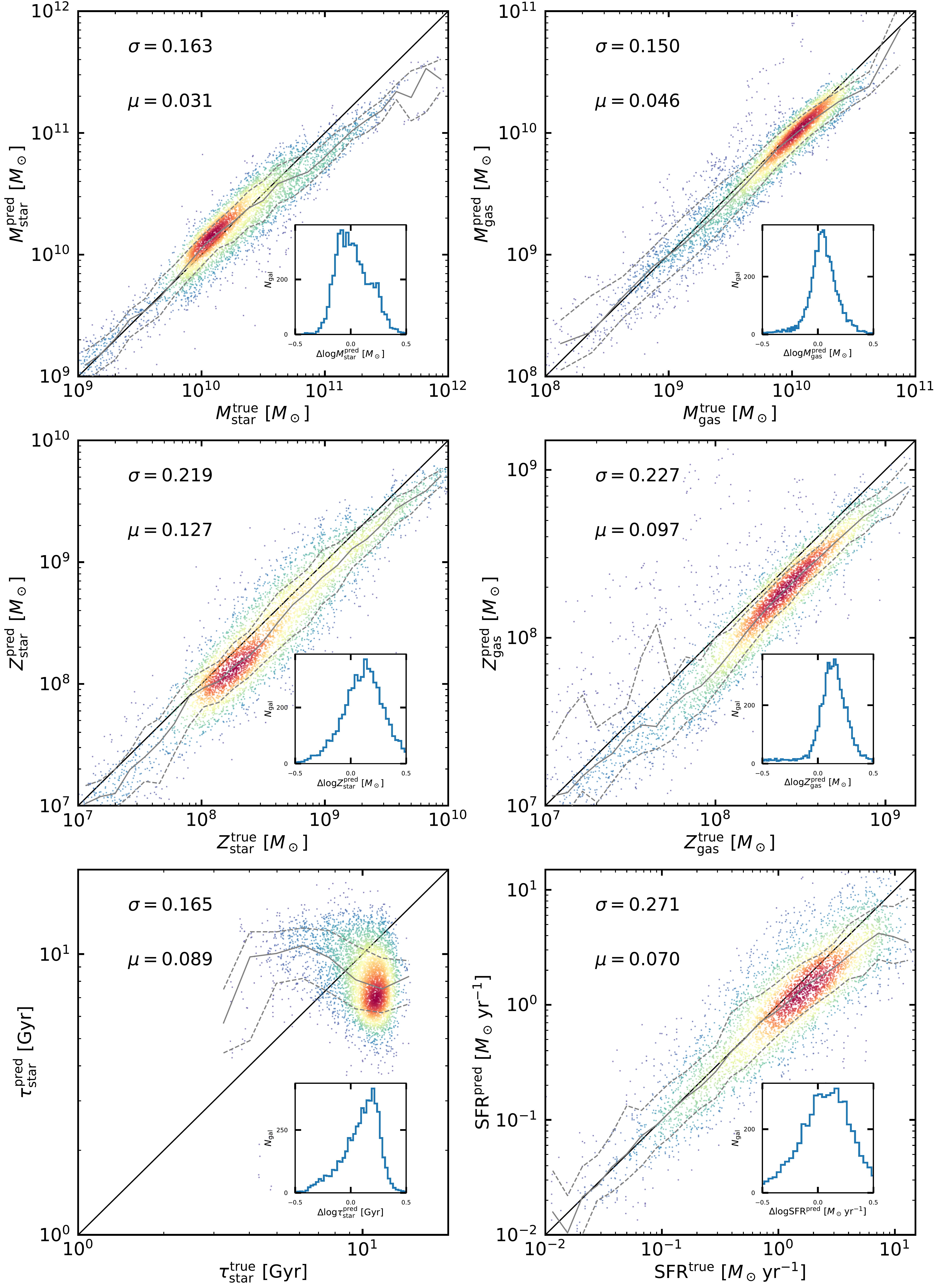}
\end{center}
\vspace{-.5cm}
\oscaption{Analysis_ugri}{Same as Fig. \ref{fig:true_vs_pred} but for training only on the 4 SDSS \textit{ugri} bands.}
\label{fig:true_vs_pred2}
\end{figure}

\begin{figure}
\vspace{-.4cm}
\begin{center}
\includegraphics[height=.5\textheight]{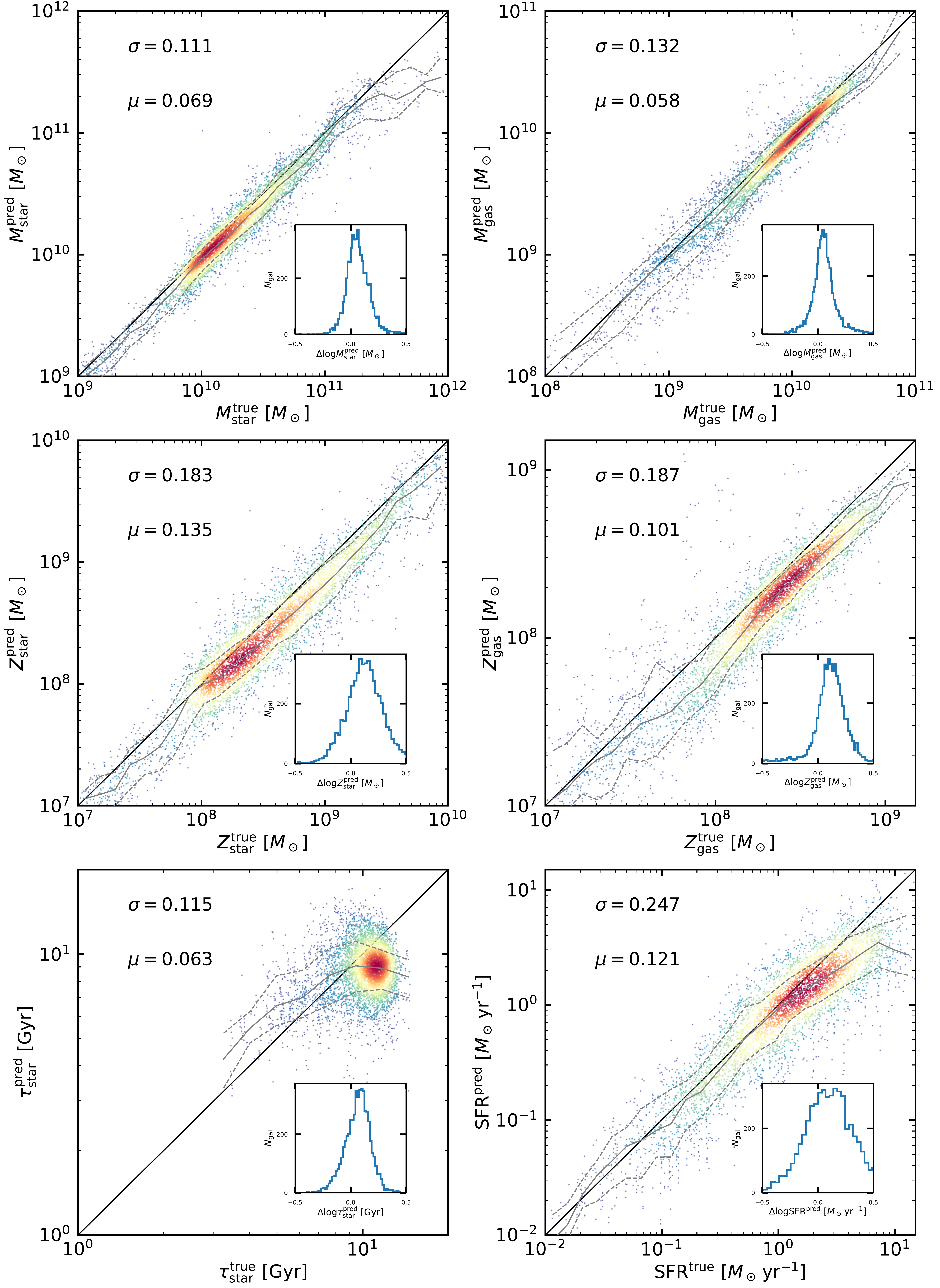}
\end{center}
\vspace{-.5cm}
\oscaption{Analysis_gr}{Same as Fig. \ref{fig:true_vs_pred} but for training only on the 2 SDSS \textit{gr} bands.}
\label{fig:true_vs_pred3}
\end{figure}

\begin{figure}
\vspace{-.4cm}
\begin{center}
\includegraphics[height=.5\textheight]{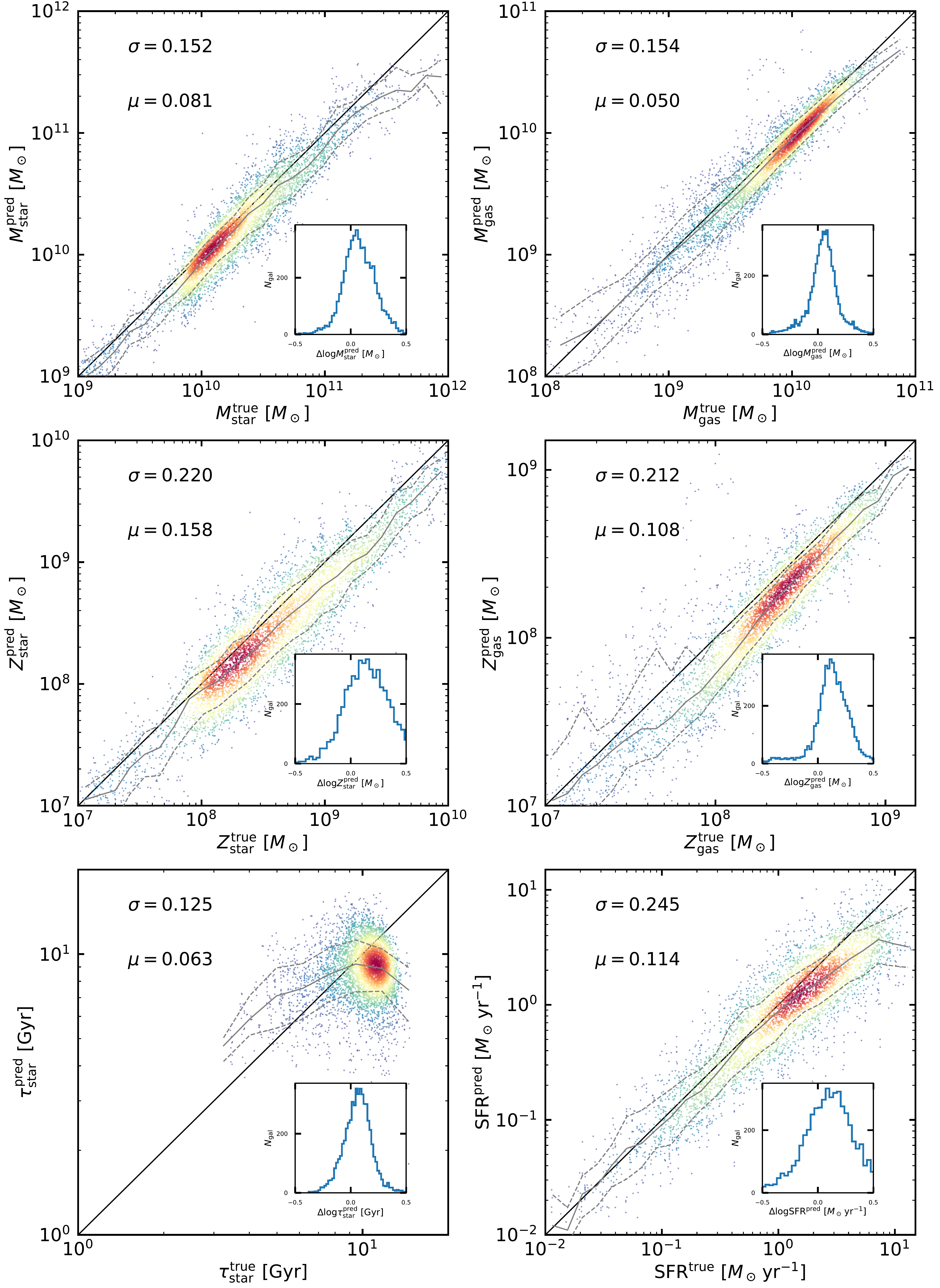}
\end{center}
\vspace{-.5cm}
\oscaption{Analysis_r}{Same as Fig. \ref{fig:true_vs_pred} but for training only on the 1 SDSS \textit{r} bands.}
\label{fig:true_vs_pred4}
\end{figure}

\section{Reconstruction with lower image resolution}
\label{sec:ap_res}

\begin{figure*}
\vspace{-.25cm}
\begin{center}
\includegraphics[width=.49\textwidth]{./plots/2096_masked.pdf}
\includegraphics[width=.49\textwidth]{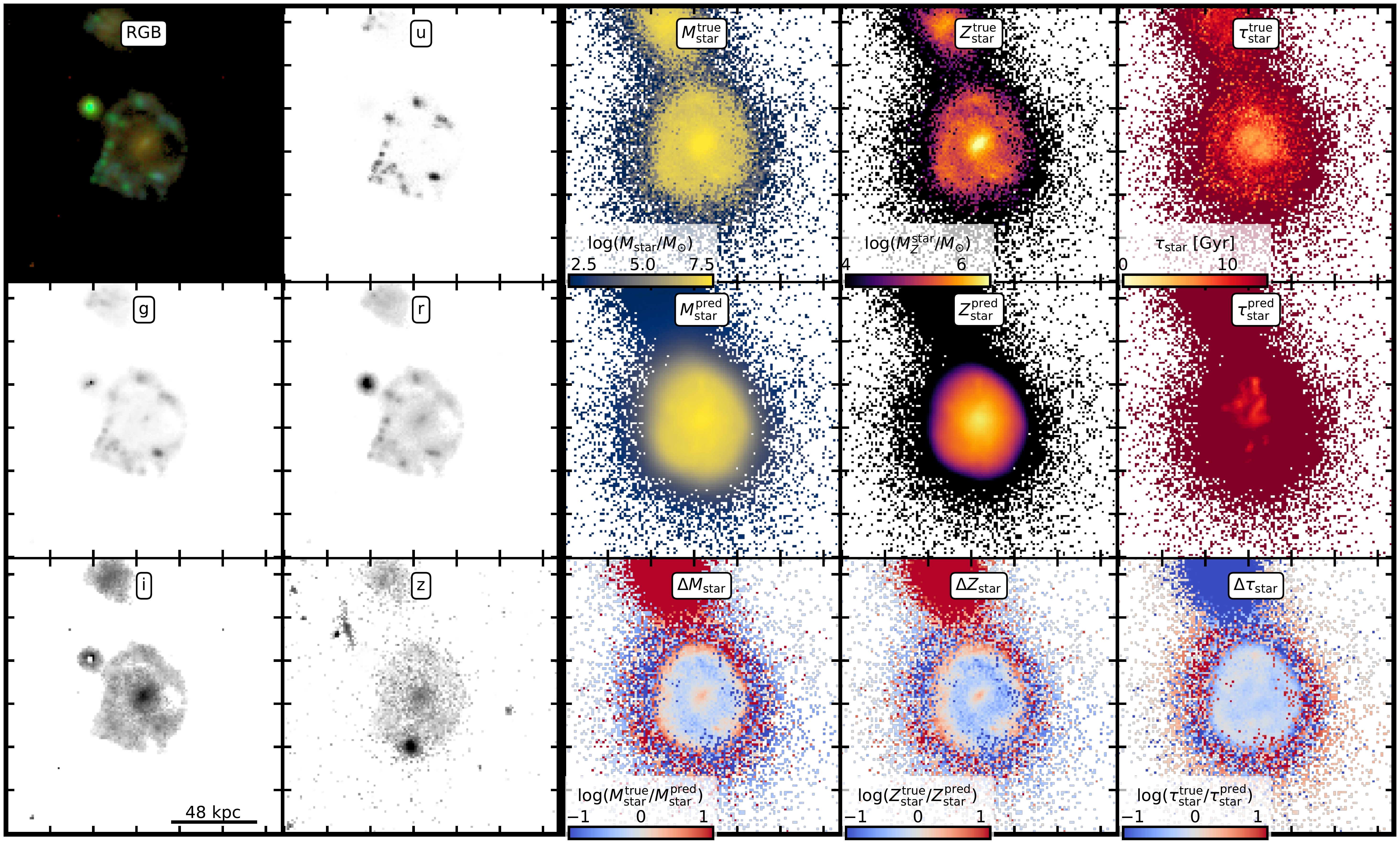}
\includegraphics[width=.49\textwidth]{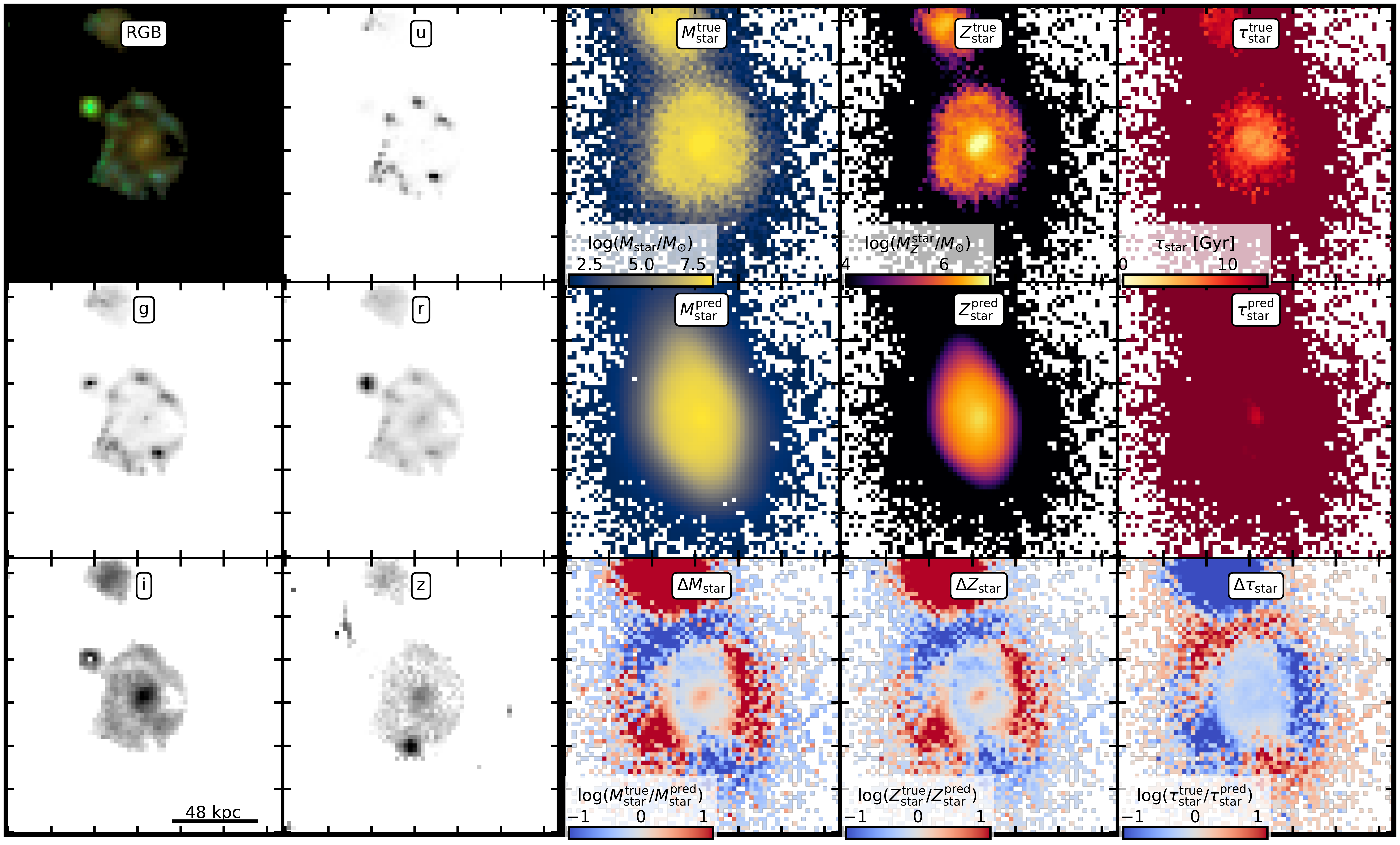}
\includegraphics[width=.49\textwidth]{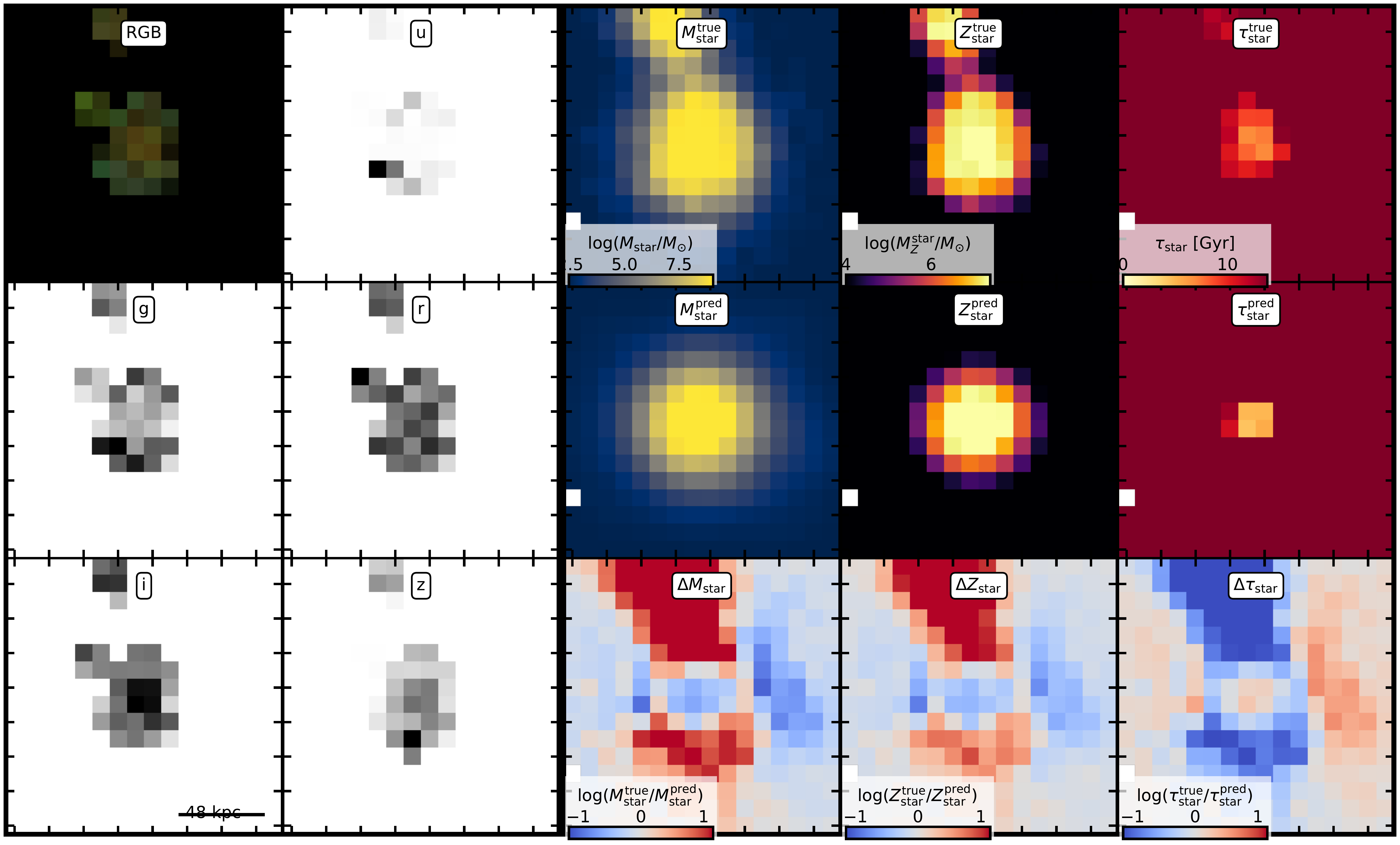}
\end{center}
\vspace{-.35cm}
\oscaption{Analysis_2}{Same as Fig. \ref{fig:images} but comparing different resolutions. The upper left panels show the fiducial resolution of 256x256 pixels, the upper right panels have 128x128 pixels and the lower left and right panels have 64x64 and 16x16 pixels, respectively.
}
\label{fig:app_res_comp}
\end{figure*}

\begin{figure*}
\vspace{-.25cm}
\begin{center}
\includegraphics[width=.49\textwidth]{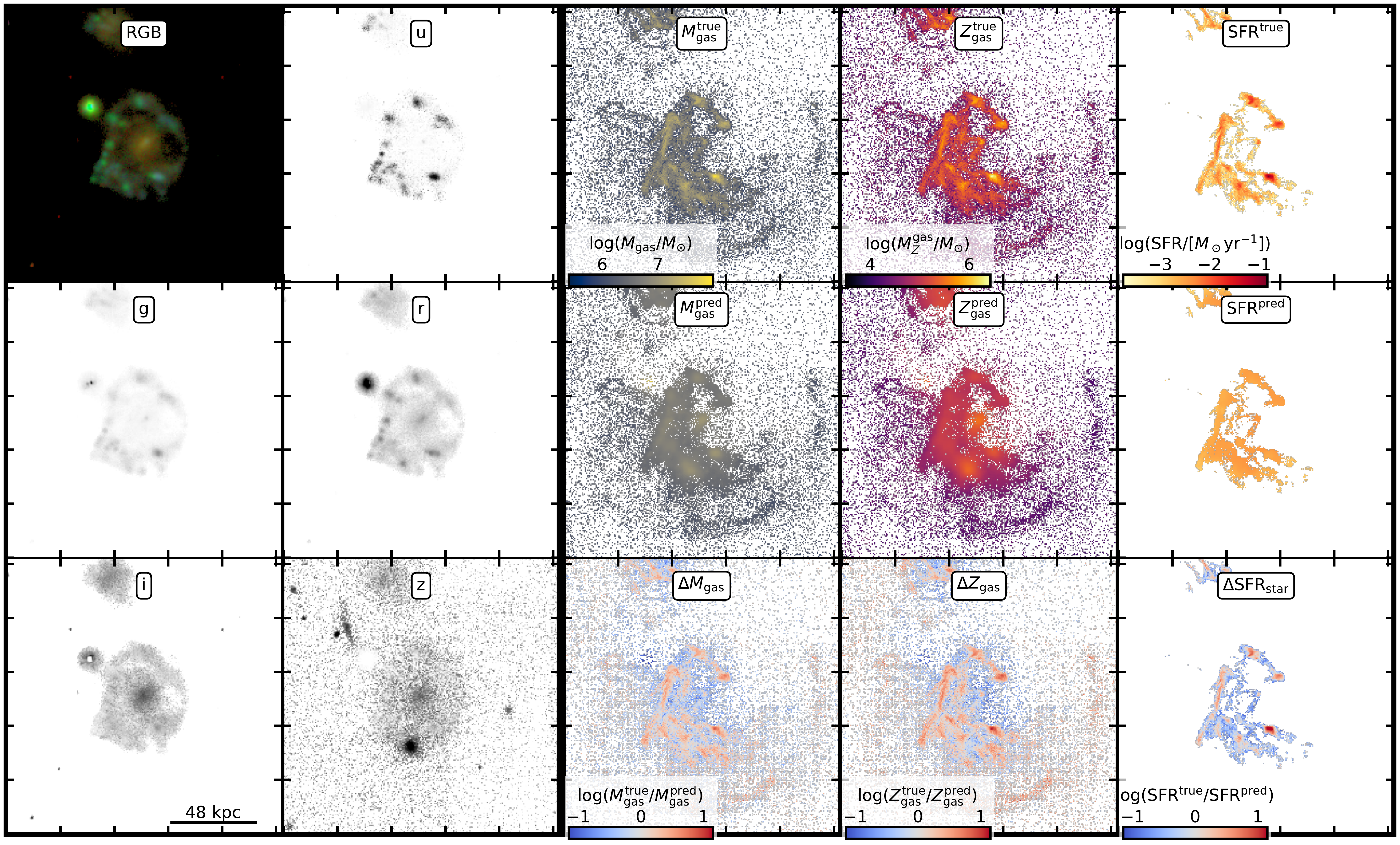}
\includegraphics[width=.49\textwidth]{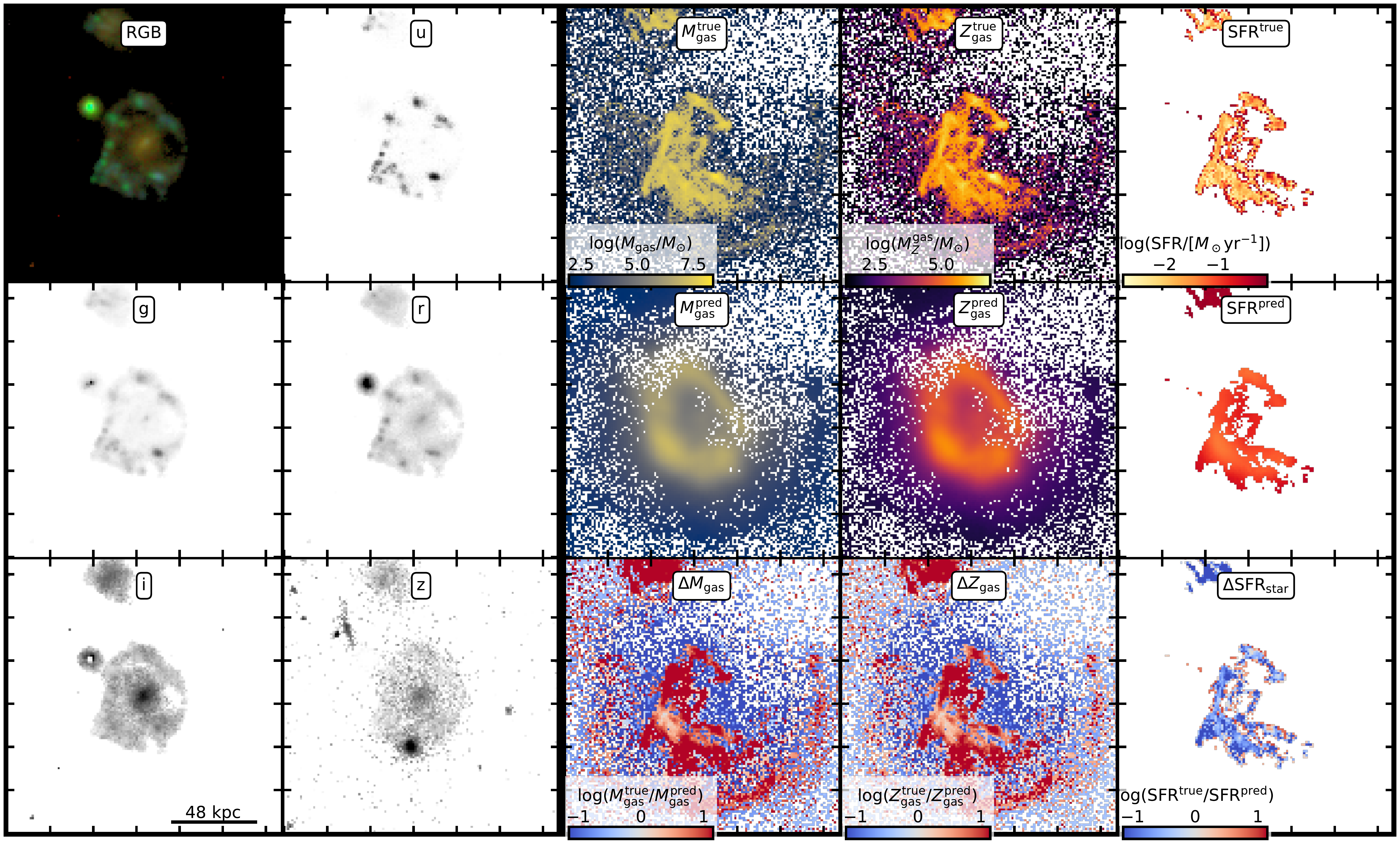}
\includegraphics[width=.49\textwidth]{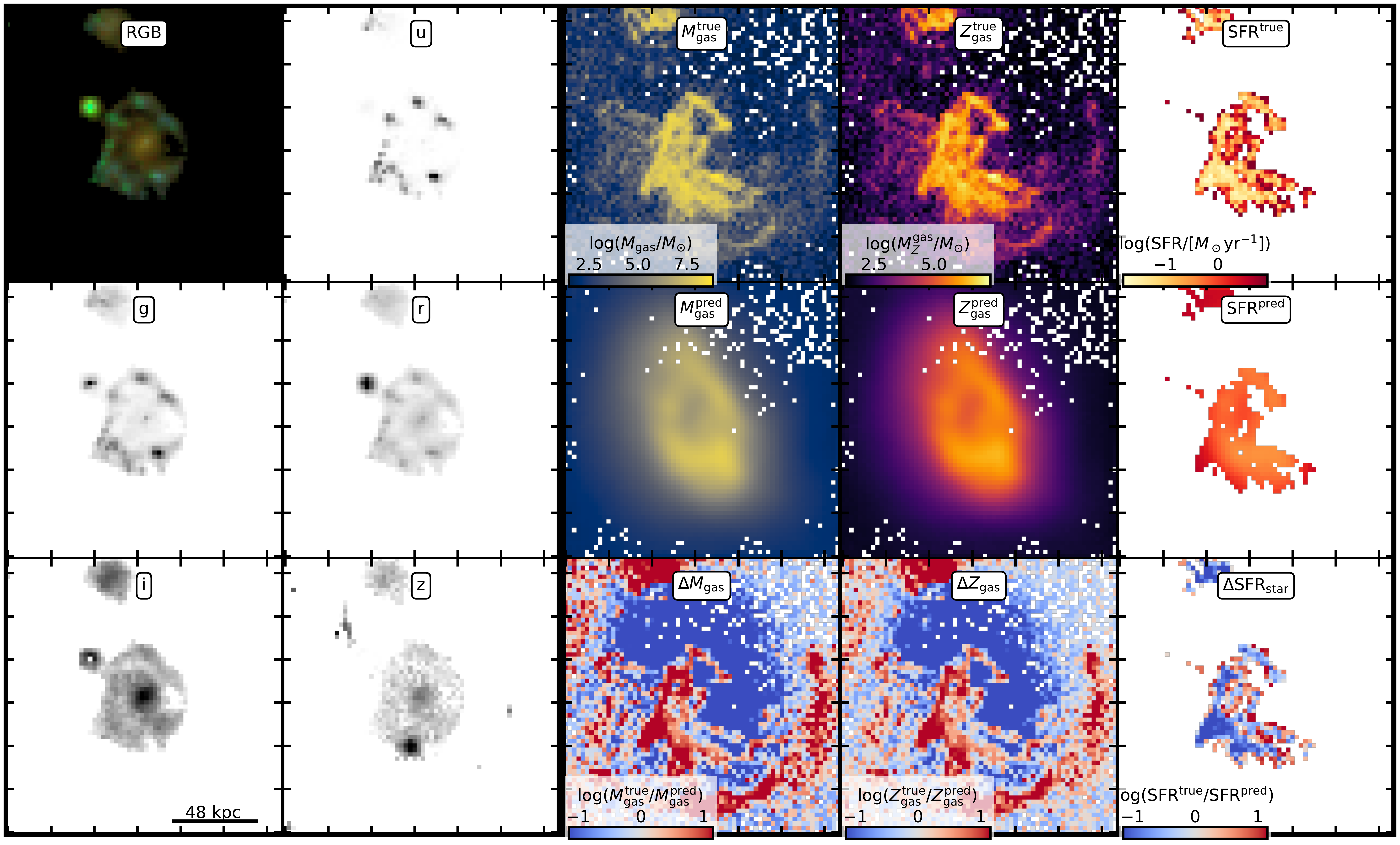}
\includegraphics[width=.49\textwidth]{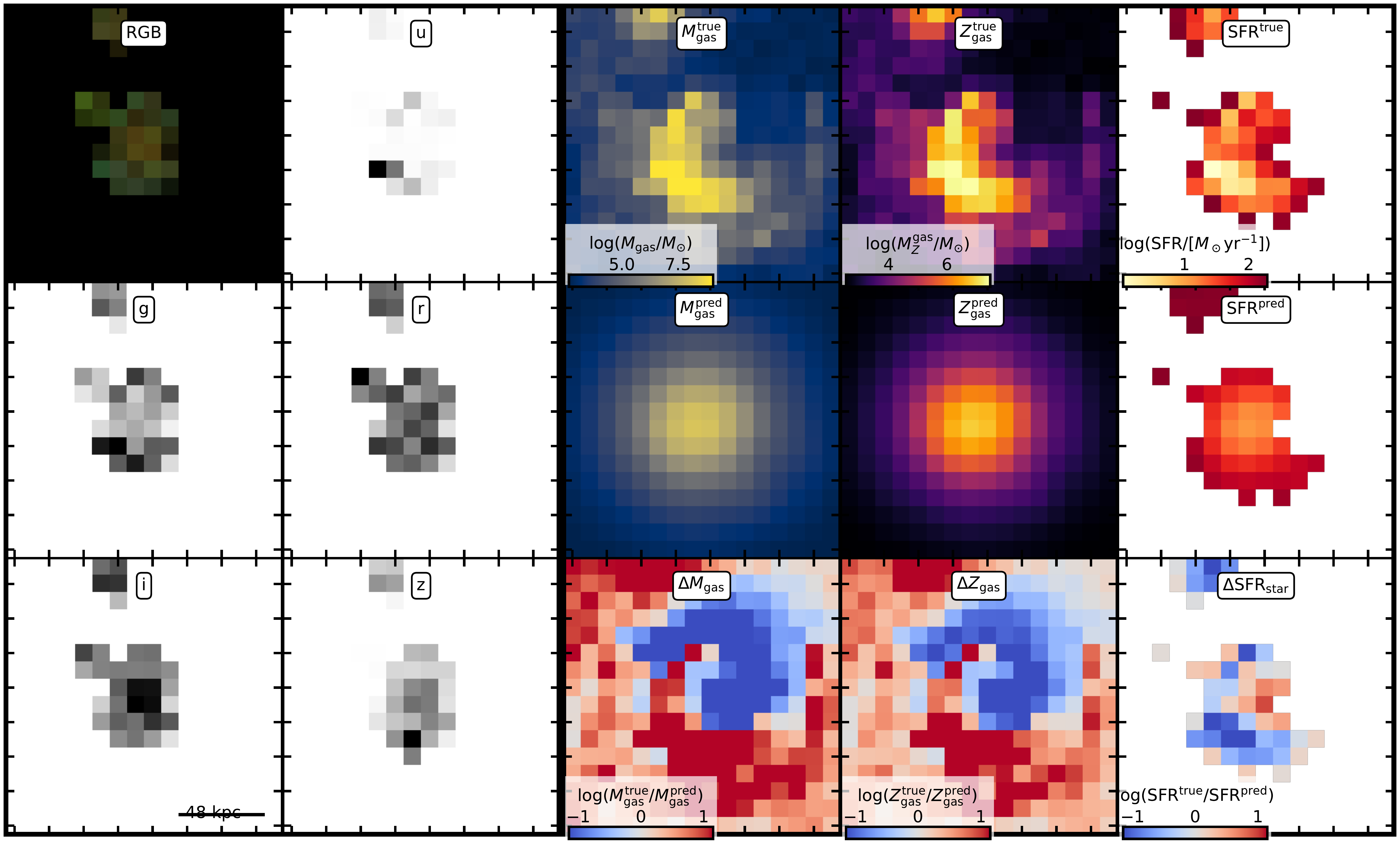}
\end{center}
\vspace{-.35cm}
\oscaptionpy{maps_gas}{Similar to Fig. \ref{fig:images} but this time showing gas properties instead of stellar properties as indicated by the labels. This figure exemplifies that the more complex gas morphology in comparison to the stellar body is less well predicted if image resolution decreases.
}
\label{fig:app_res_comp_gas}
\end{figure*}

The neural network is able to identify morphological features of galaxies and might be able to use those for the reconstruction of galaxy properties. Thus, the accuracy of the galaxy reconstruction can depend on the image resolution. In Fig. \ref{fig:app_res_comp} we show a visual comparison between the four different resolutions tested in this work. The upper left panels show the fiducial resolution of 256x256 pixels, the upper right panels have 128x128 pixels and the lower left and right panels have 64x64 and 16x16 pixels, respectively. Additionally, in Fig. \ref{fig:app_res_comp_gas} we show for the same example galaxy gas properties for the different images resolutions.

\label{lastpage}

\end{document}